%% file: main.tex
\documentclass[letterpaper,11pt]{article}
\pdfoutput=1

\usepackage{xcolor}
\usepackage{jheppub}
\usepackage{mathtools}
\usepackage{dsfont}
\usepackage{braket}
\usepackage{appendix}
\usepackage{subcaption}
\usepackage{placeins}
\usepackage{tikz}
\usetikzlibrary{decorations.markings,decorations.pathmorphing,decorations.pathreplacing}
\tikzset{
	one loop probe/.style={
		line width=0.75pt
	},
	one loop momentum/.style={
		postaction={decorate},
		decoration={
			markings,
			mark=at position 0.50 with {\arrow{>}}
		}
	},
	one loop outgoing momentum/.style={
		postaction={decorate},
		decoration={
			markings,
			mark=at position 0.58 with {\arrow{>}}
		}
	},
	one loop photon/.style={
		decorate,
		decoration={
			snake,
			amplitude=1pt,
			segment length=4.5pt
		},
		line width=0.55pt
	},
	one loop source/.style={
		font=\small,
		inner sep=0pt
	},
	one loop coupling/.style={
		font=\normalsize,
		inner sep=1pt
	},
	one loop momentum label/.style={
		font=\large,
		inner sep=1pt
	}
}
\newcommand{\ICSdiagram}{%
	\begin{tikzpicture}[scale=0.55,transform shape,
		baseline={(current bounding box.center)}]
		\coordinate (left)  at (-1.15,0);
		\coordinate (right) at ( 1.15,0);
		\draw[one loop probe,one loop momentum] (-2.15,0.95) -- (left);
		\draw[one loop probe,one loop momentum] (left) -- (right);
		\draw[one loop probe,one loop outgoing momentum] (right) -- (2.15,0.95);
		\node[one loop momentum label,font=\Large,above left=1pt]  at (-2.15,0.95) {$\vec p_1$};
		\node[one loop momentum label,font=\Large,above=3pt]       at (0,0) {$\vec\ell$};
		\node[one loop momentum label,font=\Large,above right=1pt] at (2.15,0.95) {$\vec p_2$};
		\fill (left) circle (1.5pt);
		\node[one loop coupling,font=\LARGE,anchor=south] at (-0.88,0.14) {$\beta^2$};
		\node[one loop source] (sla) at (-1.42,-1.35) {$\otimes$};
		\node[one loop source] (slb) at (-0.88,-1.35) {$\otimes$};
		\node[one loop source] (sr)  at ( 1.15,-1.35) {$\otimes$};
		\draw[one loop photon] (left)  -- (sla.north);
		\draw[one loop photon] (left)  -- (slb.north);
		\draw[one loop photon] (right) -- (sr.north);
\end{tikzpicture}}

\usepackage[linewidth=1pt]{mdframed}

\newcommand{\DD}{\mathrm{d}}

\newcommand{\defined}{\coloneqq}

\renewcommand{\t}[1]{\text{#1}}

\title{Unitarity and the Forward Direction in Theories with Long-Range Forces}

\author[a]{Luke Lippstreu}

\affiliation[a]{Higgs Centre, School of Physics \& Astronomy \\
	University of Edinburgh, EH9 3FD, United Kingdom}
\emailAdd{llippstr@ed.ac.uk}

\abstract{Integrating scattering amplitudes over the forward direction, and consequently the unitarity constraints one extracts from such integrals, can appear ambiguous in theories with long-range forces. Standard techniques for bounding EFT couplings then typically produce bounds that depend on an arbitrary infrared scale. We study a non-relativistic model in which the standard techniques produce such an ambiguous bound, but which is simple enough that the exact bound can also be derived non-perturbatively and shown to involve no infrared scale. We then show how to derive bounds perturbatively, with no infrared scale entering at any stage. This requires two ingredients: accounting for the modified distributional structure of long-range amplitudes, and using distorted-wave perturbation theory (DWPT), which treats the Coulomb dynamics exactly. Together they give amplitudes with well-defined partial-wave projections and no spurious infrared divergences. Interpreting the model as an EFT with an ultraviolet cutoff $\Lambda_{\rm EFT}$, we derive cutoff-dependent bounds valid for arbitrary UV completions. Taking $\Lambda_{\rm EFT}\to\infty$ at each order yields bounds that rapidly converge to the exact bound, reaching $0.002\%$ accuracy at fourth order in the expansion. Finally, we demonstrate that every order of the DWPT expansion resums infinitely many Feynman diagrams of short-range perturbation theory, which individually evaluate to multiple polylogarithms and complete elliptic integrals, into a compact expression. At the orders we compute, no integration is even required, suggesting that DWPT may offer a simpler representation of scattering amplitudes than standard perturbation theory.}

\begin{document}

	\maketitle
	
	\section{Introduction}

	Causality and unitarity are key structural principles of relativistic quantum field theory. In theories with a mass gap, these principles imply strong constraints on the analytic properties of scattering amplitudes, which in turn place bounds on the low-energy couplings of effective field theories (EFTs) that admit causal, unitary, and crossing-symmetric UV completions \cite{Pham:1985cr,Ananthanarayan:1994hf,Adams:2006sv, Camanho:2014apa,Bellazzini:2014waa,Cheung:2016yqr, deRham:2017avq,deRham:2017zjm,Bellazzini:2020cot, Tolley:2020gtv,Caron-Huot:2020cmc,Arkani-Hamed:2020blm, Caron-Huot:2021rmr,Caron-Huot:2022ugt}; see \cite{deRham:2022hpx} for a review. However, in theories that are most relevant to nature, namely those with long-range forces such as four-dimensional QED and gravity, these principles are less well understood than their short-range counterparts. In these theories, massless exchanges produce a $1/t$ pole that is non-integrable in the forward direction along with infrared divergences. These features obstruct a direct application of the standard S-matrix arguments leading to bounds on EFT couplings.

In this paper, we show, in a controlled non-relativistic model, how unitarity bounds can be derived in the presence of a long-range force without introducing arbitrary infrared scales or relying on ill-defined mathematical steps. Achieving this requires two modifications to standard techniques. First, one needs to account for the modified distributional structure of long-range amplitudes. Second, one needs to employ a perturbative framework that incorporates the asymptotic dynamics; in our case, this is distorted-wave perturbation theory (DWPT).
    
	\subsection{Summary}
	We study one of the simplest models that exhibit some of the central problems encountered when standard techniques for deriving EFT bounds are applied to theories with long-range forces: the Coulomb Hamiltonian in three spatial dimensions, supplemented by an attractive inverse-square interaction,
	\begin{gather}
		\hat{H}
		=
		\frac{\hat{p}^2}{2m}
		+
		\frac{\alpha}{\hat{r}}
		-
		\frac{\beta^2}{2m}\frac{1}{\hat{r}^2}.
		\label{hamil intro}
	\end{gather}
	The theory is specified not only by the Hamiltonian \eqref{hamil intro} but
	also by the boundary condition at the origin; we impose the standard
	regular condition, discussed in
	Section~\ref{sect:more details on non-perturbative bound}. A different
	boundary condition would define a different theory, with different low-energy EFTs
	and scattering amplitudes. This model provides a particularly sharp
	testing ground because its exact amplitude and its non-perturbative
	unitarity bound are known by means independent of the perturbative
	techniques we wish to test. In particular, when the
	Hamiltonian \eqref{hamil intro} is treated as the full UV theory, the exact
	unitarity bound can be derived by a simple non-perturbative argument
	(Section~\ref{sect:more details on non-perturbative bound}):
	\begin{gather}
		\boxed{\beta^2\leq\frac14 .}
		\label{eq: non pert unitarity bound intro}
	\end{gather}
	If the Hamiltonian is instead treated as an EFT with a finite UV
	cutoff, the bound is weaker. The non-perturbative derivation shows that
	the bound is controlled by short-distance physics and therefore should
	not depend on an arbitrary infrared scale. By contrast, when we apply standard short-range perturbation theory in Section~\ref{sect: naive}, the amplitudes are infrared divergent and the partial-wave projections require an arbitrary infrared regulator. The resulting bounds on $\beta^2$ inherit this regulator dependence, as \eqref{first bound naive short} shows.
	
	We resolve this problem using two ingredients. The first is the correct
	distributional treatment of the forward direction, reviewed in
	Section~\ref{sect: forward direction}. In non-relativistic scattering
	by long-range potentials---the setting considered in this
	paper---rigorous mathematical results show that the scattering matrix
	does not contain the diagonal delta-function singularity associated
	with the identity term in the short-range decomposition
	$S=\mathds{1}+i\mathbb{T}$
	\cite{Yafaev:1998LongRangeAmplitude,Yafaev:2000ScatteringTheory}. For
	non-relativistic Coulomb scattering, the $\epsilon_+$ prescription in
	\eqref{NRQMamplitudeeps} provides the forward distributional data
	normally carried by this delta function. In this setting, Herbst
	proved that this
	prescription is the \emph{unique} completion of the Coulomb amplitude
	at the forward point consistent with unitarity
	\cite{herbst1974connectedness}. It makes angular integrals, including
	partial-wave projections, well defined and consistent with unitarity.
	
	The second ingredient is distorted-wave perturbation theory
	(DWPT), reviewed in Section~\ref{sect:DWPT}. Rather than expanding
	around free states, DWPT treats the Coulomb interaction exactly and
	perturbs only in the inverse-square interaction. The resulting
	amplitudes have the correct long-range distributional structure and
	contain no spurious infrared divergences or arbitrary infrared scales.

	Combining these two ingredients, in
	Section~\ref{sect: unitarity bounds} we derive unitarity bounds on
	$\beta^2$ directly from perturbative amplitude data, with no arbitrary
	infrared scales entering the bounds. Table~\ref{tab:rapid-summary} shows
	how quickly these bounds converge, order by order, to the exact value
	$\beta^2_{\rm crit}=1/4$ in the limit where \eqref{hamil intro} is treated as the full UV theory.
	
	\begin{table}[ht]
		\centering
		\begin{tabular}{c|c|c}
			perturbative truncation
			&
			upper limit on $\beta^2$
			&
			relative error
			\\
			\hline
			$\mathcal{O}(\beta^2)$ & 0.3183 & 27.3\% \\
			$\mathcal{O}(\beta^4)$ & 0.2539 & 1.56\% \\
			$\mathcal{O}(\beta^6)$ & 0.25017 & 0.068\% \\
			$\mathcal{O}(\beta^8)$ & 0.25000465 & 0.00186\%
		\end{tabular}
		\caption{Infrared-scale-independent upper limits on $\beta^2$ using
			successive perturbative truncations of DWPT in the limit where the Hamiltonian \eqref{hamil intro} is treated as the full UV theory. No photon mass, detector resolution, or arbitrary
			infrared regulator enters any entry. The final column displays the
			rapid convergence to the exact critical value
			$\beta^2_{\rm crit}=1/4$.}
		\label{tab:rapid-summary}
	\end{table}
	\FloatBarrier
	When the Hamiltonian is instead treated as an EFT valid up to a
	physical cutoff $\Lambda_{\rm EFT}$, the same analysis yields
	corresponding finite-cutoff bounds on $\beta^2$; see, for example, the
	first-order result \eqref{EFT bound}.

To derive these bounds, we impose a unitarity condition of the form $|\operatorname{Re}a_l|\leq\tfrac{1}{2}$ on the partial waves, rather than the more common positivity bound $\operatorname{Im}a_l\geq 0$. Neither this choice nor the fact that our bound is derived at a coupling $\beta^2$ of order one is essential to the method. In each case the IR scales are eliminated the same way, by accounting for the modified distributional structure and working in DWPT.
	
	Standard techniques for deriving EFT bounds in long-range theories produce bounds that depend on an infrared scale, raising two questions: whether this dependence is intrinsic to the theory, and whether the scale should be interpreted as an experimental resolution. In this model, the answer to both is no. The exact bound is infrared-scale independent; the scale introduced by naive short-range perturbation theory is an artifact of that perturbative scheme, not a property of the theory or its measurement.
	
	Although our results are obtained in a non-relativistic setting, this is also the model's advantage: it isolates several forward-scattering difficulties that also arise in relativistic theories, yet is simple enough that their resolution can be established with mathematical rigor rather than by an ad hoc prescription. This is the level of confidence one would like in hand before attempting analogous constructions in QED and gravity.
	
	The model also provides a benchmark for other approaches. Any proposed method for deriving bounds in theories with long-range forces, if it can be formulated in this setting, can be compared directly with the exact bound \eqref{eq: non pert unitarity bound intro}. The comparison shows quantitatively how accurate the method is, and which features of the resulting bounds are artifacts of the method rather than properties of the theory.
	
	A further, rather striking payoff of DWPT is that its first two
	orders already resum infinitely many Feynman diagrams of short-range
	perturbation theory---the highly nontrivial combinations of
	multiple polylogarithms and complete elliptic integrals to which these
	diagrams evaluate---into compact expressions
	(Section~\ref{sect:comparison short range}). Remarkably, \emph{no
		integration is even required}: at fixed partial wave, the first-order
	integrand is a total derivative, and the amplitude collapses to
	boundary data at infinity (Appendix~\ref{app:first order DWPT}). This
	simplicity is invisible diagram by diagram, suggesting that DWPT may
	offer a simpler representation of scattering amplitudes than standard
	perturbation theory.

	\subsection{Three obstructions and their resolution}
	Deriving unitarity bounds in long-range theories requires addressing three related issues that arise when the standard short-range framework is applied to a theory with long-range forces: spurious infrared divergences, forward singularities that render integration over the forward direction ill-defined, and enhanced forward logarithms that make fixed-order perturbation theory non-uniform.\footnote{A further issue, which we do not discuss, is the absence of a Froissart--Martin bound for four-dimensional theories without a mass gap; see \cite{Haring:2022cyf} for its current status. Our bounds are fixed-energy partial-wave statements and do not rely on high-energy behavior.} We first explain how these issues obstruct the usual arguments, and then show how they are resolved in the exactly solvable model studied below using distorted-wave perturbation theory (DWPT) \cite{Bethe,Glauber,CROTHERS1992287,Lippstreu:2023vvg,Lippstreu:2025jit}.
	\paragraph{Spurious IR divergences.}
	Infrared divergences arise artificially when short-range perturbation theory, such as \eqref{short pert}, is applied to a theory with long-range interactions. In the model studied in this paper, defined by \eqref{hamil intro}, ordinary short-range perturbation theory produces the familiar IR divergences, and the resulting unitarity bounds depend on the IR regulator scale, as shown in Section~\ref{sect:tree unitarity bounds naive}. By contrast, the exact non-perturbative amplitude \eqref{exact amp} contains no IR divergence or arbitrary IR scale, and the corresponding non-perturbative unitarity bound \eqref{eq: non pert unitarity bound} is IR-scale independent. Moreover, resumming the short-range perturbative expansion does not recover the exact amplitude, since the IR scale remains present. The first ingredient in our resolution is therefore to use a perturbative expansion that does not introduce spurious infrared divergences and, in the solvable model studied here, reproduces the exact long-range amplitude.
	
	\paragraph{Ambiguous forward angular integrals.}
	Unitarity requires integration over all scattering angles, including the
	forward direction. For potential scattering in non-relativistic quantum
	mechanics, define the full scattering amplitude by
	\begin{gather}
		\mathcal{A}(\vec p_i\rightarrow\vec p_f)
		\defined
		\langle \vec p_f,\mathrm{out}\,|\,\vec p_i,\mathrm{in}\rangle .
	\end{gather}
	Elastic unitarity is then the requirement that (see Sec.~8.1, especially Eqs.~(8.1.14)--(8.1.15), of
	\cite{weinberg_2015})
	\begin{gather}
		\int\frac{\DD^3\vec\ell}{(2\pi)^3}\,
		\mathcal{A}(\vec p_1\rightarrow\vec\ell)\,
		\mathcal{A}^{\star}(\vec p_2\rightarrow\vec\ell)
		=
		(2\pi)^3\delta^{(3)}(\vec p_1-\vec p_2).
		\label{unitarity}
	\end{gather}
	Because the integral runs over all intermediate momenta $\vec\ell$, it
	necessarily includes the forward regions $\vec\ell=\vec p_1$ and
	$\vec\ell=\vec p_2$. In a theory with long-range forces, ordinary perturbation theory already gives, at tree level,
	\begin{gather}
		\frac{\alpha}{t},\label{tree amp}
	\end{gather}
	where $\alpha$ is the coupling and $t=|\vec{p}_1-\vec{p}_2|^2$ is the momentum transfer. Inserted into a unitarity relation such as \eqref{unitarity}, this tree amplitude yields a forward singularity that is not integrable against the phase-space measure, so the angular integrals are not well defined without an additional prescription. Thus, even if the usual amplitude-level infrared divergences are removed, for instance by factoring out the universal infrared-divergent exponential \cite{Weinberg:1965nx}, a distinct infrared ambiguity remains: the angular integrals required by unitarity still probe the forward direction, and the forward singularity is non-integrable without an additional prescription. A particularly important angular integral is the projection onto the partial-wave basis, since this is where unitarity constraints are most directly imposed. However, even the tree-level amplitude \eqref{tree amp} has no well-defined partial-wave projection without an additional prescription, and the usual partial-wave derivation of unitarity bounds becomes ambiguous already at tree level.
	
	\paragraph{Breakdown of fixed-order perturbation theory in the forward
		direction.}
	Even after the IR-divergent universal Coulomb phase is removed, at $\ell$ loops the
	amplitude contains terms of the schematic form
	\begin{gather}
		\frac{\alpha}{t}
		\left[
		\alpha \log\!\left(4p^2/|t|\right)
		\right]^{\ell}.
	\end{gather}
	At any fixed nonzero scattering angle, the logarithm is finite, and
	higher-loop terms remain suppressed by additional powers of $\alpha$.
	As $t\to0$, however, the logarithm grows without bound. In the region
	where
	\begin{gather}
		\left|
		\alpha \log\!\left(4p^2/|t|\right)
		\right|
		\gtrsim 1,
	\end{gather}
	higher-loop terms are no longer smaller than lower-loop terms. Fixed-order
	perturbation theory therefore remains reliable at generic scattering
	angles but is not uniform in the forward direction. Because unitarity
	requires integration over all angles, it necessarily probes the region
	where the fixed-order expansion breaks down.

	This should be contrasted with short-range theories, where the relevant low-$t$ logarithms remain finite as $t\to0$. In that case the forward region does not by itself invalidate unitarity bounds derived from low orders of perturbation theory. For long-range interactions, by contrast, the singular small-$t$ behavior must be captured from the outset. In loose terms, the powers of $\alpha^L\log^L(|t|/\lambda_{\text{I.R.}}^2)$ appearing in the short-range expansion must be replaced by the power-law behavior in $t$ exhibited by the exact amplitude.\footnote{This should not be understood as an ordinary resummation of short-range perturbation theory. The short-range expansion is built on incorrect asymptotic assumptions: even after resummation, it does not reproduce the correct distributional structure of the long-range amplitude, and the infrared regulator remains present. The possible exception is dimensional regularization for which it is potentially plausible that it may be able to reproduce the correct distributional structure if one uses the results of \cite{Chang:2025cxc}; see the discussion at the end of
		Section~\ref{sect: forward direction}.} More precisely, we will use a different perturbative expansion whose leading term already contains the correct power-law behavior in the forward direction.

	\paragraph{The resolution.}
	We now summarize how the three issues above are resolved in the model studied below. Instead of using the short-range perturbative expansion \eqref{short pert}, we use distorted-wave perturbation theory (DWPT), defined in \eqref{long pert}--\eqref{long pert 3}: the Coulomb interaction is treated exactly, and only the $\beta^2$ interaction is expanded. In short-range perturbation theory, the perturbative expansion of the amplitude takes the schematic form
	\begin{gather}
		\mathcal{A}^{\text{short pert.}}
		=(2\pi)^3\delta^{(3)}(\vec{p}_1-\vec{p}_2)+
		\frac{\alpha}{t}
		\left(1+\mathcal{O}(\text{loops})\right) .
		\label{short pert first}
	\end{gather}
	By contrast, the DWPT expansion takes the form
	\begin{gather}
		\mathcal{A}^{\text{DWPT}}
		=
		\lim_{\epsilon_+\to 0^+}
		\left(\frac{4|\vec{p}|^2}{t}\right)^{1+i\gamma-\epsilon_+}
		\Big(1+\mathcal{O}(\beta^2)\Big) ,
		\label{DWPT amp}
	\end{gather}
	where $\gamma=\alpha m/|\vec{p}|$ is the real Sommerfeld parameter. Comparing the two expansions exhibits the resolution of each issue in turn.
	
	First, no infrared regulator or arbitrary infrared scale appears in \eqref{DWPT amp}. Because DWPT expands about the correct long-range asymptotic dynamics from the outset, it introduces no artificial infrared divergences and, in the solvable model studied here, reproduces the correct long-range amplitude.
	
	Second, the forward-direction integrals of \eqref{DWPT amp} are well defined, unlike those of \eqref{short pert first}. The positive infinitesimal $\epsilon_+$ specifies how the forward singularity is completed as a distribution: one performs angular integrals at fixed $\epsilon_+>0$ and takes $\epsilon_+\to0^+$ only at the end. With this prescription, angular integrals over the forward direction, including partial-wave projections, are finite and unambiguous at each order of the DWPT expansion. Note also that \eqref{DWPT amp} contains no disconnected $\delta^{(3)}(\vec{p}_1-\vec{p}_2)$: the scattering operator of a long-range theory contains no such term, and the forward-direction distributional data carried by the $\mathds{1}$ in the short-range decomposition $S=\mathds{1}+i\mathbb{T}$ is instead supplied by the $\epsilon_+$ prescription.
	
	Third, the leading term of \eqref{DWPT amp} already contains the power-law behavior $t^{-i\gamma}$ that, as discussed above, replaces the forward logarithms $\alpha^L\log^L(|t|/\lambda_{\text{I.R.}}^2)$ of the short-range expansion. Successive DWPT orders are progressively less singular in the forward direction (Section~\ref{sect:forward behavior DWPT}), so higher orders do not become parametrically larger than lower orders in the forward region.
	
	The parameter $\epsilon_+$ is not an infrared regulator. Unlike regulating with a photon mass, detector resolution, or dimensional regulator, the limit $\epsilon_+\to0^+$ leaves no arbitrary infrared scale behind. Nor is the prescription an ad hoc choice: Herbst proved that it is the unique distributional completion of the Coulomb amplitude at the forward point consistent with unitarity \cite{herbst1974connectedness}. The same prescription was recently used in \cite{FuentesZamoro:2025exp} to compute partial waves for longitudinal $W^+W^-$ scattering, where $t$-channel photon exchange produces the same non-integrable forward singularity, and to derive unitarity bounds on the Higgs mass. We explain the $\epsilon_+$ prescription in detail in Section~\ref{sect: forward direction}.
	
	Finally, the modified distributional structure changes how unitarity is imposed: with no disconnected $\mathds{1}$, the general optical theorem is replaced by the modified optical theorem \cite{Lippstreu:2025jit}, reviewed in Section~\ref{sect: modified optical}. In Section~\ref{sect: unitarity bounds} we use it, together with the DWPT amplitudes, to derive unitarity bounds on the coupling $\beta^2$ that contain no arbitrary infrared scale.

	\subsection{Comparison to existing literature}
	In relativistic QFT, existing approaches address analogous infrared
	obstructions in several ways. One possibility is to define amplitudes with the universal infrared-divergent exponential \cite{Weinberg:1965nx} factored out \cite{bellazzini2026positivitylongrangeinteractions}; however, the resulting bounds depend on an infrared scale $\mathcal{E}$, which may be interpreted as the energy resolution or observation timescale of the detector. Another approach is to extract bounds by acting on the dispersion relations with more general functionals, rather than taking the strict forward limit, thereby controlling the $t$-channel poles \cite{Caron-Huot:2021rmr,Caron-Huot:2022ugt}. In four-dimensional applications the resulting bounds depend on an infrared cutoff \cite{Henriksson_2022}. Other approaches include studying pole-subtracted amplitudes \cite{Alberte:2020bdz,Alberte_2020}, working in spacetime dimensions $d>4$ \cite{Caron-Huot:2021rmr}, placing the theory in AdS so that the curvature scale regulates the infrared \cite{Caron-Huot:2021enk}, or compactifying to three dimensions and subtracting the calculable infrared modes \cite{Bellazzini:2019xts}. In \cite{Chang:2025cxc}, assuming that the eikonal formula captures the all-orders forward behavior, the leading infrared-enhanced terms were resummed before removing the dimensional regulator, yielding a formally finite, regulator-independent bound. The resulting constraint is independent of Newton's constant and therefore has a puzzling non-decoupling limit; it was already noted in \cite{Chang:2025cxc} not to admit a straightforward interpretation as a bound on a Wilson coefficient, and \cite{bellazzini2026positivitylongrangeinteractions} subsequently argued that, once the universal infrared suppression of the full amplitude is taken into account, the resulting inequality no longer constrains the Wilson coefficient.
	
	Oller and collaborators' program instead removes the Weinberg phase from plane-wave amplitudes and unitarizes the resulting infrared-finite partial waves \cite{blas2021unitarizationinfiniterangeforcesgravitongraviton,Oller:2022Coulomb,Oller:2022Review}. In \cite{FuentesZamoro:2025exp}, a relativistic analog of the Herbst $\epsilon_{+}$ prescription was argued for and used to derive a unitarity bound on the Higgs mass from longitudinal $W$-boson scattering. Our non-relativistic analysis provides further support for the prescription underlying their treatment. In \cite{plestid2026partialwaveunitaritylongrangeinteractions}, the Coulomb amplitude (without the $\epsilon_{+}$ prescription) is subtracted from the full amplitude, defining a Coulomb-subtracted amplitude whose partial-wave projections are well defined and whose unitarity properties can therefore be analyzed. It would be interesting to investigate whether this subtraction is equivalent to Herbst's subtracted representation \eqref{herbst subtracted representation} of the $\epsilon_{+}$ prescription.

	\subsection{Outline}
	Section~\ref{sect: review} reviews unitarity in theories with long-range forces: the modified distributional structure of the amplitude, the $\epsilon_{+}$ prescription that makes forward-direction integrals and partial-wave projections well defined, and the modified optical theorem. Section~\ref{Sect: setup} introduces the exactly solvable model, Coulomb scattering deformed by a short-range attractive inverse-square interaction, and derives the exact non-perturbative unitarity bound $\beta^2\leq1/4$. Section~\ref{sect: naive} applies ordinary short-range perturbation theory to this model; its purpose is diagnostic, exhibiting concretely the three obstructions described above. Section~\ref{sect:DWPT} introduces distorted-wave perturbation theory, in which the Coulomb dynamics is treated exactly and the inverse-square interaction perturbatively, and computes its leading and first subleading terms; the expansion is free of spurious infrared divergences, its angular integrals are well defined, and successive orders become less singular in the forward direction. Section~\ref{sect: unitarity bounds} derives unitarity bounds from the Coulomb-stripped partial waves; these contain no arbitrary infrared scale and rapidly converge to the exact bound. Section~\ref{sect:conc} summarizes the lessons, discusses whether EFT bounds should contain infrared scales, and outlines generalizations toward QED and gravity. Appendix~\ref{sect:bench} derives the unitarity bounds of the purely short-range theory, to which the long-range bounds reduce at high energies; Appendices~\ref{app:1loop amplitudes} and~\ref{app:first order DWPT} collect technical details.

	\section{Review: Unitarity in theories with long-range forces}\label{sect: review}
	A primary obstacle to deriving the consequences of unitarity in theories with long-range forces is that integrating scattering amplitudes over the forward direction can appear ill defined or ambiguous. In this section we review Section~5 of \cite{Lippstreu:2025jit}, which collected classic results \cite{herbst1974connectedness,Taylor,Yafaev:1998LongRangeAmplitude,Yafaev:2000ScatteringTheory} showing that in non-relativistic quantum mechanics, where these questions can be studied with mathematical rigor, no such ambiguity exists. The resolution lies in the distributional structure of the amplitude, which is modified relative to the short-range case: there is no disconnected $\mathds{1}$, and unitarity instead fixes a unique distributional completion of the amplitude in the forward direction, the $\epsilon_{+}$ prescription of \eqref{NRQMamplitudeeps}. We use this prescription to demonstrate unitarity in the momentum basis and to project the amplitude onto partial waves (Section~\ref{sect: forward direction}); we explain why the non-convergence of the partial-wave expansion is harmless (Section~\ref{sect: nonconvergence}); and we derive the modified optical theorem that replaces the general optical theorem when the amplitude has no disconnected piece (Section~\ref{sect: modified optical}). The lesson of this section is that, in order to analyze the consequences of unitarity in theories with long-range forces, one should account for the modified distributional structure of the amplitude. For the quantum-mechanical models studied in this paper this suffices to remove all ambiguities.
	\subsection{Integrating amplitudes over the forward direction}\label{sect: forward direction}
	To make the difficulty associated with the forward direction concrete, consider the canonical example of a theory with a long-range force: non-relativistic Coulomb scattering in three spatial dimensions,
	\begin{gather}
		\hat{H}=\frac{\hat{p}^2}{2m}+\frac{\alpha}{\hat{r}}\, .
		\label{hamil nr}
	\end{gather}
	This model is exactly solvable, and its amplitude is stated in textbooks as
	\cite{Landau:1991wop}\footnote{Explicit position-space expressions for the
		Coulomb in- and out-wavefunctions are given later in \eqref{coulomb in} and
		\eqref{coulomb out}.}
	\begin{align}
		\mathcal{A}(\vec{p}_1\rightarrow\vec{p}_2)
		&\defined \braket{\phi_{\t{out}}(\vec{p}_2)|\phi_{\t{in}}(\vec{p}_1)}
		\nonumber\\
		&=-i\,\frac{2\pi^2\gamma}{mp}\,
		\delta(E_1-E_2)\,
		\frac{\Gamma(1+i\gamma)}{\Gamma(1-i\gamma)}
		\left(
		\frac{2}{1-\hat p_1\cdot\hat p_2}
		\right)^{1+i\gamma} ,
		\label{NRQMamplitude}
	\end{align}
	where, setting $p_1=p_2\equiv p$ on the support of $\delta(E_1-E_2)$, we define
	\begin{gather}
		\gamma=\frac{m\alpha}{p},\qquad
		E=\frac{p^2}{2m},\qquad
		p=|\vec{p}|,\qquad
		\hat{p}=\frac{\vec{p}}{p}\, .
	\end{gather}
	Two features of \eqref{NRQMamplitude} deserve emphasis. First, the amplitude does not take the short-range form $\mathds{1}+i\mathbb{T}$: there is no disconnected $\delta^{(3)}(\vec{p}_1-\vec{p}_2)$ term. Second, the expression \eqref{NRQMamplitude} is incomplete as written: it defines the amplitude only away from the forward point $\hat{p}_1\cdot\hat{p}_2=1$ and does not specify its distributional behavior there. We now show that unitarity forces a unique completion.
	
	Unitarity is the requirement that
	\begin{gather}
		\int\frac{\DD^3\vec\ell}{(2\pi)^3}\,
		\mathcal A(\vec p_1\rightarrow\vec\ell)\,
		\mathcal A^\star(\vec p_2\rightarrow\vec\ell)
		=
		(2\pi)^3\delta^{(3)}(\vec p_1-\vec p_2)\, .
		\label{does unitarity hold}
	\end{gather}
	For an attractive Coulomb potential, the complete set of intermediate states also contains bound states; they do not contribute to \eqref{does unitarity hold} because the bound and scattering eigenstates are orthogonal.
	
	Inserting the textbook amplitude \eqref{NRQMamplitude} into the unitarity relation \eqref{does unitarity hold} gives an ill-defined integral: the integrand is non-integrable at the forward points $\hat{\ell}=\hat{p}_1$ and $\hat{\ell}=\hat{p}_2$. Herbst \cite{herbst1974connectedness} proved that the Coulomb amplitude admits a unique distributional completion at the forward point consistent with unitarity, namely
	\begin{gather}
		\mathcal{A}(\vec{p}_1\rightarrow\vec{p}_2)
		=\lim_{\epsilon_{+}\rightarrow 0^+}\left[-i\,\frac{2\pi^2\gamma}{mp}\,
		\delta(E_1-E_2)\,
		\frac{\Gamma(1+i\gamma)}{\Gamma(1-i\gamma)}
		\left(
		\frac{2}{1-\hat p_1\cdot\hat p_2}
		\right)^{1+i\gamma-\epsilon_{+}}\right] .
		\label{NRQMamplitudeeps}
	\end{gather}
	The limit is understood in the sense of distributions: in any integral involving the amplitude, one keeps $\epsilon_{+}>0$ fixed, evaluates the integral, and only then takes $\epsilon_{+}\rightarrow 0^{+}$.
	
	The $\epsilon_{+}$ prescription supplies the forward-direction distributional data that, in short-range scattering, is carried by the disconnected $\mathds{1}$. That the usual connectedness structure is modified in the presence of Coulomb forces and massless particles was noted long ago \cite{Eden:1966dnq,Taylor,Zwanziger}, and in non-relativistic long-range scattering it is a theorem: the scattering operator contains no disconnected $\delta$-function term \cite{Yafaev:1998LongRangeAmplitude,Yafaev:2000ScatteringTheory}. One may therefore loosely regard the $\epsilon_{+}$ as the long-range analog of the $\mathds{1}$ in $S=\mathds{1}+i\mathbb{T}$, since both matter only in the forward direction. The analogy should not be taken too literally, however: the $\epsilon_{+}$ completion is not a delta function supported in the forward direction and, as emphasized in \cite{herbst1974connectedness}, is in a certain sense more singular than one.\footnote{Herbst also gives an equivalent representation of the distributional action of the Coulomb amplitude in Eq.~(13) of \cite{herbst1974connectedness}. For any continuously differentiable, square-integrable function $f$,
		\begin{equation}
			\begin{aligned}
				(S_Cf)(\vec k)
				={}&\frac{\Gamma(1+i\gamma)}{\Gamma(1-i\gamma)}
				\Bigg\{
				f(\vec k)
				+\frac{\gamma}{2\pi i k}
				\int\DD^3\vec k'\,
				\delta(k^2-k'^2)
				\left(
				\frac{1-\hat k\cdot\hat k'}{2}
				\right)^{-1-i\gamma}
				\left[f(\vec k')-f(\vec k)\right]
				\Bigg\}.
				\label{herbst subtracted representation}
			\end{aligned}
		\end{equation} The subtraction of the forward value makes the integral well defined at $\hat k'=\hat k$ and is equivalent to the $\epsilon_+$ prescription. The explicit $f(\vec k)$ term in this representation should not be interpreted as a separate delta-function component of $S_C$.}
	
	We now illustrate the two most important uses of the $\epsilon_{+}$ prescription: verifying that the amplitude \eqref{NRQMamplitudeeps} satisfies unitarity, and projecting the amplitude onto partial waves. 
	
	\paragraph{Unitarity of the amplitude.}
	We first verify that the completed amplitude \eqref{NRQMamplitudeeps} satisfies the unitarity relation \eqref{does unitarity hold}. On the left-hand side of \eqref{does unitarity hold}, the ratio of Gamma functions in \eqref{NRQMamplitudeeps} is a pure phase and cancels against its complex conjugate, while the energy-conserving delta functions set $|\vec{\ell}\,|=p$ and allow the radial part of the $\vec{\ell}$ integral to be performed. This gives
	\begin{align}
		&\int\frac{\DD^3\vec\ell}{(2\pi)^3}\,
		\mathcal A(\vec p_1\rightarrow\vec\ell)\,
		\mathcal A^{\star}(\vec p_2\rightarrow\vec\ell)
		=
		\frac{\pi\gamma^2}{2mp}\,
		\delta(E_1-E_2)\,
		I(\hat p_1,\hat p_2),
		\label{first eqn}
	\end{align}
	where the remaining integral runs over the direction $\hat\ell$ only,
	\begin{align}
		I(\hat p_1,\hat p_2)
		\defined
		\lim_{\epsilon_+\rightarrow0^+}
		\int \DD^2\Omega_\ell\,
		\left(
		\frac{2}{1-\hat p_1\cdot\hat\ell}
		\right)^{1+i\gamma-\epsilon_+}
		\left(
		\frac{2}{1-\hat p_2\cdot\hat\ell}
		\right)^{1-i\gamma-\epsilon_+} .
		\label{2evaluate}
	\end{align}
	The angular integral \eqref{2evaluate} has a hidden conformal symmetry,\footnote{This conformal symmetry originates from the hidden Runge--Lenz symmetry, which combines with rotations to furnish a unitary representation of $SO(1,3)$, the two-dimensional Euclidean conformal group; the scattering states carry the principal-series dimension $\Delta=1+i\gamma$. See Section~2.4 of \cite{Lippstreu:2023vvg}.} which is made manifest by parameterizing all unit vectors in stereographic coordinates,
	\begin{gather}
		\hat{k} = \frac{1}{1 + |z|^2}
		\left( z + \bar{z}, i(\bar{z} - z), 1 - |z|^2 \right),
		\qquad
		z = e^{i\phi} \tan \frac{\theta}{2}\, ,
	\end{gather}
	in which the measure and the forward factors become
	\begin{gather}
		\int \DD^2\Omega
		=
		4\int \frac{\DD^2z}{(1+|z|^2)^2}\, ,
		\qquad
		\frac{2}{1-\hat p\cdot\hat \ell}
		=
		\frac{(1+|z_p|^2)(1+|z_\ell|^2)}{|z_p-z_\ell|^2}\, .
	\end{gather}
	The integral \eqref{2evaluate} then takes the form
	\begin{align}
		I(\hat p_1,\hat p_2)
		&=
		4\lim_{\epsilon_+\rightarrow0^+}
		(1+|z_1|^2)^{1+i\gamma-\epsilon_+}
		(1+|z_2|^2)^{1-i\gamma-\epsilon_+}
		\nonumber\\
		&\quad\times
		\int \DD^2z_\ell\,
		\frac{(1+|z_\ell|^2)^{-2\epsilon_+}}
		{|z_\ell-z_1|^{2+2i\gamma-2\epsilon_+}
			|z_\ell-z_2|^{2-2i\gamma-2\epsilon_+}}
		\label{def of i}\\
		&=\frac{4\pi^2}{\gamma^2}
		(1+|z_1|^2)^2
		\delta^{(2)}(z_1-z_2)\, .
		\label{angular integral result}
	\end{align}
	The integral in \eqref{def of i} is a standard one in two-dimensional conformal field theory and is evaluated in Appendix~A of \cite{dolan2012conformalpartialwavesmathematical}, building on the classic conformal integrals of \cite{Symanzik:1972wj} (see also Appendix~A of \cite{guevara2021celestialopeblocks}). In that setting, too, the integral is defined by analytic continuation in the conformal dimensions \cite{dolan2012conformalpartialwavesmathematical}.\footnote{The factor $(1+|z_\ell|^2)^{-2\epsilon_+}$ introduces no new singularities in the integration region and therefore does not affect the $\epsilon_+\rightarrow0^+$ limit.} Substituting \eqref{angular integral result} into \eqref{first eqn} and recombining the two delta functions into a three-dimensional one, we find
	\begin{align}
		\int\frac{\DD^3\vec\ell}{(2\pi)^3}\,
		\mathcal A(\vec p_1\rightarrow\vec\ell)\,
		\mathcal A^{\star}(\vec p_2\rightarrow\vec\ell)&=
		\frac{\pi\gamma^2}{2mp}\,
		\delta(E_1-E_2)\,
		\frac{4\pi^2}{\gamma^2}
		(1+|z_1|^2)^2
		\delta^{(2)}(z_1-z_2)
		\nonumber\\
		&=
		(2\pi)^3\delta^{(3)}(\vec p_1-\vec p_2)\, ,
	\end{align}
	which is precisely the unitarity relation \eqref{does unitarity hold}. Note the role played by the $\epsilon_{+}$: at $\epsilon_{+}=0$ the angular integral \eqref{2evaluate} is not convergent, while for $\epsilon_{+}>0$ it converges, and the $\epsilon_{+}\rightarrow0^{+}$ limit produces precisely the forward delta function that unitarity requires, with no leftover infrared scale. This verification also shows that we were not missing a disconnected piece: the completed amplitude \eqref{NRQMamplitudeeps} satisfies the unitarity relation \eqref{does unitarity hold} entirely on its own. Supplementing it with a disconnected $(2\pi)^3\delta^{(3)}(\vec{p}_1-\vec{p}_2)$ term would instead violate unitarity.

	\paragraph{Partial-wave projection.}
	Textbooks state the exact Coulomb amplitude both in the momentum basis \eqref{NRQMamplitude} and in the partial-wave basis \eqref{coulomb pw series} \cite{Landau:1991wop}. What is usually left unstated is the prescription required to pass between the two: the projection onto partial waves integrates through the forward point, and this integral is only defined once the distributional completion \eqref{NRQMamplitudeeps} is specified. Because in this exactly solvable model the answer is known independently in both bases, the omission is harmless. In more general settings, however, the partial-wave coefficients must be obtained by actually performing the projection, and there the $\epsilon_{+}$ prescription is essential: it is what makes the projection integral well defined. We therefore use this example, where the result can be checked against the known answer, to demonstrate how the projection is carried out.
	
	Stripping off the energy-conserving delta function, we define the reduced amplitude $A(x)$ and its partial-wave coefficients $S_l$ by
	\begin{align}
		\mathcal{A}(\vec p_1\rightarrow\vec p_2)
		&=
		\frac{2\pi^2}{mp}\,
		\delta(E_1-E_2)\,
		A(x),
		&
		A(x)
		&=
		\sum_{l=0}^{\infty}(2l+1)S_lP_l(x),
		\qquad
		x=\hat p_1\cdot\hat p_2 .
		\label{exact partial nrwm}
	\end{align}
	Comparing with \eqref{NRQMamplitudeeps}, the reduced Coulomb amplitude at fixed $\epsilon_+>0$ is
	\begin{gather}
		A_{\epsilon_+}(x)
		=
		-i\gamma\,
		\frac{\Gamma(1+i\gamma)}{\Gamma(1-i\gamma)}
		\left(\frac{2}{1-x}\right)^{1+i\gamma-\epsilon_+},
	\end{gather}
	and orthogonality of the Legendre polynomials gives
	\begin{gather}
		S_l
		=
		\frac{1}{2}
		\lim_{\epsilon_+\rightarrow0^+}
		\int_{-1}^{1}\DD x\,
		P_l(x)A_{\epsilon_+}(x) .
		\label{legendre projection}
	\end{gather}
	The potentially ambiguous part of the projection \eqref{legendre projection} is
	\begin{gather}
		\lim_{\epsilon_+\rightarrow0^+}
		\int_{-1}^{1}\DD x\,
		(1-x)^{-1-i\gamma+\epsilon_+}P_l(x) .
		\label{partial wave integral eps}
	\end{gather}
	At $\epsilon_+=0$ the integral does not converge at the forward point: since $P_l(1)=1$, the integrand behaves as $(1-x)^{-1-i\gamma}$ near $x=1$, which is not absolutely integrable. For $\epsilon_+>0$, the integral converges, and we can use Eq.~7.127 of \cite{gradshteyn2014table},
	\begin{gather}
		\int_{-1}^{1}\DD x\,
		(1+x)^{\sigma}P_{\nu}(x)
		=
		\frac{2^{\sigma+1}\Gamma(\sigma+1)^2}
		{\Gamma(\sigma+\nu+2)\Gamma(1+\sigma-\nu)},
		\qquad
		\operatorname{Re}\sigma>-1 .
		\label{GR Legendre integral}
	\end{gather}
	Substituting $x\rightarrow-x$ and using $P_l(-x)=(-1)^lP_l(x)$ brings \eqref{partial wave integral eps} to the form \eqref{GR Legendre integral} with $\sigma=-1-i\gamma+\epsilon_+$ and $\nu=l$. The convergence condition $\operatorname{Re}\sigma=-1+\epsilon_+>-1$ is satisfied precisely because of the $\epsilon_+$. We therefore obtain
	\begin{align}
		&\lim_{\epsilon_+\rightarrow0^+}
		\int_{-1}^{1}\DD x\,
		(1-x)^{-1-i\gamma+\epsilon_+}P_l(x)
		\nonumber\\
		&\qquad=
		\lim_{\epsilon_+\rightarrow0^+}
		(-1)^l
		\frac{2^{-i\gamma+\epsilon_+}\Gamma(-i\gamma+\epsilon_+)^2}
		{\Gamma(l+1-i\gamma+\epsilon_+)\Gamma(-i\gamma-l+\epsilon_+)} ,
	\end{align}
	in which the limit $\epsilon_+\rightarrow0^+$ is now smooth and can simply be taken. Substituting the result into the projection \eqref{legendre projection} and simplifying with the reflection formula $\Gamma(z)\Gamma(1-z)=\pi/\sin(\pi z)$ gives
	\begin{align}
		S_l
		&=
		-i\gamma(-1)^l
		\frac{\Gamma(1+i\gamma)}{\Gamma(1-i\gamma)}
		\frac{\Gamma(-i\gamma)^2}
		{\Gamma(l+1-i\gamma)\Gamma(-i\gamma-l)}
		\nonumber\\
		&=
		\frac{\Gamma(l+1+i\gamma)}
		{\Gamma(l+1-i\gamma)} .
		\label{partial wave Coulomb result}
	\end{align}
	These are the standard Coulomb partial waves. The $\epsilon_+$ prescription has thus made the partial-wave projection well defined without introducing an infrared regulator, and no arbitrary scale remains in the result. The same prescription was recently employed in \cite{FuentesZamoro:2025exp}, where a direct projection of the exact Coulomb amplitude was likewise shown to reproduce \eqref{partial wave Coulomb result}.
	
	\paragraph{How does the $\epsilon_{+}$ arise?}
	The exact Coulomb scattering amplitude is the overlap of an
	in-state with an out-state,\footnote{We use ``amplitude'' for the
		full in--out overlap. In short-range scattering this overlap
		decomposes as $\hat S=\mathds{1}+i\mathbb{T}$, and the term
		``scattering amplitude'' is often reserved for the connected
		contribution $\mathbb{T}$. Because the pure Coulomb scattering
		operator admits no analogous separation, we use ``amplitude'' for
		the full overlap.}
	\begin{align}
		\mathcal{A}(\vec p_1\rightarrow\vec p_2)
		&\defined
		\braket{\phi_{\t{out}}(\vec p_2)|\phi_{\t{in}}(\vec p_1)}
		\nonumber\\
		&=
		\int\DD^3\vec{x}\,
		\phi_{\t{out}}^\star(\vec p_2,\vec{x})\,
		\phi_{\t{in}}(\vec p_1,\vec{x}) .
		\label{coulomb overlap integral}
	\end{align}
	As an ordinary integral, the second line is neither absolutely nor
	conditionally convergent. This does not signal an infrared
	divergence: the integral evaluates to a distribution rather than an
	ordinary function. Indeed, even in a genuinely short-range theory
	one can write
	\begin{gather}
		\braket{\psi_{\t{out}}(\vec p_2)|\psi_{\t{in}}(\vec p_1)}
		=
		(2\pi)^3\delta^{(3)}(\vec p_1-\vec p_2)
		+
		i\,\delta(E_1-E_2)\,
		\mathcal{T}(\vec p_1\rightarrow\vec p_2).
	\end{gather}
	The disconnected term already contains the familiar plane-wave
	identity
	\begin{gather}
		\int\DD^3\vec{x}\,
		e^{i(\vec p_1-\vec p_2)\cdot\vec{x}}
		=
		(2\pi)^3\delta^{(3)}(\vec p_1-\vec p_2),
		\label{plane wave overlap}
	\end{gather}
	which is an equality of distributions rather than of ordinary
	convergent integrals. Such overlaps are therefore naturally
	understood through their action on momentum-space wave packets.
	
	Herbst \cite{herbst1974connectedness} made the distributional
	content of the Coulomb overlap explicit by giving the following
	Abel representation:
	\begin{gather}
		\mathcal{A}(\vec{p}_1\rightarrow\vec{p}_2)
		=
		\lim_{\lambda\rightarrow 0^{+}}
		\int\DD^3\vec{x}\,
		e^{-\lambda|\vec{x}|}\,
		\phi_{\t{out}}^{\star}(\vec{p}_2,\vec{x})\,
		\phi_{\t{in}}(\vec{p}_1,\vec{x})\, .
		\label{herbst definition}
	\end{gather}
	The parameter $\lambda$ should not be interpreted as a physical
	infrared regulator. Indeed, for the plane-wave identity \eqref{plane wave overlap}, the
	same Abel convergence factor gives
	\begin{gather}
		\lim_{\lambda\rightarrow0^+}
		\int\DD^3\vec{x}\,
		e^{-\lambda|\vec{x}|}e^{i\vec q\cdot\vec{x}}
		=
		\lim_{\lambda\rightarrow0^+}
		\frac{8\pi\lambda}{(\vec q^{\,2}+\lambda^2)^2}
		=
		(2\pi)^3\delta^{(3)}(\vec q),
	\end{gather}
	where the limit is distributional. Just as the preceding Abel limit
	provides a distributional representation of the delta function,
	\eqref{herbst definition} provides a distributional representation
	of the Coulomb amplitude. Once the amplitude has acted on wave packets
	and the limit is performed, all dependence on $\lambda$ disappears.
	Thus $\lambda$ introduces no arbitrary infrared scale into the
	amplitude.
	
	The dimensionful parameter $\lambda$ should not be identified with
	the dimensionless $\epsilon_{+}$. Nevertheless, Herbst showed that
	the Abel prescription \eqref{herbst definition} defines the same
	distribution as the $\epsilon_{+}$ prescription in
	\eqref{NRQMamplitudeeps}. In this sense, the position-space Abel
	construction provides one natural way in which the $\epsilon_{+}$
	distribution arises.
	
	We close this subsection with two remarks. First, Herbst's analysis of relativistic scalar QED at first order in $\alpha$
\cite{herbst1974connectedness} suggests that the forward delta function is absent
there too. Second, it may be worth investigating whether dimensional regularization naturally produces the $\epsilon_{+}$ completion. The exact Coulomb amplitude in $d$ spatial dimensions behaves as \cite{Yafaev:2000ScatteringTheory}
	\begin{gather}
		\mathcal{A}_d(x)
		\propto
		(1-x)^{-\frac{d-1}{2}-i\gamma},
		\qquad x=\hat{p}_1\cdot\hat{p}_2\, ,
	\end{gather}
	so the spatial dimension enters precisely in the real part of the singular exponent; formally, continuing $d=3-2\epsilon$ reproduces the exponent $-1+\epsilon-i\gamma$ of the completed amplitude \eqref{NRQMamplitudeeps}. Relatedly, the dimensionally regularized eikonal amplitude of \cite{Chang:2025cxc} contains the factor
	\begin{gather}
		\left(-\frac{t}{\mu^2}\right)^{\frac{D-4}{2}},
	\end{gather}
	so that the dimensional regulator again shifts the power of the forward momentum transfer. These observations are suggestive but not conclusive: in a dimensionally regularized integral the angular measure is also $d$-dependent, and the $d$-dependence of the measure and of the exponent combine in a way that does not reproduce Herbst's $\epsilon_{+}$ prescription.

	\subsection{Non-convergence of the partial-wave expansion}\label{sect: nonconvergence}
	In the partial-wave basis \eqref{exact partial nrwm}, with the coefficients \eqref{partial wave Coulomb result}, the reduced Coulomb amplitude reads
	\begin{gather}
		A(x)
		=
		\sum_{l=0}^{\infty}(2l+1)\,
		\frac{\Gamma(l+1+i\gamma)}{\Gamma(l+1-i\gamma)}\,
		P_l(x)\, .
		\label{coulomb pw series}
	\end{gather}
	At large $l$ the coefficients behave as
	\begin{gather}
		\frac{\Gamma(l+1+i\gamma)}{\Gamma(l+1-i\gamma)}
		\sim
		l^{2i\gamma}\, ,
	\end{gather}
	and are pure phases. For any fixed $|x|<1$, the Legendre
	polynomials have an oscillatory envelope proportional to
	$l^{-1/2}$. The full summand in \eqref{coulomb pw series} therefore
	has an envelope proportional to $\sqrt{l}$ and does not tend to
	zero. Consequently, the series is neither conditionally nor
	absolutely convergent for any fixed $|x|<1$. One might worry that
	this non-convergence signals that the Coulomb problem is ill defined,
	or that it is yet another manifestation of an infrared divergence.
	Neither is the case. Amplitudes are always understood as
	distributions, and, as emphasized in \cite{Taylor}, the series
	\eqref{coulomb pw series} is perfectly well defined as a
	distribution: it has an unambiguous action on any function that
	admits a partial-wave expansion.
	
	This action is simple to state. Let
	\begin{gather}
		f(\hat p)
		=
		\sum_{l,m}f_{lm}Y_{lm}(\hat p)
	\end{gather}
	be a function on the sphere. At fixed energy, the kernel \eqref{coulomb pw series} acts by
	\begin{align}
		(\mathsf{S}_c f)(\hat p_2)
		&\defined
		\frac{1}{4\pi}
		\int\DD^2\Omega_{p_1}\,
		A(\hat p_1\cdot\hat p_2)\,f(\hat p_1)
		=
		\sum_{l,m}S_l\,f_{lm}\,Y_{lm}(\hat p_2)\, ,
	\end{align}
	i.e., the Coulomb amplitude multiplies each partial-wave coefficient by the phase $S_l=\Gamma(l+1+i\gamma)/\Gamma(l+1-i\gamma)$, with $|S_l|=1$. This action is manifestly norm preserving,
	\begin{gather}
		\|\mathsf{S}_cf\|^2
		=
		\sum_{l,m}|S_lf_{lm}|^2
		=
		\sum_{l,m}|f_{lm}|^2
		=
		\|f\|^2\, ,
	\end{gather}
	so the distribution defined by the series \eqref{coulomb pw series} is not only well defined but unitary. Moreover, its action on wave packets agrees with that of the momentum-space distribution \eqref{NRQMamplitudeeps}: the two are the same distribution expressed in different bases.
	
	In fact, non-convergent partial-wave expansions are already familiar from short-range scattering. The disconnected piece $\mathds{1}$ of $S=\mathds{1}+i\mathbb{T}$ has partial-wave coefficients $S_l=1$, and its expansion
	\begin{gather}
		\sum_{l=0}^{\infty}(2l+1)P_l(x)
		=
		2\delta(1-x)
	\end{gather}
	also fails to converge pointwise; the sum instead defines the delta distribution. The Coulomb series \eqref{coulomb pw series} should be understood in exactly the same way; the only difference is that the distribution it defines is the $\epsilon_{+}$ completion \eqref{NRQMamplitudeeps}, rather than a forward delta function.
	
	\subsection{A modified optical theorem}\label{sect: modified optical}
	For short-range interactions, inserting $S=\mathds{1}+i\mathbb{T}$ into $S^{\dagger}S=\mathds{1}$ gives the general optical theorem
	\begin{gather}
		\mathbb{T}-\mathbb{T}^{\dagger}
		=
		i\mathbb{T}^{\dagger}\mathbb{T}\, .
		\label{usual optical}
	\end{gather}
	This relation is how unitarity is imposed on perturbative amplitudes: the left-hand side is linear in $\mathbb{T}$ while the right-hand side is quadratic, so \eqref{usual optical} relates each order of perturbation theory to products of lower orders. Its partial-wave form, $\operatorname{Im}a_l=|a_l|^2$, is the starting point for deriving unitarity bounds on couplings. We have just seen, however, that in theories with long-range forces there is no disconnected $\mathds{1}$, and that the distributional structure of the amplitude is modified. How, then, is unitarity to be imposed on perturbative amplitudes? To answer this, we derive the analogs of \eqref{usual optical} and of $\operatorname{Im}a_l=|a_l|^2$ for the model studied in this paper,
	\begin{gather}
		\hat{H}
		=
		\frac{\hat{p}^2}{2m}
		+\frac{\alpha}{\hat{r}}
		-\frac{\beta^2}{2m}\frac{1}{\hat{r}^2}\, ,
		\label{hamil beta}
	\end{gather}
	which deforms the Coulomb Hamiltonian \eqref{hamil nr} by a short-range inverse-square interaction.
	
	Let $\mathcal{A}_c$ denote the exact Coulomb amplitude \eqref{NRQMamplitudeeps}, and decompose the full amplitude of the model \eqref{hamil beta} as
	\begin{gather}
		\mathcal{A}(\vec p_1\rightarrow\vec p_2)
		=
		\mathcal{A}_c(\vec p_1\rightarrow\vec p_2)
		+
		i\,\mathcal{T}(\vec p_1\rightarrow\vec p_2)\, ,
		\label{def of T}
	\end{gather}
	so that $\mathcal{T}$ contains all corrections due to the short-range $\beta^2$ interaction. Inserting \eqref{def of T} into the unitarity relation \eqref{does unitarity hold} and using the fact, verified above, that $\mathcal{A}_c$ satisfies \eqref{does unitarity hold} by itself, the $\mathcal{A}_c\mathcal{A}_c^{\star}$ term cancels against the right-hand side, leaving the modified optical theorem \cite{Lippstreu:2025jit}
	\begin{align}
		&\int\frac{\DD^3\vec{k}}{(2\pi)^3}
		\left[
		\mathcal{T}(\vec p_1\rightarrow\vec k)\,
		\mathcal{A}_c^{\star}(\vec p_2\rightarrow\vec k)
		-
		\mathcal{A}_c(\vec p_1\rightarrow\vec k)\,
		\mathcal{T}^{\star}(\vec p_2\rightarrow\vec k)
		\right]
		\nonumber\\
		&\qquad=
		i\int\frac{\DD^3\vec{k}}{(2\pi)^3}\,
		\mathcal{T}(\vec p_1\rightarrow\vec k)\,
		\mathcal{T}^{\star}(\vec p_2\rightarrow\vec k)\, .
		\label{modified optical momentum}
	\end{align}
	Like the general optical theorem, this relation is linear in $\mathcal{T}$ on one side and quadratic on the other, and it therefore relates successive orders of perturbation theory. It also reduces to the general optical theorem when the long-range force is switched off: for $\alpha\rightarrow0$, $\mathcal{A}_c\rightarrow(2\pi)^3\delta^{(3)}(\vec p_1-\vec p_2)$, and \eqref{modified optical momentum} becomes \eqref{usual optical}.\footnote{The Coulomb amplitude is proportional to $\alpha$ and hence vanishes for generic kinematics as $\alpha\rightarrow0$. However, expanding the amplitude about the forward limit gives $(1-x)^{-1-i\gamma}=\delta(1-x)/(-i\gamma)+\ldots$, and this pole cancels the vanishing prefactor, producing the $\delta^{(3)}(\vec{p}_1-\vec{p}_2)$. The order of limits matters: $\epsilon_{+}$ is taken to zero before $\gamma$.}
	
	Just as short-range perturbation theory is hardwired to satisfy the general optical theorem order by order, distorted-wave perturbation theory is hardwired to satisfy the modified optical theorem \eqref{modified optical momentum}. In short-range perturbation theory the discontinuity $\mathbb{T}-\mathbb{T}^{\dagger}$ originates from the energy denominators $(E_1-E_k+i\epsilon)^{-1}$, whose imaginary parts place intermediate states on shell and thereby generate $\mathbb{T}^{\dagger}\mathbb{T}$. In distorted-wave perturbation theory the same energy denominators appear, but the exact Coulomb amplitude always enters in convolution with the short-range corrections, so the on-shell intermediate states generate precisely the $\mathcal{A}_c$--$\mathcal{T}$ convolutions appearing in \eqref{modified optical momentum}. We will see this convolution structure arise explicitly in the factorized representation \eqref{first-order-factorized} of distorted-wave perturbation theory (Section~\ref{sect:DWPT}).
	
	Finally, we convert \eqref{modified optical momentum} into a condition on partial-wave coefficients. To this end, we parameterize the full amplitude by factoring out the Coulomb phases,
	\begin{gather}
		\mathcal{A}(\vec p_1\rightarrow\vec p_2)
		=
		\frac{2\pi^2}{mp}\,
		\delta(E_1-E_2)
		\sum_{l=0}^{\infty}(2l+1)\,
		S_{c,l}\left(1+2i\widehat{a}_l\right)
		P_l(\hat p_1\cdot\hat p_2)\, ,
		\label{Coulomb stripped partial waves}
	\end{gather}
	where $S_{c,l}$ are the Coulomb partial waves \eqref{partial wave Coulomb result}; this defines the Coulomb-stripped coefficients $\widehat{a}_l$, which encode the short-range corrections. Comparing \eqref{Coulomb stripped partial waves} with \eqref{def of T}, the corresponding expansion of $\mathcal{T}$ has coefficients $2S_{c,l}\widehat{a}_l$. We substitute these expansions into \eqref{modified optical momentum}, perform the radial integral using the energy-conserving delta functions as before, and perform the angular integral using
	\begin{gather}
		\int\DD^2\Omega_k\,
		P_l(\hat p_1\cdot\hat k)\,P_{l'}(\hat p_2\cdot\hat k)
		=
		\delta_{ll'}\,
		\frac{4\pi}{2l+1}\,
		P_l(\hat p_1\cdot\hat p_2)\, .
	\end{gather}
	The relation then decouples into an independent condition for each $l$, and since $|S_{c,l}|=1$ the Coulomb phases drop out entirely, leaving
	\begin{gather}
		\left(1+2i\widehat{a}_l\right)
		\left(1-2i\widehat{a}_l^{\star}\right)
		=
		1\, ,
	\end{gather}
	or equivalently
	\begin{gather}
		\boxed{
			\operatorname{Im}\widehat{a}_l
			=
			|\widehat{a}_l|^2}
		\, .
		\label{modified optical partial wave}
	\end{gather}
	This has the same form as the usual partial-wave unitarity condition, but it applies to the Coulomb-stripped coefficients. In perturbation theory, let $\widehat{a}_l^{(n)}$ denote the complete order-$\beta^{2n}$ contribution, including its coupling dependence, so that
	\begin{gather}
		\widehat{a}_l
		=
		\widehat{a}_l^{(1)}
		+
		\widehat{a}_l^{(2)}
		+\cdots,
		\qquad
		\widehat{a}_l^{(n)}
		=
		\mathcal O(\beta^{2n}).
	\end{gather}
	Expanding \eqref{modified optical partial wave} order by order in $\beta^2$ gives
	\begin{gather}
		\operatorname{Im}\widehat{a}_l^{(1)}=0,
		\qquad
		\operatorname{Im}\widehat{a}_l^{(2)}
		=
		|\widehat{a}_l^{(1)}|^2 .
	\end{gather}

	\section{Exactly solvable model}\label{Sect: setup}
	
	Throughout this paper, we study the following model from non-relativistic quantum mechanics. We consider a low-energy EFT for a particle governed by the Hamiltonian
	\begin{gather}
		\hat{H}=\frac{\hat{p}^2}{2m}+\frac{\alpha}{\hat{r}}-\frac{\beta^2}{2m}\frac{1}{\hat{r}^2},\label{hamil}
	\end{gather}
	which combines a long-range Coulomb potential $\alpha/\hat{r}$ with a short-range\footnote{Many definitions of long-range and short-range interactions exist in the literature. Throughout this paper we take short-range to mean that the asymptotic Hamiltonian is the free Hamiltonian and hence standard perturbation theory applies.} inverse-square potential $-\beta^2/(2m\hat{r}^2)$. To define the quantum system, we must also specify the domain of the Hamiltonian, or equivalently the boundary condition imposed on wavefunctions at the origin.
	Throughout this paper, we choose the regular-at-the-origin domain, corresponding to the regular branch of the radial wavefunctions near $r=0$ (we explain this point more fully in Section~\ref{sect:more details on non-perturbative bound}). This is one of the simplest models exhibiting the core difficulties outlined in the introduction. For example, standard short-range perturbation theory applied to \eqref{hamil} produces infrared divergences, and the resulting unitarity bounds on the coupling $\beta^2$ inherit a spurious dependence on the infrared regulator. At the same time, the model is exactly solvable, so every claim we make can be verified explicitly against the non-perturbative answer.

	To make the connection with the EFT bounds literature explicit, we will use the Hamiltonian \eqref{hamil} in two distinct ways. First, we can treat \eqref{hamil} as a UV-complete theory in its own right. In this case we can derive a non-perturbative unitarity bound on $\beta^2$. This is not the usual EFT-bounds situation, since the UV completion is known. Its role is instead to provide a benchmark: it lets us check whether the EFT bounds derived below are consistent with a known non-perturbative bound in one possible UV completion consistent with unitarity. Second, and more directly analogous to the EFT bounds literature, we can treat \eqref{hamil} as a low-energy EFT, valid only below a physical scale $\Lambda_{\rm EFT}$. Throughout we will consider \eqref{hamil} as an EFT description for scattering on a charged nucleus at energy scales low enough that the nuclear structure is not resolved. In this case the UV completion is not specified, and the bounds should be understood as necessary constraints on $\beta^2$ implied by the existence of some unitary short-distance completion. The first interpretation gives one possible completion of the effective model, while in a realistic nuclear setting the completion would be supplied by the shorter-distance physics that resolves the nucleus.

	\subsection{An exact amplitude}
	
	The scattering amplitude is unambiguously defined as the inner product of the in- and out-states,
	\begin{align}
		\mathcal{A}(\vec{p}_1\rightarrow\vec{p}_2)
		&\,\defined\,\braket{\psi_{\mathrm{out}}(\vec p_2)\,|\,\psi_{\mathrm{in}}(\vec p_1)},
	\end{align}
	which by the axioms of quantum mechanics is the probability amplitude for the transition $\vec{p}_1\to\vec{p}_2$. We stress this point because infrared divergences encountered in short-range perturbation theory can create the false impression that the amplitude is ambiguous or ill-defined. It is not: the inner product is well-defined non-perturbatively, and the divergences are an artifact of applying short-range perturbation theory. Stripping the overall energy-conserving delta function, we write
	\begin{align}
		\mathcal{A}(\vec{p}_1\rightarrow\vec{p}_2) \defined \frac{2\pi^2}{m|\vec{p}|}\delta(E_1-E_2)\,A(\hat{p}_1\cdot\hat{p}_2),
	\end{align}
	where $A(x)$ is the reduced amplitude. Treating the Hamiltonian \eqref{hamil} as the full UV-complete theory, we can solve for the exact in- and out-wavefunctions of \eqref{hamil} (see Appendix~A of \cite{Lippstreu:2023vvg}, and also \cite{kang1962higher,pilkuhn2013relativistic}), and then compute the inner product of these states to obtain the exact partial-wave expansion
	\begin{gather}
		A(x)=\sum_{l=0}^{\infty}(2l+1)\,S_l
		P_{l}(x),\label{exact amp}\\
		S_l=\frac{\Gamma\!\left(\sqrt{(l+\tfrac{1}{2})^{2}-\beta^{2}}+\tfrac{1}{2}+i\gamma\right)}
		{\Gamma\!\left(\sqrt{(l+\tfrac{1}{2})^{2}-\beta^{2}}+\tfrac{1}{2}-i\gamma\right)}
		\exp\!\left[i\pi\!\left(l+\tfrac{1}{2}-\sqrt{(l+\tfrac{1}{2})^{2}-\beta^{2}}\right)\right]\label{Sl exact def}
	\end{gather}
	where $\gamma = m\alpha/|\vec{p}\,|$ is the Sommerfeld parameter and $P_l$ are the Legendre polynomials. This exact result serves as the benchmark against which all perturbative approximations in this paper are tested.
	
	The partial-wave series in \eqref{exact amp} does not converge as an ordinary function. This is not a problem: amplitudes are always understood as distributions, and \eqref{exact amp} makes perfect sense as a distribution \cite{Taylor}. This is analogous to the familiar free-theory identity
	\begin{gather}
		\sum_{l=0}^{\infty}
		(2l+1)P_l(\hat p_1\cdot \hat p_2)
		=
		4\pi\,\delta^{(2)}(\hat p_1-\hat p_2).
		\label{forwards}
	\end{gather}
	The series on the left-hand side does not converge as an ordinary function, but it nevertheless has a well-defined meaning as a distribution.
	More generally, \eqref{exact amp} is defined by its action on normalizable wave packets. If $f(\hat{p})$ is a normalizable function on the sphere that admits a partial-wave expansion
	\begin{gather}
		f(\hat p)
		=
		\sum_{l,m}
		f_{lm}Y_{lm}(\hat p),
	\end{gather}
	then the exact scattering operator acts diagonally in the partial-wave basis,
	\begin{gather}
		f_{lm}
		\longmapsto
		S_l f_{lm},
	\end{gather}
	where $S_l$ is defined in \eqref{Sl exact def}. Thus the amplitude is a well-defined distributional kernel, even though the partial-wave sum need not converge pointwise. This distributional interpretation was discussed in Section~\ref{sect: nonconvergence}; see also Appendix~B.3 of \cite{Lippstreu:2023vvg} for a more detailed discussion.
	
	\subsection{The non-perturbative unitarity bound}
	The exact amplitude \eqref{exact amp} satisfies momentum-space unitarity \eqref{unitarity} provided each partial-wave coefficient has unit modulus,
	\begin{gather}
		|S_l|=1.
	\end{gather}
	For the partial-wave coefficients \eqref{Sl exact def}, this holds when
	\begin{gather}
		\gamma\in\mathbb{R},
		\qquad
		\beta^2\leq\frac{1}{4}.
	\end{gather}
	The first obstruction occurs in the $s$-wave. When $\beta^2>\frac{1}{4}$, the square roots in \eqref{Sl exact def} become imaginary for $l=0$, and the corresponding $S_0$ is no longer a pure phase. More generally, any partial wave satisfying
	\begin{gather}
		\beta^2>
		\left(l+\frac{1}{2}\right)^2
	\end{gather}
	has $|S_l|\neq 1$. Thus elastic unitarity is lost unless
	\begin{gather}
		\boxed{\beta^2\leq\frac{1}{4}.}\label{eq: non pert unitarity bound}
	\end{gather}
	This is an exact bound if the Hamiltonian \eqref{hamil}, with no additional short-distance data, is treated as the full UV-complete point-particle theory\footnote{One can construct UV-complete theories that exceed this bound, but these theories require additional UV data, such as a different boundary condition at the origin \cite{EssinGriffiths2006,BawinCoon2003,BarfordBirse2005}. Correspondingly, the resulting amplitudes and low-energy EFTs for such theories will differ from those generated by \eqref{hamil} with the standard boundary condition at the origin. See Section~\ref{sect:more details on non-perturbative bound} for more details.}. In the rest of this paper, however, we will treat \eqref{hamil} as an EFT valid up to a cutoff scale $\Lambda_{\rm EFT}$, for example the inverse size of a nucleus. We will derive perturbative bounds on $\beta^2$ that depend on $\Lambda_{\rm EFT}$. In the limit $\Lambda_{\rm EFT}\to\infty$, these bounds will also recover the point-particle result \eqref{eq: non pert unitarity bound}, without introducing any infrared divergences or arbitrary infrared scales at any intermediate step, thus providing a consistency check on our methods. In other words, we will derive bounds on $\beta^2$ treating \eqref{hamil} as an EFT with many possible UV completions, and our bounds will be valid for any UV completion; treating \eqref{hamil} as the full UV theory is one possible UV completion.

    For more details on the non-relativistic quantum-mechanical analysis of the $1/r^2$ potential, see \cite{Case1950,EssinGriffiths2006,BawinCoon2003}.

	As an indication of how robust the bound is, consider one possible embedding of \eqref{hamil} in a UV theory: relativistic scalar QED with a heavy charged nucleus treated as an external source. For a charged scalar in the Coulomb field of an infinitely heavy point source of charge $Ze$, the stationary Klein--Gordon equation takes exactly the radial form of \eqref{hamil}, with the inverse-square term generated by the square of the Coulomb potential, so that $\beta=Z\alpha_{\rm em}$. In this embedding $\beta^2$ is not a free coupling but is fixed by the charge of the source, and the bound \eqref{eq: non pert unitarity bound} becomes a bound on that charge, $Z\alpha_{\rm em}\le\frac{1}{2}$. Naively, one could then violate \eqref{eq: non pert unitarity bound} simply by increasing $Z$. However, this is precisely where the scalar QED vacuum becomes unstable. Once $Z$ exceeds $1/(2\alpha_{\rm em})\approx68$, the strong field of the nucleus will create a scalar--antiscalar pair. The oppositely charged member of the pair is attracted to the nucleus and becomes bound to it, thereby screening the charge back below the critical value. This relativistic embedding therefore does not provide a stable nucleus with a supercritical charge. Similar conclusions are reached for the scattering of a spin-$\frac{1}{2}$ particle on a nucleus, where the critical coupling is instead $Z\alpha_{\rm em}=1$ \cite{Zeldovich:1972SuperheavyAtoms,Pieper:1969InteriorShells,GreinerMuellerRafelski1985}. Thus QED does not allow stable pointlike charged particles or nuclei with charge greater than $1/\alpha_{\rm em}\approx137$, while scalar QED does not allow stable point charges with charge greater than $1/(2\alpha_{\rm em})\approx68$ \cite{GreinerMuellerRafelski1985}.

	Of course, real nuclei have a finite radius. Accounting for this finite size increases the critical atomic number relative to the point-particle estimate \cite{GreinerMuellerRafelski1985,bawin1981instability}. For example, for fermions the critical charge increases to $Z_{\text{crit}}\sim 173$ for extended nuclei. This physical effect will be reflected in our EFT bounds on $\beta^2$: as the EFT cutoff $\Lambda_{\rm EFT}$ is lowered, the allowed values of $\beta^2$ increase for a repulsive potential $\alpha>0$. In this sense, the EFT bounds track the way in which the critical atomic number increases as the nuclear radius is increased \cite{bawin1981instability}.

	This physical interpretation of the bound highlights an important reason to study this model. The physics of the bound, namely the instability of the vacuum in a supercritical Coulomb field, has nothing to do with the energy resolution of the detector used to measure the nucleus. By contrast, if one uses standard short-range perturbation theory to derive approximate bounds on $\beta^2$, as in Section~\ref{sect: naive}, infrared scales enter the result. This could suggest that the bound depends on experimental details, such as the timescale of the measurement\footnote{In QED, spontaneous electron--positron pair production in the Coulomb field of a supercritical nucleus occurs on a timescale of about $10^{-19}\,\mathrm{s}$ \cite{GreinerMuellerRafelski1985}.}. In this example, at least, that appears to be the wrong physical conclusion.
	
	As we will see in Section~\ref{sect:more details on non-perturbative bound}, the bound can be seen more directly from the radial Schrödinger equation: when $\beta^2>\frac{1}{4}$, the corresponding radial wavefunction requires additional short-distance boundary data at the origin to define a unitary theory. Our goal is not to give the shortest derivation of the bound, but to show that such bounds can be detected from perturbative amplitude data alone, mirroring the logic of modern derivations of bounds on EFT couplings \cite{deRham:2022hpx}.

	\subsection{More details on the non-perturbative bound}\label{sect:more details on non-perturbative bound}
	A theory is specified not only by its Hamiltonian but also by the domain of that Hamiltonian, or equivalently by the boundary condition imposed on wavefunctions at the origin.
	Standard perturbation theory \eqref{short pert} implicitly makes a choice of boundary condition at the origin, and the non-perturbative bound \eqref{eq: non pert unitarity bound} applies to this choice. Because this is important for understanding the bound, we now explain it in more detail.

	Consider the radial Schr\"{o}dinger equation near the origin. Writing the wavefunction as $\psi(r,\Omega)=u_l(r)Y_{lm}(\Omega)/r$, the terms that dominate as $r\to0$ are
	\begin{gather}
		\left[
		-\frac{d^2}{dr^2}
		+
		\frac{l(l+1)-\beta^2}{r^2}
		\right]u_l(r)
		\simeq 0 .
	\end{gather}
	The two independent solutions behave as
	\begin{gather}
		u_l(r)\sim A_l r^{1/2+\nu_l}+B_l r^{1/2-\nu_l},
		\qquad
		\nu_l=
		\sqrt{\left(l+\frac12\right)^2-\beta^2}.
		\label{eq:near-origin-solutions}
	\end{gather}
	For subcritical couplings, $\beta^2<\frac{1}{4}$, $\nu_l$ is real and positive. We will choose the boundary condition at the origin $B_l=0$
	\begin{gather}
		u_l(r)\sim A_l r^{1/2+\nu_l},
		\qquad r\to0,
		\label{eq:standard-boundary-condition}
	\end{gather}
    which is the boundary condition implicitly selected by standard perturbation theory. By standard perturbation theory we mean \eqref{short pert} together with the usual free Green's function\footnote{For a fixed partial wave, the usual outgoing free radial Green's function has the form $g_{0l}^{(+)}(r,r';p)\propto [r_<j_l(pr_<)][r_>h_l^{(1)}(pr_>)]$, so its leg approaching the origin behaves as $r_<j_l(pr_<)\sim r_<^{l+1}$ and imposes the regular boundary condition. Since the Born series starts from the regular free partial wave and uses this same Green's function at every order, it cannot generate an independent irregular homogeneous component; its resummation therefore selects $u_l\sim r^{1/2+\nu_l}$, or $B_l=0$.}. A nonzero $B_l$ would require additional short-distance boundary data, represented in the Green's function by an appropriate homogeneous term. This $B_l=0$ boundary condition is the short-distance boundary condition implicitly used in the exact amplitude \eqref{exact amp}.
	The $s$-wave $B_l=0$ branch remains unitary at $\beta^2=\frac{1}{4}$, so the bound is $\beta^2\leq\frac{1}{4}$ rather than $\beta^2<\frac{1}{4}$.
    
However, in any partial wave for which
\begin{gather}
    \beta^2>
    \left(l+\frac12\right)^2,
    \label{eq:supercritical-condition}
\end{gather}
the exponent $\nu_l$ becomes complex. Writing
	\begin{gather}
		\sigma_l
		\defined
		\sqrt{
			\beta^2
			-
			\left(l+\frac12\right)^2
		}
		\label{eq:supercritical-sigma}
	\end{gather}
	the two solutions become
	\begin{gather}
		u_l(r)
		\sim
		A_l r^{1/2+i\sigma_l}
		+
		B_l r^{1/2-i\sigma_l}.
		\label{eq:supercritical-solutions}
	\end{gather}
	The two modes in \eqref{eq:supercritical-solutions}, with $\sigma_l$ defined
	in \eqref{eq:supercritical-sigma}, carry opposite radial probability flux.
	Computing amplitudes using standard perturbation theory in the supercritical
	regime \eqref{eq:supercritical-condition} implicitly sets $B_l$ to zero, and
	the resulting theory is non-unitary because the origin acts as a source of probability.

	One can, however, define unitary theories that exceed the critical coupling \eqref{eq: non pert unitarity bound}; we refer to these as supercritical theories. In such theories, the boundary condition at the origin is different. One may instead impose a self-adjoint boundary condition of the form
	\begin{gather}
		B_l
		=
		e^{i\theta_l}
		A_l
		\label{eq:self-adjoint-boundary-condition}
	\end{gather}
	with real $\theta_l$.\footnote{A reference scale is implicit in the
	complex powers $r^{\pm i\sigma_l}$, and changing this reference shifts
	$\theta_l$. Equivalently, the single parameter specifying the
	self-adjoint extension can be expressed as a physical boundary-condition
	scale. Its appearance is an instance of dimensional transmutation in the
	supercritical inverse-square problem \cite{HammerSwingle2006}; this scale
	is not an arbitrary infrared regulator.}
	The boundary condition \eqref{eq:self-adjoint-boundary-condition} enforces zero net probability flux through the origin.\footnote{The outward probability flux through a small sphere is
\[
r^2\int\mathrm d\Omega\,\hat r\cdot\vec{\jmath}
=
\frac{1}{m}\operatorname{Im}(u_l^*u_l')
=
\frac{\sigma_l}{m}\left(|A_l|^2-|B_l|^2\right),
\qquad
\vec{\jmath}=\frac{1}{m}\operatorname{Im}(\psi^*\vec\nabla\psi).
\]
Since \(B_l=e^{i\theta_l}A_l\) implies \(|A_l|=|B_l|\), the flux vanishes.} Physically, it represents elastic reflection from unresolved short-distance physics. The boundary condition introduces one additional short-distance parameter, which encodes the physics that replaces the singular point-particle potential \cite{Case1950,EssinGriffiths2006,BawinCoon2003}.

	Such supercritical quantum systems are useful in real physical problems. A standard example is Efimov three-body physics, where the low-energy dynamics is governed by a supercritical inverse-square potential; see \cite{BawinCoon2003} for further applications.

	In summary, the precise statement of the bound is that \textit{the theory with $\beta^2>\frac{1}{4}$ and the standard boundary condition at the origin, $B_l=0$, is not unitary}. Thus, whether it is treated as a low-energy EFT or as a UV-complete theory, it violates unitarity. Supercritical unitary theories exist, but they have a different boundary condition at the origin and, consequently, different low-energy scattering amplitudes. Furthermore, a supercritical unitary theory will never flow down to a supercritical $B_l=0$ theory, because unitarity should be preserved along the renormalization-group flow. See \cite{BarfordBirse2003,BarfordBirse2005} for the renormalization-group flow of supercritical theories.

	Using the radial Schr\"{o}dinger equation, the non-perturbative bound is simple and unambiguous to derive, and we do not claim that the $\beta^2 \leq \frac{1}{4}$ bound is new or nontrivial. Rather, this is perhaps the simplest example in which analyzing a theory's low-loop amplitude data, in the manner of the modern EFT bounds literature, leads to ambiguous bounds: the amplitudes carry infrared divergences, and, furthermore, integrating the amplitude over the forward direction is ambiguous. As we show in Section~\ref{sect: naive}, bounds derived by these traditional techniques inherit these ambiguities, even though the non-perturbative bound is known to be unambiguous. The purpose of this paper is to show how to analyze perturbative amplitude data so as to infer this bound in a well-defined, unambiguous way.

	This derivation of the bound using the radial Schr\"{o}dinger equation also makes clear why the bound is insensitive to infrared deformations. For example, replacing the Coulomb potential with a Yukawa potential (giving the photon a mass $\lambda_{\text{I.R.}}$) changes the large-distance behavior but leaves the short-distance behavior unchanged:
	\begin{gather}
		\frac{\alpha e^{-\lambda_{\text{I.R.}} r}}{r}
		=
		\frac{\alpha}{r}
		+
		\mathcal{O}(\lambda_{\text{I.R.}}),
		\qquad r\to0.
	\end{gather}
	The critical behavior is controlled by the inverse-square term in the near-origin radial equation and is therefore unaffected by a small photon mass. By contrast, low orders in ordinary short-range perturbation theory introduce spurious dependence on the infrared regulator or photon mass (see, for example, \eqref{first bound naive short}). Bounds derived directly from such an expansion can therefore appear to depend on an arbitrary infrared scale, even though the exact non-perturbative bound does not.
	
\subsection{More singular potentials and UV renormalization}
For the $\frac{1}{r^2}$ interaction, both ordinary perturbation theory and DWPT are UV finite order by order (although their perturbative series fail to converge in the supercritical regime, $\beta^2 > \frac{1}{4}$). The situation changes when the Hamiltonian contains interactions more singular than $1/r^2$. For example, consider
\begin{gather}
	\hat{H}
	=
	\frac{\hat{p}^2}{2m}
	+\frac{\alpha}{\hat{r}}
	-\frac{\beta^2}{2m\hat{r}^2}
	+\frac{c_1}{m\hat{r}^3\Lambda}
	+\frac{c_2}{m\hat{r}^4\Lambda^2}+\cdots,
	\label{higher order}
\end{gather}
where $\Lambda$ is the EFT cutoff. Using standard perturbation theory with these additional terms generally produces UV divergences because of their singular behavior near the origin. The isolated inverse-square interaction therefore provides one of the simplest settings to study the appearance of IR scales in perturbative unitarity bounds without also having to deal with the additional complications of UV renormalization. See \cite{Long_2008} for an explicit treatment of the renormalization of singular potentials.

UV divergences signal that additional short-distance input, in the form of a boundary condition at the origin, is required to specify the low-energy EFT. Thus, unlike the inverse-square potential, the conflict between the boundary condition at the origin chosen by standard perturbation theory and a given range of Wilson coefficients is more immediate. In this case, one would first perform the non-perturbative analysis of the near-origin behavior and choose a boundary condition at the origin that is consistent with unitarity and build a perturbation theory and renormalization procedure around that unitarity condition. Therefore, if the set of Wilson coefficients $c_i$ is constrained by unitarity, the allowed values will be more subtle to derive than the non-perturbative analysis above.

Nonetheless, we expect the standard methods for deriving such perturbative unitarity bounds on these couplings to encounter the same IR problems---ill-defined partial-wave projections and IR-divergent amplitudes---regardless of the renormalization prescription or boundary condition imposed at the origin. As before, resolving these problems should require accounting for the modified distributional structure and using Coulomb wavefunctions to describe the asymptotic dynamics. In this case, however, imposing the chosen boundary condition at the origin generally requires a linear combination of the regular and irregular solutions of the radial Coulomb equation.

	\section{Naive short-range unitarity bounds}\label{sect: naive}
	In this section we use low orders of short-range perturbation theory to derive partial-wave unitarity bounds on the $\beta^2$ coupling. The purpose is to highlight the problems that arise when one tries to derive unitarity bounds on couplings in theories with long-range forces using short-range perturbation theory. In particular, we will see that (a) the amplitudes cannot be integrated unambiguously over the forward direction, and consequently the partial-wave coefficients are not well defined without a further prescription; (b) the resulting unitarity bounds depend on the infrared regulator; and (c) higher orders in perturbation theory can dominate lower orders in the bounds.

	\subsection{Tree amplitudes using short-range perturbation theory}
	
	If one applies short-range perturbation theory to the Hamiltonian \eqref{hamil}, one obtains the formal amplitude
	\begin{align}
		\mathcal{A}_{\rm short}(\vec{p}_1\rightarrow \vec{p}_2)
		&=
		\braket{\vec p_2|T\!\left(\exp\!\left[-i\int_{-\infty}^{\infty}\DD t\, V_{I}(t)\right]\right)|\vec p_1}.
		\label{short pert}
	\end{align}
	Here $\ket{\vec p_i}$ are free states and $V_{I}(t)$ is the potential in the interaction picture,
	\begin{gather}
		V_I(t)
		=
		e^{i\hat{H}_0t}
		\left(
		\frac{\alpha}{\hat{r}}
		-
		\frac{\beta^2}{2m}\frac{1}{\hat{r}^2}
		\right)
		e^{-i\hat{H}_0t}.
	\end{gather}
	Short-range perturbation theory is constructed to contain the free disconnected component,
	\begin{gather}
		\mathcal{A}_{\rm short}(\vec{p}_1\rightarrow \vec{p}_2)
		=
		(2\pi)^3\delta^{(3)}(\vec{p}_1-\vec{p}_2)
		+
		\sum_{n=0}^{\infty}
		\mathcal{A}^{(n)}_{\rm short}(\vec{p}_1\rightarrow \vec{p}_2),
	\end{gather}
	where $n$ indicates the Born order. This is already different from the exact long-range amplitude: the exact amplitude does not contain this particular disconnected component \cite{herbst1974connectedness,Yafaev:2000ScatteringTheory,Yafaev:1998LongRangeAmplitude}. The issue is that \eqref{short pert} is not equal to
	\begin{gather}
		\braket{\psi_{\mathrm{out}}(\vec p_2)|\psi_{\mathrm{in}}(\vec p_1)}.
	\end{gather}
	The point is that short-range perturbation theory does not compute the exact inner product of the full in-state with the full out-state, where these states are defined as eigenstates of the full Hamiltonian that look as much as possible like free plane waves at early and late times, respectively. The exact amplitude does contain a disconnected component that is only important in the forward direction, similar to $\delta^{(3)}(\vec{p}_1-\vec{p}_2)$, but it is distinct from this particular Dirac delta distribution.
	
	Using short-range perturbation theory, the tree-level term is, with $\vec q=\vec p_2-\vec p_1$,
	
	\begin{align}
		\mathcal{A}^{(0)}_{\rm short}(\vec p_1\rightarrow \vec p_2)
		&=
		-2\pi i\delta(E_1-E_2)
		\int \DD^3\vec{x}\,
		e^{-i\vec{q}\cdot\vec{x}}
		\left(
		\frac{\alpha}{|\vec{x}|}
		-
		\frac{\beta^2}{2m}\frac{1}{|\vec{x}|^2}
		\right)
		\\
		&=
		-2\pi i\delta(E_1-E_2)
		\left(
		\frac{4\pi\alpha}{|\vec{q}|^{\,2}}
		-
		\frac{\pi^2\beta^2}{m|\vec{q}|}
		\right).
		\label{tree short}
	\end{align}

	\subsection{Tree-level unitarity bounds}\label{sect:tree unitarity bounds naive}
	
	To project onto partial waves, we define the reduced angular kernel by
	\begin{gather}
		\mathcal{A}(\vec p_1\rightarrow \vec p_2)
		\defined
		\frac{2\pi^2}{mp}\,
		\delta(E_1-E_2)\,
		A(\hat p_1\cdot \hat p_2),
		\qquad
		p=|\vec p_1|=|\vec p_2|.
	\end{gather}
	We then expand
	\begin{gather}
		A(x)
		=
		\sum_{l=0}^{\infty}
		(2l+1)
		\left(
		1+2i a_l
		\right)
		P_l(x),
		\qquad
		x=\hat p_1\cdot\hat p_2.
	\end{gather}
	Equivalently, for the connected part,
	\begin{gather}
		A_{\rm conn}(x)
		=
		2i
		\sum_{l=0}^{\infty}
		(2l+1)a_l P_l(x),
	\end{gather}
	so that
	\begin{gather}
		a_l
		=
		\frac{1}{4i}
		\int_{-1}^{1}\DD x\,
		A_{\rm conn}(x)P_l(x).
		\label{projection formula}
	\end{gather}
	Using the tree-level amplitude \eqref{tree short} and the projection formula \eqref{projection formula}, the $\beta^2$ term gives
	\begin{gather}
		a_l^{(0),\beta^2}
		=
		\frac{\pi\beta^2}{2(2l+1)},
	\end{gather}
	which is indeed consistent with \eqref{consistent betasquared}.
	
	The tree-level Coulomb term in \eqref{tree short} proportional to $1/|\vec q|^{\,2}$ is not
	integrable over the forward direction and hence does not have a well-defined
	partial-wave projection. We regulate it by giving the photon a mass $\lambda_{\text{I.R.}}$,
	so that
	\begin{gather}
		\frac{1}{|\vec q|^{\,2}}
		\longrightarrow
		\frac{1}{|\vec q|^{\,2}+\lambda_{\text{I.R.}}^2}.
	\end{gather}
	Using $|\vec q|^2=2p^2(1-x)$, the regulated Coulomb contribution
	to the partial-wave coefficient is
	\begin{gather}
		a_l^{(0),\alpha}
		=
		-\frac{\gamma}{2}
		\int_{-1}^{1}\DD x\,
		\frac{P_l(x)}
		{1-x+\frac{\lambda_{\text{I.R.}}^2}{2p^2}}
		=
		-\gamma\,
		Q_l\!\left(1+\frac{\lambda_{\text{I.R.}}^2}{2p^2}\right),
	\end{gather}
	where $\gamma=\alpha m/|\vec p|$ is the Sommerfeld parameter and $Q_l$ is the
	Legendre function of the second kind.
	
	For small $\lambda_{\text{I.R.}}$, this gives
	\begin{gather}
		a_l^{(0)}
		=
		-\gamma
		\left[
		\log\!\left(\frac{2p}{\lambda_{\text{I.R.}}}\right)
		-
		H_l
		\right]
		+
		\frac{\pi\beta^2}{2(2l+1)}
		+
		\mathcal{O}(\lambda_{\text{I.R.}}^2\log\lambda_{\text{I.R.}}),
	\end{gather}
	where $H_l$ is the harmonic number of order $l$.
	We are therefore tempted to impose the approximate tree-level unitarity bound
	\begin{gather}
		\left|
		-\gamma
		\left[
		\log\!\left(\frac{2p}{\lambda_{\text{I.R.}}}\right)
		-
		H_l
		\right]
		+
		\frac{\pi\beta^2}{2(2l+1)}
		\right|
		\lesssim
		\frac{1}{2}.\label{first bound naive short}
	\end{gather}
	This already displays two basic problems with applying short-range
	partial-wave bounds in a theory with a long-range force:
	\begin{itemize}
		\item The bound is ambiguous, since it depends on the infrared regulator, here the photon mass $\lambda_{\text{I.R.}}$. Taken at face value, \eqref{first bound naive short} becomes pathological as the regulator scale is taken to zero: the logarithm $\log(2p/\lambda_{\text{I.R.}})$ grows without bound, so no fixed finite value of $\beta^2$ can satisfy the inequality for sufficiently small $\lambda_{\text{I.R.}}$. This highlights the need to resum these logarithms as in \cite{Chang:2025cxc}.
		One might be tempted to interpret the presence of an infrared scale in the bound as saying that the bound on $\beta^2$ depends on the infrared resolution of the measurement device used to extract the coupling. In the present model, however, this interpretation is likely incorrect. The exact non-perturbative bound \eqref{eq: non pert unitarity bound} contains no infrared scale. Furthermore, the physical interpretation of the non-perturbative bound, namely the instability of sufficiently highly charged nuclei, is not related to the IR resolution with which one measures the nucleus. Thus the $\lambda_{\text{I.R.}}$-dependence in \eqref{first bound naive short} is not physical detector dependence; it is a symptom of applying short-range perturbation theory to a long-range problem.

		\item The regulator also makes the large-$l$ behavior ambiguous. A
		partial wave with angular momentum $l$ probes impact parameters
		$b\sim l/p$, while the photon mass introduces the screening length
		$1/\lambda_{\text{I.R.}}$. Thus the partial waves depend on the ratio
		$l\lambda_{\text{I.R.}}/p$. If one first takes $\lambda_{\text{I.R.}}\to0$ at fixed $l$, then
		\begin{gather}
			Q_l\!\left(1+\frac{\lambda_{\text{I.R.}}^2}{2p^2}\right)
			=
			\log\!\left(\frac{2p}{\lambda_{\text{I.R.}}}\right)-H_l+\cdots .
		\end{gather}
		But this expansion is only appropriate when $l\lambda_{\text{I.R.}}/p\ll1$. For
		$l\lambda_{\text{I.R.}}/p\gtrsim1$, the Yukawa screening of the Coulomb tail is already
		being resolved by the partial wave, and the large-$l$ behavior is
		different. Hence there is no regulator-independent large-$l$ bound: the
		answer depends on how the limits $l\to\infty$ and $\lambda_{\text{I.R.}}\to0$ are taken.
	\end{itemize}

	\subsection{One-loop amplitudes}
	
	The same obstruction appears at the next order in short-range perturbation
	theory; see also \cite{Oller:2022Coulomb}. In
	Appendix~\ref{app:1loop amplitudes} we compute the second Born
	terms obtained from \eqref{short pert}. The infrared-sensitive pieces contain
	the characteristic Coulomb logarithms
	\begin{gather}
		I_{CC}
		\sim
		-\frac{4\pi i m\alpha^2}{p|\vec q|^2}
		\log\!\left(\frac{|\vec q|^2}{\lambda_{\text{I.R.}}^2}\right),
		\qquad
		I_{CS}
		\sim
		\frac{2\pi\alpha\beta^2}{p|\vec q|}
		i\pi
		\log\!\left(\frac{|\vec q|}{\lambda_{\text{I.R.}}}\right),
	\end{gather}
	where $\lambda_{\text{I.R.}}$ is the infrared regulator. These terms illustrate three related
	problems.
	
	First, the amplitudes depend explicitly on the regulator $\lambda_{\text{I.R.}}$. Therefore
	any unitarity bound obtained from these amplitudes will
	inherit this unphysical dependence.
	
	Second, the loop expansion is not uniformly perturbative in the forward region.
	The Coulomb--Coulomb contribution $I_{CC}$ contains the tree-level Coulomb singularity
	multiplied by an additional logarithm,
	\begin{gather}
		\frac{\alpha}{|\vec q|^2}
		\longrightarrow
		\frac{\alpha}{|\vec q|^2}
		\left[
		1
		+
		\mathcal{O}\!\left(
		\gamma\log\!\frac{|\vec q|^2}{\lambda_{\text{I.R.}}^2}
		\right)
		+
		\cdots
		\right].
	\end{gather}
	Thus, for sufficiently small $|\vec q|$, higher orders are not parametrically
	suppressed relative to lower orders. 
	Because unitarity involves integrals over all scattering angles, including the
	forward region, a fixed-order loop expansion is unreliable due to these logarithms.

	Third, because these fixed-order amplitudes are not integrable over the forward
	direction, their partial-wave projections are not defined without an infrared
	prescription. After introducing a regulator, the resulting partial waves and
	the bounds derived from them depend on the regulator. 
	
	We now turn to distorted-wave
	perturbation theory, where these problems are absent.

	\section{Distorted-wave perturbation theory}\label{sect:DWPT}

	Our goal is to compute the inner product of the in-state with the out-state for the full Hamiltonian \eqref{hamil}. Usually, one assumes that the in-state looks like a free plane wave at early times. This assumption is incorrect in the presence of long-range forces, and building it into perturbation theory is precisely what produces infrared divergences. A perturbative framework that avoids this incorrect assumption is distorted-wave perturbation theory \cite{Bethe,Glauber,CROTHERS1992287,Lippstreu:2023vvg,Lippstreu:2025jit}:
	\begin{align}
		\mathcal{A}(\vec{p}_1\rightarrow \vec{p}_2)
		&= \braket{\psi_{\t{out}}(\vec p_2)|\psi_{\t{in}}(\vec p_1)} \\
		&=
		\left\langle
		\phi_{\t{out}}(\vec p_2)
		\left|
		T\!\left(
		\exp\!\left[-i\int_{-\infty}^{\infty}\DD t\, \tilde{V}_{I}(t)\right]
		\right)
		\right|
		\phi_{\t{in}}(\vec p_1)
		\right\rangle .
		\label{long pert}
	\end{align}
	Here $\tilde{V}_{I}(t)$ is the short-range interaction in the following interaction-like picture:
	\begin{gather}
		\tilde{V}_{I}(t)
		=
		-e^{i\hat{H}_ct}\,
		\frac{\beta^2}{2m}\frac{1}{\hat r^2}\,
		e^{-i\hat{H}_ct},
		\label{long pert 2}\\
		\hat{H}_c
		=
		\frac{\hat p^2}{2m}
		+
		\frac{\alpha}{\hat r}.
		\label{long pert 3}
	\end{gather}
	The states $\ket{\phi_{\t{in/out}}(\vec p)}$ are eigenstates of the Coulomb Hamiltonian,
	\begin{gather}
		\hat{H}_c\ket{\phi_{\t{in/out}}(\vec p)}
		=
		\frac{p^2}{2m}\ket{\phi_{\t{in/out}}(\vec p)} .
	\end{gather}
	Their explicit expressions are
	\begin{align}
		\phi_{\mathrm{in}}(\vec{x},\vec{p})
		&=
		\Gamma(1+i\gamma)e^{-\frac{\pi\gamma}{2}}
		e^{-iEt+i\vec{p}\cdot\vec{x}}\,
		{}_1F_1\!\left(
		-i\gamma,
		1,
		i\bigl(|\vec{p}|\,|\vec{x}|-\vec{p}\cdot\vec{x}\bigr)
		\right)\label{coulomb in}, \\[4pt]
		\phi_{\mathrm{out}}(\vec{x},\vec{p})
		&=
		\Gamma(1-i\gamma)e^{-\frac{\pi\gamma}{2}}
		e^{-iEt+i\vec{p}\cdot\vec{x}}\,
		{}_1F_1\!\left(
		i\gamma,
		1,
		-i|\vec{p}|\,|\vec{x}|-i\vec{p}\cdot\vec{x}
		\right).\label{coulomb out}
	\end{align}
	The in- and out-wavefunctions each form a complete orthonormal basis \cite{Mukunda:1978CoulombCompleteness,Mukhamedzhanov:2008CoulombCompleteness,michel2008direct}.\footnote{For attractive potentials $\alpha<0$, one must also include the bound states in the completeness relation. These states are straightforward to include \cite{Lippstreu:2023vvg} and do not significantly alter the amplitudes. For simplicity, however, we work with the repulsive potential $\alpha>0$.} The crucial point in \eqref{long pert} is that, unlike short-range perturbation theory \eqref{short pert}, DWPT does compute the inner product of the in-state with the out-state.

	\subsection{Tree-level amplitude}
	The leading, or ``tree-level'', term in this perturbation theory is the inner product of the Coulomb wavefunctions \eqref{coulomb in} and \eqref{coulomb out} \cite{Lippstreu:2023vvg},
	\begin{align}
		\mathcal{A}^{(0)}(\vec{p}_1,\vec{p}_2)
		&=
		\langle \phi_{\mathrm{out}}(\vec{p}_2) | \phi_{\mathrm{in}}(\vec{p}_1) \rangle \\
		&=
		\int\DD^3\vec{x}\,
		\phi_{\mathrm{out}}^*(\vec{x},\vec{p}_2)\,
		\phi_{\mathrm{in}}(\vec{x},\vec{p}_1) \\
		&=
		-\lim_{\epsilon_{+}\rightarrow 0^+}
		\frac{8i\pi^2\alpha}{|\vec{p}_1-\vec{p}_2|^2}\,
		\delta(E_1-E_2)\,
		\frac{\Gamma(1+i\gamma)}{\Gamma(1-i\gamma)}\,
		\left(
		\frac{2|\vec{p}_1|}{|\vec{p}_1-\vec{p}_2|}
		\right)^{2i\gamma-2\epsilon_{+}} .
		\label{trees}
	\end{align}
	This closely resembles the familiar eikonal formula \cite{Kabat_1992}, but with two important differences. First, there is no infrared divergence and no infrared regulator: for example, there is no photon mass, no dimensional-regularization scale or parameter, and no arbitrary cutoff scale $\Lambda_{\t{IR}}$. Second, the exponent contains the $\epsilon_{+}$ prescription, which is absent from the usual eikonal formula but is essential for unitarity. This distributional prescription was discussed in detail in Section~\ref{sect: forward direction}. The presence of the $\epsilon_{+}$ prescription allows us to integrate over the forward direction correctly and unambiguously.

	Although distorted-wave perturbation theory has been used many times before, our emphasis is that it does not make the incorrect assumption that the asymptotic dynamics is free. Consequently, it gives a perturbative expansion without artificial infrared divergences. The finiteness properties of the expansion are analyzed in Section~\ref{sect:finiteness DWPT}. Since the infrared divergence has been eliminated, the resulting unitarity bounds will not depend on an infrared regulator.
	
	Let us emphasize that, although the DWPT expansion \eqref{long pert} is written as a time-ordered exponential, its zeroth-order term is not a $\delta^{(3)}(\vec{p}_1-\vec{p}_2)$. The reason is that the states between which the exponential is sandwiched are not free states, but Coulomb in- and out-states. Thus DWPT already incorporates the nontrivial long-range part of the scattering at zeroth order, and therefore faithfully captures the distributional structure of the exact non-perturbative amplitude.
	
	\subsection{First-order correction}
	
	At the next order in the DWPT expansion \eqref{long pert}, we have
	\begin{align}
		\mathcal{A}^{(1)}
		&=
		2\pi i\frac{\beta^2}{2m}
		\delta(E_1-E_2)
		\left\langle
		\phi_{\t{out}}(\vec p_2)
		\left|
		\frac{1}{\hat r^2}
		\right|
		\phi_{\t{in}}(\vec p_1)
		\right\rangle
		\label{first-order-dwpt}
		\\
		&=
		2\pi i\frac{\beta^2}{2m}
		\delta(E_1-E_2)
		\int\DD^3\vec{x}\,
		\phi_{\t{out}}^*(\vec{x},\vec{p}_2)
		\frac{1}{|\vec{x}|^2}
		\phi_{\t{in}}(\vec{x},\vec{p}_1) .\label{rep i used}
	\end{align}
	In the second line, we inserted a complete set of position states. Although this integral can be evaluated directly (see Appendix~B of \cite{Lippstreu:2023vvg}), it is useful to rewrite it in a form that makes the structure of DWPT more transparent. Inserting a complete set of Coulomb in-states gives\footnote{For an attractive Coulomb potential, the completeness relation also includes bound states. These states are orthogonal to the scattering in- and out-states and therefore do not contribute here.}
	\begin{align}
		\mathcal{A}^{(1)}(\vec{p}_1\rightarrow \vec{p}_2)
		&=
		2\pi i\frac{\beta^2}{2m}
		\delta(E_1-E_2)
		\int\frac{\DD^3\vec\ell}{(2\pi)^3}\,
		\braket{\phi_{\t{out}}(\vec p_2)|\phi_{\t{in}}(\vec\ell)}
		\left\langle
		\phi_{\t{in}}(\vec\ell)
		\left|
		\frac{1}{\hat r^2}
		\right|
		\phi_{\t{in}}(\vec p_1)
		\right\rangle
		\label{first-order-insert-complete}
		\\
		&=
		2\pi i\frac{\beta^2}{2m}
		\delta(E_1-E_2)
		\int\frac{\DD^3\vec\ell}{(2\pi)^3}\,
		\mathcal{A}^{(0)}(\vec\ell\rightarrow \vec{p}_2)
		\left\langle
		\phi_{\t{in}}(\vec\ell)
		\left|
		\frac{1}{\hat r^2}
		\right|
		\phi_{\t{in}}(\vec p_1)
		\right\rangle .
		\label{first-order-factorized}
	\end{align}
	This form exhibits a useful structural feature of DWPT. At any order in the DWPT expansion, one can insert a complete set of Coulomb in-states next to the final out-state, thereby factoring out the pure Coulomb amplitude as a kernel in a convolution---precisely the $\mathcal{A}_c$--$\mathcal{T}$ convolution structure that, as anticipated in Section~\ref{sect: modified optical}, hardwires DWPT to satisfy the modified optical theorem \eqref{modified optical momentum} order by order. This gives the factor
	$\mathcal{A}^{(0)}(\vec\ell\rightarrow \vec{p}_2)$ in \eqref{first-order-factorized}, which carries the characteristic
	$t^{-1-i\gamma+\epsilon_{+}}$ behavior. Thus the forward behavior responsible for the non-uniformity of short-range perturbation theory is already built into the kernel that appears at every order. In this sense, DWPT gives a controlled perturbative expansion around the eikonal amplitude \eqref{trees}. This structure is reminiscent of Reggeization: the exchange no longer behaves like a fixed integer-power pole, but instead acquires an anomalous exponent controlled by the long-range interaction.
	
	We evaluate the integral \eqref{rep i used} in
	Appendix~\ref{app:first order DWPT} and find
	\begin{align}
		A^{(1)}(x)
		=
		\beta^2
		\sum_{l=0}^{\infty}
		\frac{\Gamma(l+1+i\gamma)}
		{\Gamma(l+1-i\gamma)}
		\left(
		i\pi
		-
		H_{l+i\gamma}
		+
		H_{l-i\gamma}
		\right)
		P_l(x).
		\label{first order distorted}
	\end{align}
	This exactly reproduces the order-$\beta^2$ term in the expansion of the exact amplitude \eqref{exact amp}--\eqref{Sl exact def}. Here $H_z$ denotes the analytic continuation of the harmonic numbers. It is striking that evaluating this matrix element requires no direct radial integration: in the partial-wave basis the integrand is the total derivative of a Wronskian, so the integral collapses to asymptotic boundary data at radial infinity; see Appendix~\ref{app:first order DWPT}.
	
	\subsection{Comparison to short-range perturbation theory}\label{sect:comparison short range}
	Although the first-order DWPT amplitude \eqref{first order distorted} is more complicated than the first-order result in standard
	short-range perturbation theory, it provides a very compact expression for
	the infinite sum of Feynman diagrams that it encodes. Indeed,
	\eqref{first order distorted} captures the sum of all Feynman diagrams
	with one insertion of the quartic vertex
	$\frac{\beta^2}{2m}\frac{1}{r^2}$ and any number of
	$\frac{\alpha}{r}$ insertions, as depicted in
	Figure~\ref{fig:DWPT resummed diagram}.
	
	\input{Figures/dwpt_resummed_diagram}

	For example, the one-loop diagram in short-range perturbation theory with
	one quartic insertion contributes
	\begin{align}
		\mathcal{A}^{\text{short}}_{\alpha\beta^{2}}
		&=
		2\,\times\;
		\ICSdiagram
		\nonumber\\
		&=
		4\pi i\,\delta(E_1-E_2)\,
		(4\pi\alpha)\,
		\frac{\pi^{2}\beta^{2}}{m}
		\int \frac{\DD^3\vec{\ell}}{(2\pi)^3}\,
		\frac{1}{E_p-\frac{|\vec\ell|^2}{2m}+i\epsilon}\,
		\frac{1}{|\vec\ell-\vec p_1|}\,
		\frac{1}{|\vec p_2-\vec\ell|^2+\lambda_{\text{I.R.}}^2}
		\nonumber\\
		&=
		-2\pi i\,\delta(E_1-E_2)\,
		\frac{\pi\alpha\beta^{2}}{p^2\sin(\theta/2)}
		\Bigg[
		i\pi\log\!\left(\frac{2p\sin(\theta/2)}{\lambda_{\text{I.R.}}}\right)
		+
		i\pi\log\!\left(\frac{4}{1+\sin\tfrac{\theta}{2}}\right)
		\nonumber\\
		&\hspace{2cm}
		+
		\log\!\left(\sin\tfrac{\theta}{2}\right)
		\log\!\left(
		\frac{1-\sin\tfrac{\theta}{2}}
		{1+\sin\tfrac{\theta}{2}}
		\right)
		+
		\operatorname{Li}_2\!\left(\sin\tfrac{\theta}{2}\right)
		-
		\operatorname{Li}_2\!\left(-\sin\tfrac{\theta}{2}\right)
		\Bigg]
		+
		\mathcal{O}(\lambda_{\text{I.R.}}).
		\label{eq:A short alpha beta2}
	\end{align}
	The factor of two counts the two orderings of the Coulomb and
	inverse-square insertions. Here $\theta$ is the scattering angle, with
	$\sin(\theta/2)=|\vec q|/(2p)$, and $\lambda_{\text{I.R.}}$ is a photon mass introduced
	to regulate the infrared divergence of short-range perturbation theory.
	The function in \eqref{eq:A short alpha beta2} is a combination of
	multiple polylogarithms of uniform transcendental weight two.
	
	We can compare this result with the DWPT result at the same order, $\alpha\beta^2$.
	In Appendix~\ref{app:first order DWPT}, we sum the partial-wave series
	\eqref{first order distorted} to obtain a momentum-space representation
	of the first-order DWPT amplitude as a sum of three hypergeometric
	functions, \eqref{app summed first order partial wave series}. Expanding
	these hypergeometric functions, we find
	\begin{align}
		\mathcal{A}_{\alpha\beta^2}^{\rm DWPT}
		=
		-2\pi i\,\delta(E_1-E_2)\,
		\frac{\pi\alpha\beta^{2}}{p^2\sin(\theta/2)}
		\Bigg[
		i\pi
		\left(
		\gamma_{\rm E}
		+
		\log\!\left(\sin\tfrac{\theta}{2}\right)
		+
		\log\!\left(\frac{4}{1+\sin\tfrac{\theta}{2}}\right)
		\right)
		\nonumber\\
		\hspace{2cm}
		+
		\log\!\left(\sin\tfrac{\theta}{2}\right)
		\log\!\left(
		\frac{1-\sin\tfrac{\theta}{2}}
		{1+\sin\tfrac{\theta}{2}}
		\right)
		+
		\operatorname{Li}_2\!\left(\sin\tfrac{\theta}{2}\right)
		-
		\operatorname{Li}_2\!\left(-\sin\tfrac{\theta}{2}\right)
		\Bigg].
		\label{eq:A DWPT alpha beta2}
	\end{align}
	Comparing \eqref{eq:A short alpha beta2} and
	\eqref{eq:A DWPT alpha beta2} shows that DWPT reproduces the full
	angle-dependent, weight-two multiple-polylogarithmic structure of the
	short-range result. The remaining difference is precisely the universal
	Coulomb screening phase displayed in \eqref{short range and DWPT amplitude relation}.
	
	More generally, expanding the hypergeometric functions in
	\eqref{app summed first order partial wave series} at order
	$\beta^2\alpha^n$ gives multiple polylogarithms of uniform transcendental
	weight $n+1$. This provides strong evidence that DWPT resums nontrivial
	expressions into a compact object whose existence is obscured by
	short-range perturbation theory. At order $\beta^4$, the short-range
	amplitude in Appendix~\ref{app:1loop amplitudes} contains elliptic
	integrals\footnote{This is easy to see from the identity $\sum_{l}\frac{1}{2l+1}P_l(x)=\frac{1}{2}K\left(\frac{1+x}{2}\right)$, where $K$ is the complete elliptic integral of the first kind.}. Correspondingly, the second-order DWPT result
	\eqref{second order DWPT partial wave amplitude} encodes these elliptic
	and multiple-polylogarithmic structures in a compact partial-wave
	expression.
	
	More generally, away from the forward point, we find evidence in
	Appendix~\ref{app:first order DWPT} that DWPT and short-range perturbation theory are related by the universal Coulomb phase:
	\begin{gather}
		\mathcal A_{{\rm short},\lambda_{\text{I.R.}}}(\theta)
		=
		\exp\left[
		-2i\gamma
		\left(
		\log\frac{2p}{\lambda_{\text{I.R.}}}
		-
		\gamma_{\rm E}
		\right)
		\right]
		\mathcal A_{\rm DWPT}(\theta)
		+
		\mathcal O(\lambda_{\text{I.R.}}),
		\qquad
		\theta\neq0.
		\label{short range and DWPT amplitude relation}
	\end{gather}
	This is the familiar screening-phase relation
	\cite{Weinberg:1965nx,blas2021unitarizationinfiniterangeforcesgravitongraviton,Oller:2022Coulomb}.

	This gives a second reason to study DWPT beyond the elimination of
	infrared divergences: it resums highly nontrivial functions from
	short-range perturbation theory into compact expressions whose existence
	is obscured in the short-range expansion.
	
	\subsection{Finiteness of DWPT}
	\label{sect:finiteness DWPT}
	
	\paragraph{UV and IR finiteness.}
	DWPT is infrared finite order by order in each partial wave. The
	long-range Coulomb dynamics is already incorporated into the distorted
	waves, leaving insertions of the inverse-square interaction, whose
	$1/r^2$ falloff gives convergent large-radius radial integrals. This is
	illustrated explicitly at first order by the radial integral
	\eqref{app first order radial integral}; its convergence properties are
	discussed immediately below that equation in Appendix~\ref{app:first order DWPT}.
	UV finiteness is a property of the interactions considered here rather
	than of DWPT itself: interactions that do produce UV divergences can
	equally be added to theories with long-range forces, where DWPT remains
	the appropriate expansion, and the resulting renormalization procedure
	within DWPT has been studied in \cite{BarfordBirse2003,BarfordBirse2005}.
	Thus we only need to comment on the convergence of the integrals and
	partial-wave sums that represent the amplitude.
	
	\paragraph{Convergence of integrals and summations.}
	Convergence improves order by order. The zeroth-order $\beta^0$ Coulomb term is
	exceptional: neither the position-space overlap
	\eqref{coulomb overlap integral} nor the partial-wave series
	\eqref{coulomb pw series} converges in the ordinary sense. This is not a symptom of an IR divergence; it simply reflects the fact that the zeroth-order term is a distribution. This is
	exactly analogous to zeroth order in short-range perturbation theory,
	where the plane-wave overlap \eqref{plane wave overlap} and the
	partial-wave identity \eqref{forwards} are likewise distributional
	representations of delta functions rather than ordinarily convergent
	expressions.
	
	At first order $\beta^2$, each fixed-$l$ radial integral is absolutely convergent,
	while the partial-wave sum \eqref{first order distorted} is conditionally
	convergent for $-1<x<1$. The unprojected position-space matrix element
	\eqref{rep i used} is an Abel-convergent oscillatory integral, defined by
	the finite limit \eqref{app Abel limit first order matrix element}. In the
	factorized representation \eqref{first-order-factorized}, the same
	distributional completion is carried by the $\epsilon_{+}$ prescription
	of the Coulomb kernel.
	
	From second order ($\beta^4$) onward, the partial-wave sums are absolutely
	convergent. Indeed, the order-$\beta^{2n}$ correction to a partial-wave
	coefficient falls as $1/l^n$ \eqref{forward large l hierarchy}, so the
	full summand is $\mathcal{O}(l^{\frac12-n})$.
	
\paragraph{Radius of convergence.}
The radius of convergence of the perturbative expansion for $S_l$ in powers of
$\beta^2$ can be read directly from the exact partial-wave amplitudes
\eqref{Sl exact def}. For fixed $l$, the nearest singularity in the complex
$\beta^2$ plane is the square-root branch point at
$\beta^2=(l+\tfrac12)^2$,\footnote{The gamma functions in \eqref{Sl exact def}
also have poles, at $\beta^2=(l+\tfrac12)^2-[(n+\tfrac12)+i\gamma]^2$ for
$n=0,1,2,\dots$. These require the square root to have negative real part, so
they lie on the second sheet and do not limit the radius of convergence.} so
the DWPT expansion for partial wave $l$ has radius of convergence
$(l+\tfrac12)^2$. The smallest radius occurs for the $s$-wave, and the full DWPT
expansion therefore has radius $1/4$, precisely coinciding with the
non-perturbative unitarity bound \eqref{eq: non pert unitarity bound}. These are
nonetheless logically distinct statements: the radius of convergence follows from
analyticity in the complex $\beta^2$ plane, whereas the unitarity bound follows
from the failure of $S_0$ to have unit modulus for real $\beta^2>1/4$. They
coincide here because both are controlled by the same square-root branch point.
This coincidence is responsible for a non-generic feature of the model.
Generically, one would expect the expansion to break down well before the coupling
reaches the unitarity bound, placing the onset of unitarity violation beyond the
reach of perturbation theory. Here, instead, the series converges for every
$\beta^2$ at which the model is unitary, and diverges for every $\beta^2$ at
which it is not.
	
	\subsection{Forward behavior of DWPT}
	\label{sect:forward behavior DWPT}
	
	We now examine the forward behavior of the DWPT expansion. There are
	two closely related concerns. The first is
	whether higher orders could contain new forward singularities whose
	angular integrals require an additional distributional prescription.
	The second is whether higher orders can dominate lower orders in the
	forward direction, as occurs in short-range perturbation theory. This
	matters because the unitarity integrals necessarily probe the forward
	region. We will show that no new forward distributions appear beyond
	the zeroth-order Coulomb term. We will also show that successive DWPT
	corrections are less singular and therefore less dominant in the
	forward region.
	
	Both results follow from the large-$l$ structure of the exact
	partial-wave coefficients \eqref{Sl exact def},
	\begin{gather}
		S_l
		=
		S_{c,l}
		\left[
		1
		+
		\beta^2\mathcal O\left(\frac1l\right)
		+
		\beta^4\mathcal O\left(\frac1{l^2}\right)
		+
		\beta^6\mathcal O\left(\frac1{l^3}\right)
		+\cdots
		\right],
		\qquad
		S_{c,l}
		=
		\frac{\Gamma(l+1+i\gamma)}
		{\Gamma(l+1-i\gamma)}
		\sim l^{2i\gamma}.
		\label{forward large l hierarchy}
	\end{gather}
	The important point is that each power of $\beta^2$ is accompanied by
	an inverse power of $l$.  Each successive order is therefore less
	singular at large $l$, and hence less singular in the forward
	direction.  This statement can be made more precise using the large-$l$
	Bessel approximation
	\begin{gather}
		P_l(\cos\theta)
		\simeq
		J_0(l\theta).
	\end{gather}
	At order $\beta^{2n}$ the singular large-$l$ part of the partial-wave
	sum therefore behaves schematically as
	\begin{align}
		\mathcal A_{\mathrm{sing}}^{(n)}(\theta)
		&\sim
		\beta^{2n}
		\int^\infty\DD l\,
		l^{1-n+2i\gamma}J_0(l\theta)
		\nonumber\\
		&\sim
		\beta^{2n}\theta^{\,n-2-2i\gamma}.
		\label{forward DWPT power counting}
	\end{align}
	Thus each additional order in $\beta^2$ softens the singular large-$l$
	contribution by one power of the forward scattering angle.  The
	resulting hierarchy is
	\begin{center}
		\begin{tabular}{c|c|c}
			order
			& forward behavior
			& angular integrability
			\\
			\hline
			$\beta^0$
			& $\theta^{-2-2i\gamma}$
			& requires the Coulomb $\epsilon_+$ prescription
			\\
			$\beta^2$
			& $\theta^{-1-2i\gamma}$
			& integrable
			\\
			$\beta^4$
			& at most logarithmic
			& integrable
			\\
			$\beta^{2n}$, $n\geq3$
			& finite at $\theta=0$
			& integrable
		\end{tabular}
	\end{center}
	
	This power counting is confirmed directly by the first-order
	hypergeometric expression \eqref{app summed first order partial wave series}:
	expanding it in the forward limit gives
	\begin{gather}
		\mathcal A^{(1)}
		\sim
		\sin^{-1-2i\gamma}\left(\frac{\theta}{2}\right),
		\qquad
		\theta\rightarrow0,
		\label{forward hypergeometric check}
	\end{gather}
	in agreement with \eqref{forward DWPT power counting} at $n=1$.
	
	The order-$\beta^0$ Coulomb amplitude therefore contains the only
	non-integrable forward singularity and must be treated with its
	$\epsilon_+$ prescription.  Beyond leading order there are no
	additional forward-supported distributional terms or prescriptions:
	all DWPT corrections are locally integrable against the angular
	measure, and their partial-wave projections are unambiguous.  Moreover,
	the forward regions of the higher-order terms do not dominate those of
	lower orders.

	\section{Unitarity bounds}\label{sect: unitarity bounds}
	
	As is manifest in the factorized form of DWPT in \eqref{first-order-factorized}, the pure Coulomb contribution can always be separated from the remaining short-range dynamics in the partial-wave basis. We therefore write the amplitude as
	\begin{gather}
		\mathcal{A}(\vec{p}_1\rightarrow \vec{p}_2)
		=
		\frac{2\pi^2}{mp}\delta(E_1-E_2)
		\sum_{l=0}^{\infty}(2l+1)
		S_{c,l}
		\left(
		1+2i\widehat{a}_l
		\right)
		P_l(\hat{p}_1\cdot\hat p_2),
		\label{unitarity bounds Coulomb stripped expansion}
	\end{gather}
	where
	\begin{gather}
		S_{c,l}
		=
		\frac{\Gamma(1+l+i\gamma)}
		{\Gamma(1+l-i\gamma)}
	\end{gather}
	is the pure Coulomb phase. The decomposition
	\eqref{unitarity bounds Coulomb stripped expansion} defines the
	Coulomb-stripped partial-wave coefficients $\widehat{a}_l$.
	
	Since $S_{c,l}$ is a pure phase,
	\begin{gather}
		S_{c,l}S_{c,l}^{\star}=1,
	\end{gather}
	elastic unitarity for the Coulomb-stripped coefficients has exactly the form derived in \eqref{modified optical partial wave}, which is also the form of the purely short-range condition \eqref{eq: pure short elastic unitarity}. In particular,
	\begin{gather}
		\left(
		1+2i\widehat a_l
		\right)
		\left(
		1-2i\widehat a_l^{\star}
		\right)
		=
		1
		\qquad
		\Longrightarrow
		\qquad
		|\operatorname{Re}\widehat a_l|
		\leq
		\frac12 ,
		\label{distorted unitarity}
	\end{gather}
	which will be the condition we impose to derive unitarity bounds. 
	\subsection{First-order bound}
	
	Using the first-order DWPT amplitude \eqref{first order distorted}, the order-$\beta^2$ contribution to the Coulomb-stripped partial-wave coefficient is
	\begin{gather}
		\widehat{a}_l^{(1)}
		=
		\frac{\beta^2}{2(2l+1)}
		\left(
		\pi
		-
		2\,\operatorname{Im}H_{l+i\gamma}
		\right),
		\qquad
		\gamma=\alpha\frac{m}{|\vec p|}.
	\end{gather}
	The elastic unitarity condition \eqref{distorted unitarity} therefore gives the first-order bound
	\begin{gather}
		\beta^2
		\lesssim
		\frac{2l+1}
		{
			\left|
			\pi
			-
			2\,\operatorname{Im} H_{l+i\gamma}
			\right|
		} .
		\label{constraint 1}
	\end{gather}
	This bound still depends on the scattering momentum $|\vec{p}|$ through $\gamma$. This dependence is not an infrared ambiguity; it reflects the fact that the partial-wave unitarity condition is imposed at fixed momentum. To obtain a bound on $\beta^2$, one must specify the range of momenta over which the Hamiltonian is meant to be valid.

	We now interpret the Hamiltonian \eqref{hamil}, with its regular-at-the-origin domain, as an EFT valid only up to a momentum scale $\Lambda_{\rm EFT}$. For scattering on a charged nucleus, this cutoff may be set by the inverse nuclear radius, since the pointlike potentials in \eqref{hamil} no longer describe the physics once distances of order the nuclear size are resolved. Unitarity should therefore be imposed only within the regime of validity of the EFT,
	\begin{gather}
		|\vec p|\lesssim \Lambda_{\rm EFT}.
	\end{gather}

	For a repulsive Coulomb field, the first-order bound \eqref{constraint 1} becomes more restrictive as the scattering momentum is increased. This is because the Coulomb parameter
	\begin{gather}
		\gamma=\frac{\alpha m}{|\vec p|}
	\end{gather}
	decreases with increasing $|\vec p|$, and the right-hand side of \eqref{constraint 1} decreases as $\gamma$ is taken toward zero. Therefore, within the EFT domain, the strongest first-order constraint is obtained at the largest momentum for which the effective Hamiltonian is valid, together with the $s$-wave. Defining
	\begin{gather}
		\gamma_{\rm EFT}
		=
		\frac{\alpha m}{\Lambda_{\rm EFT}},
	\end{gather}
	the first-order EFT bound is therefore
	\begin{gather}
		\boxed{\beta^2
			\lesssim
			\frac{1}
			{
				\left|
				\pi
				-
				2\,\operatorname{Im}H_{i\gamma_{\rm EFT}}
				\right|
			} .}\label{EFT bound}
	\end{gather}
	It is useful to compare this finite-cutoff EFT bound with the point-particle limit. If \eqref{hamil} is treated as an exact point-particle theory, rather than as an EFT with a finite cutoff, then unitarity should hold at arbitrarily large momentum. Equivalently, one should take
	\begin{gather}
		\Lambda_{\rm EFT}\to\infty,
		\qquad
		\gamma_{\rm EFT}=
		\frac{\alpha m}{\Lambda_{\rm EFT}}
		\to0 .
	\end{gather}
	In this limit the strongest first-order constraint again comes from the $s$-wave. Since $H_0=0$, or equivalently $\operatorname{Im}H_{i\gamma_{\rm EFT}}\to0$ as $\gamma_{\rm EFT}\to0$, the EFT bound \eqref{EFT bound} becomes
	\begin{gather}
		\beta^2
		\lesssim
		\frac{1}{\pi}.
		\label{point particle bound}
	\end{gather}
	This is the first-order DWPT estimate of the exact point-particle bound
	\begin{gather}
		\beta^2\leq\frac14 .\label{non-pert bnd}
	\end{gather}
	The finite-cutoff bound \eqref{EFT bound} has the expected physical behavior in both limits. In the limit $\Lambda_{\rm EFT}\to\infty$, it reduces to the point-particle estimate \eqref{point particle bound}, the first-order approximation to the exact bound \eqref{non-pert bnd}. This is as it should be: sending $\Lambda_{\rm EFT}\to\infty$ trusts the point-source description to arbitrarily short distances, so the point-particle Hamiltonian is the only allowed UV completion. At finite $\Lambda_{\rm EFT}$, by contrast, many UV completions are possible, and, correspondingly, for a repulsive nucleus, the allowed values of $\beta^2$ increase as $\Lambda_{\rm EFT}$ is lowered.
	
	The relaxation of the bound with decreasing $\Lambda_{\rm EFT}$ tracks a known physical effect, reviewed in Section~\ref{Sect: setup}: the critical atomic number increases with the size of the nucleus \cite{bawin1981instability}. In the nuclear scattering example, one UV completion is to give the nucleus a finite radius, with the physics above the cutoff supplied by the finite-size structure of the nucleus, ultimately by QCD. The pointlike critical atomic number $Z_{\rm crit}=137$, set by fermion pair production from the vacuum, then rises to roughly $Z_{\rm crit}\sim 173$ for extended nuclei \cite{GreinerMuellerRafelski1985}.

	\subsection{Second order and beyond}
	We now extend the unitarity bound to second order. For simplicity we take $\alpha>0$ in what follows, so that the Coulomb Hamiltonian has no bound states. At second order, the DWPT expansion \eqref{long pert} instructs us to compute
	\begin{align}
		\mathcal A^{(2)}
		&=
		\frac{(-i)^2}{2}
		\int\DD t_1\,\DD t_2\,
		\braket{
			\phi_{\mathrm{out}}(\vec p_2)
			|
			T\!\left\{
			\tilde V_I(t_1)\tilde V_I(t_2)
			\right\}
			|
			\phi_{\mathrm{in}}(\vec p_1)
		}
		\nonumber \\[6pt]
		&=
		-2\pi i\left(\frac{\beta^2}{2m}\right)^2\delta(E_1-E_2)
		\int\frac{\DD^3\vec\ell}{(2\pi)^3}\,
		\frac{
			\braket{
				\phi_{\mathrm{out}}(\vec p_2)
				|
				\frac{1}{\hat r^2}
				|
				\phi_{\mathrm{in}}(\vec\ell)
			}
			\braket{
				\phi_{\mathrm{in}}(\vec\ell)
				|
				\frac{1}{\hat r^2}
				|
				\phi_{\mathrm{in}}(\vec p_1)
			}
		}{
			E_1-E_\ell+i0
		} ,
	\end{align}
	where one would also include bound states in the intermediate state sum if
	the potential were attractive. This second-order spectral integral can also
	be evaluated analytically in the partial-wave basis. Its on-shell part is
	fixed by the first-order radial matrix element, while its principal-value
	part can be converted, using the Coulomb Green function, into a nested radial
	integral. The same continuous-angular-momentum Wronskian identity
	\eqref{app Wronskian derivative} used at first order then reduces this
	integral to a derivative of the first-order
	boundary term. Consequently, the final result is again determined entirely
	by asymptotic data at radial infinity. Rather than presenting this derivation
	here, we will assume that DWPT reproduces the expansion in $\beta^2$
	of the exact amplitude \eqref{exact amp}, as we verified explicitly at first
	order \eqref{first order distorted}. The second-order contribution is then
	the order-$\beta^4$ term in the expansion of the exact partial-wave
	coefficients \eqref{Sl exact def}:
	\begin{align}
		\mathcal A^{(2)}
		&=
		\frac{2\pi^2}{mp}\beta^4\delta(E_1-E_2)
		\sum_{l=0}^{\infty}
		(2l+1)\,
		S_{c,l}\,
		P_l(\cos\theta)
		\nonumber\\
		&\quad\times
		\left[
		-
		\frac{K_l(\gamma)^2}{2(2l+1)^2}
		+
		i\left(
		\frac{K_l(\gamma)}{(2l+1)^3}
		+
		\frac{J_l(\gamma)}{(2l+1)^2}
		\right)
		\right] .
		\label{second order DWPT partial wave amplitude}
	\end{align}
	Here
	\begin{gather}
		K_l(\gamma)
		=
		\pi
		-
		2\,\operatorname{Im}H_{l+i\gamma},\qquad 
		J_l(\gamma)
		=
		\operatorname{Im}\psi^{(1)}(l+1+i\gamma),
	\end{gather}
	with $\psi^{(n)}(z)$ denoting the polygamma function.
	Equivalently, through second order, the Coulomb-stripped partial-wave coefficient is\footnote{With $\widehat a_l^{(n)}=\mathcal O(\beta^{2n})$, the order-$\beta^4$ part of the modified optical theorem \eqref{modified optical partial wave} gives $\operatorname{Im}\widehat a_l^{(2)}=|\widehat a_l^{(1)}|^2$.}
	\begin{align}
		1+2i\widehat a_l
		&=
		1
		+
		i\beta^2
		\frac{K_l(\gamma)}{2l+1}
		+
		\beta^4
		\left[
		-
		\frac{K_l(\gamma)^2}{2(2l+1)^2}
		+
		i\left(
		\frac{K_l(\gamma)}{(2l+1)^3}
		+
		\frac{J_l(\gamma)}{(2l+1)^2}
		\right)
		\right]
		+
		\mathcal O(\beta^6).
	\end{align}
	The real part is therefore
	\begin{align}
		\operatorname{Re}\widehat a_l
		&=
		\frac{\beta^2}{2(2l+1)}
		K_l(\gamma)
		+
		\frac{\beta^4}{2}
		\left[
		\frac{K_l(\gamma)}{(2l+1)^3}
		+
		\frac{J_l(\gamma)}{(2l+1)^2}
		\right]
		+
		\mathcal O(\beta^6).
	\end{align}
	Imposing $|\operatorname{Re}\widehat{a}_l| \lesssim \frac{1}{2}$ gives
	\begin{align}
		\left|
		\frac{\beta^2}{2(2l+1)}
		K_l(\gamma)
		+
		\frac{\beta^4}{2}
		\left[
		\frac{K_l(\gamma)}{(2l+1)^3}
		+
		\frac{J_l(\gamma)}{(2l+1)^2}
		\right]
		\right|
		\lesssim
		\frac12 ,
	\end{align}
	and solving for $\beta^2$ gives 
	\begin{gather}
		\boxed{ \beta^2
			\lesssim
			\frac{
				2(2l+1)
			}{
				K_l(\gamma)
				+
				\sqrt{
					K_l(\gamma)^2
					+
					\frac{4}{2l+1}
					\left[
					K_l(\gamma)
					+
					(2l+1)J_l(\gamma)
					\right]
				}
			} .}\label{second order bound}
	\end{gather}
	We therefore obtain a bound without any ambiguities arising from IR divergences. If the Hamiltonian is treated as an EFT, the same reasoning as in the first-order analysis applies. For a repulsive Coulomb field, the bound becomes most restrictive at the largest momentum in the EFT domain. Thus the second-order EFT bound is obtained by evaluating \eqref{second order bound} at $\gamma=\gamma_{\rm EFT}$.
	
	If the Hamiltonian is treated as an exact model rather than as an EFT, we may take the high-energy limit,
	\begin{gather}
		|\vec p|\to\infty,
		\qquad
		\gamma=\frac{m\alpha}{|\vec p|}\to0.
	\end{gather}
	In this limit
	\begin{gather}
		K_l(\gamma)\to \pi,
		\qquad
		J_l(\gamma)\to 0,
	\end{gather}
	and hence the bound becomes
	\begin{gather}
		\beta^2
		\lesssim
		\frac{
			2(2l+1)
		}{
			\pi
			+
			\sqrt{
				\pi^2
				+
				\frac{4\pi}{2l+1}
			}
		} .
	\end{gather}
	The strongest constraint again comes from the $s$-wave. Setting $l=0$ gives
	\begin{gather}
		\beta^2
		\lesssim
		\frac{2}{\pi+\sqrt{\pi^2+4\pi}}
		\simeq
		0.2539 .
	\end{gather}
	Thus, already at second order, the perturbative unitarity bound is close to the exact critical value
	\begin{gather}
		\beta^2_{\mathrm{crit}}
		=
		\frac14 .
	\end{gather}
	Taking the high-energy limit shows that the bounds obtained from the $l=0$
	mode coincide with those of the purely short-range theory. The bounds at
	successive orders in $\beta^2$ can therefore be computed directly in the
	purely short-range theory: set $\alpha=0$, expand the exact partial waves
	\eqref{short exact} through the indicated order in $\beta^2$, and impose the
	elastic unitarity condition \eqref{eq: pure short elastic unitarity} on the
	truncated $s$-wave. This computation is carried out in
	Appendix~\ref{sect:bench} and generates Table~\ref{tab:rapid}. No photon
	mass, detector resolution, or other infrared scale enters these bounds. The
	rapid agreement with the exact bound demonstrates that low-order amplitude
	data can yield accurate, infrared-scale-independent unitarity bounds even in
	the presence of a long-range force.
	\begin{table}
		\centering
		\begin{tabular}{c|c}
			truncation & estimated upper limit on $\beta^2$ \\ \hline
			$\mathcal{O}(\beta^2)$ & 0.3183 \\
			$\mathcal{O}(\beta^4)$ & 0.2539 \\
			$\mathcal{O}(\beta^6)$ & 0.25017 \\
			$\mathcal{O}(\beta^8)$ & 0.25000465
		\end{tabular}
		\caption{Upper limits on $\beta^2$ from partial-wave unitarity at successive truncations of the perturbative expansion, converging rapidly to the exact critical value $\beta^2_{\rm crit}=1/4$. The entries are computed in the purely short-range theory in Appendix~\ref{sect:bench}; the $l=0$ Coulomb-stripped bounds of this section, which contain no infrared regulator scale, coincide with them in the high-energy limit.}
		\label{tab:rapid}
	\end{table}

\subsection{Interpreting perturbative unitarity bounds}
Unlike positivity bounds, which only use the sign of the imaginary part, $\operatorname{Im}a_l\geq0$, the bounds derived above use the upper bound that unitarity places on the real part of the partial waves,
\begin{gather}
	|\operatorname{Re}a_l|\leq\frac12 .\label{v7:upper general}
\end{gather}
Saturating \eqref{v7:upper general} perturbatively requires the leading term to already be of order one, and the optical theorem \eqref{modified optical partial wave} then forces the next order to be of comparable size. Saturation therefore occurs precisely where higher orders are no longer guaranteed to be parametrically smaller than lower ones. The conservative reading of any bound of this type is that its violation signals the breakdown of weak-coupling perturbation theory and is not by itself a proof that the exact theory has become non-unitary.

The most famous bound of this kind is the Lee--Quigg--Thacker (LQT) bound on the Higgs mass \cite{Lee:1977eg}, which underpinned the expectation that the LHC would either discover a relatively light Higgs boson or probe strongly coupled electroweak-symmetry-breaking dynamics at the TeV scale \cite{Bagger:1995mk}. They imposed the slightly weaker condition $|a_l|\leq1$, and were explicit that what such a bound delimits is the class of theories in which low orders of perturbation theory remain a reliable guide\footnote{The LQT analysis was ultimately able to discard the troublesome tree-level contributions that do not admit an ordinary partial-wave projection, such as $t$-channel photon exchange in longitudinal $W$-boson scattering, because these terms do not grow with the Higgs mass and are subleading in the heavy-Higgs approximation used to derive the bound. More recently, Ref.~\cite{FuentesZamoro:2025exp} retained this contribution and defined the corresponding partial waves using a relativistic analogue of Herbst's $\epsilon_+$ prescription. Our non-relativistic analysis therefore provides further support for the prescription underlying their treatment.}.

Whether a violation of \eqref{v7:upper general} signals a genuine loss of unitarity or only the failure of the weak-coupling expansion, the methods outlined in this paper remain useful. One would like to locate the coupling at which such a bound is violated, and standard techniques leave that location ambiguous, as infrared-divergent logarithms enter the bound itself. In both scenarios, we expect the same two ingredients---the correct distributional structure in the forward direction, and a perturbation theory built on the correct asymptotic dynamics---to be central to eliminating these ambiguities. We expect these same two ingredients to be equally essential for eliminating infrared ambiguities in positivity bounds, as nothing in the proposed framework relies on imposing \eqref{v7:upper general} rather than $\operatorname{Im}a_l\geq0$.

	\section{Conclusions}\label{sect:conc}
	The central lesson of this solvable model is simple. The Coulomb infrared divergences present in the amplitude and EFT bound for this model \eqref{hamil} obtained from naive short-range perturbation theory are artifacts of the perturbative scheme, not ambiguities of the exact theory. The exact amplitude, defined as the inner product of the full in- and out-states, is finite and well defined. It has no infrared divergence, and it has a modified distributional structure. The divergences appear when short-range perturbation theory is applied to a long-range problem, because it expands around the wrong asymptotic dynamics and assumes the incorrect distributional structure of the amplitude. Partial waves, and hence unitarity bounds, derived from this expansion then acquire spurious infrared dependence.
	
	The resolution has three parts, each addressing a distinct failure of the short-range expansion.
	
	\begin{itemize}
		
		\item \textbf{No spurious IR divergences.} Distorted-wave perturbation theory (DWPT) treats the Coulomb dynamics exactly and treats only the remaining short-range interaction perturbatively. It therefore correctly computes the inner product of the in- and out-states. In the model studied here, this removes the artificial infrared divergence from the amplitude and leaves no dependence on a photon mass, detector resolution, dimensional-regulator scale, or other arbitrary infrared parameter (Section~\ref{sect:finiteness DWPT}).

		\item \textbf{Correct distributional structure.} Long-range scattering does not have the same forward-direction distributional structure as short-range scattering. In particular, the usual decomposition $S=\mathds{1}+i\mathbb{T}$ is not the correct starting point for the Coulomb problem. DWPT incorporates the long-range structure from zeroth order through the $\epsilon_+$ prescription, which gives the distributional completion of the Coulomb amplitude at the forward point required by unitarity (Section~\ref{sect: forward direction}). Accounting for this is what makes partial-wave projections well-defined.
		
		\item \textbf{Controlled forward behavior.} In the short-range expansion, higher loops generate increasingly severe forward-region logarithms, so a fixed-order amplitude cannot be reliably inserted into the angular integrals required by unitarity. In DWPT, the singular Coulomb behavior is already included at zeroth order. The remaining order-$\beta^{2n}$ corrections fall as $\mathcal{O}(1/l^n)$ at large angular momentum, so higher orders become progressively less singular in the forward direction, as shown in \eqref{forward large l hierarchy} and discussed in Section~\ref{sect:forward behavior DWPT}. This makes higher-order corrections less likely to dramatically alter the low-order unitarity bounds than their counterparts in short-range perturbation theory.
		
	\end{itemize}
	
	With these ingredients in place, we derived unitarity bounds on the coupling $\beta^2$ of the short-range inverse-square interaction in the presence of the long-range Coulomb force. As shown in Table~\ref{tab:rapid}, the bounds obtained at successive orders rapidly approach the exact critical value $\beta^2_{\rm crit}=1/4$. The essential point is that these bounds are obtained without introducing any infrared regulator or arbitrary infrared scale. Short-range perturbation theory does not have this property: it gives bounds with IR-scale dependence that is physically spurious in this model.

	\subsection{Should EFT bounds contain IR scales?}
	An important emphasis of this paper is that, at least in non-relativistic quantum mechanics, the question of whether a given Hamiltonian defines a unitary scattering problem is a completely well-defined mathematical question not plagued by infrared divergences. Once the Hamiltonian, inner product, and domain are specified, one can ask whether the corresponding in--out map is unitary and for what range of couplings this is true. This question has no intrinsic ambiguity arising from infrared divergences. Apparent infrared divergences arise only if one expands around the wrong asymptotic dynamics or imposes asymptotic boundary conditions appropriate to free states rather than to the true long-range problem. In this sense, viewed purely as a problem in mathematical physics, one should not expect unitarity bounds on the couplings to require an infrared scale.
	
	There is, however, a distinct and physically important question concerning real measurements. In a collider, gravitational-wave detector, or any other finite experimental setup, the asymptotic dynamics is not governed solely by the isolated Hamiltonian of the particles being scattered. It is also shaped by the detector environment: finite observation times, external fields, and eventual absorption by the apparatus. From this perspective, it is natural that experimentally measured couplings can involve infrared scales associated with the measurement. This is the viewpoint emphasized in \cite{bellazzini2026positivitylongrangeinteractions}, and we regard it as a physically motivated way of formulating detector-level bounds.
	
	It is not clear to us whether, in general, the EFT bounds that should be compared with experiment are the purely mathematical bounds, independent of infrared scales, or bounds formulated directly for detector-level observables, which necessarily involve an experimental infrared resolution. In the example studied here, however, the interpretation is less ambiguous. The IR-scale-dependent bound from short-range perturbation theory in \eqref{first bound naive short} should not be taken seriously as a detector-resolution-dependent bound. The condition $\beta^2 \leq 1/4$ is a non-relativistic approximation to a genuine physical statement: sufficiently highly charged nuclei are unstable \cite{bawin1981instability,GreinerMuellerRafelski1985}. The critical charge is a property of the quantum system, not of the infrared scale associated with the detector used to observe it. Thus, in this example, an infrared-scale dependence in the naive short-range perturbative bound does not encode physical detector dependence. It instead signals that the perturbative expansion has been organized around the wrong asymptotic dynamics.

	Regardless of this interpretation, DWPT provides a framework in which forward-direction integrals and partial-wave projections can be analyzed without arbitrary IR regulators, and in which unitarity bounds can be derived without the ambiguities introduced by short-range perturbation theory. Thus, even in contexts where detector-defined IR scales remain physically relevant, DWPT can still be a useful tool and can help clarify which features of a bound are intrinsic and which are artifacts of the perturbative framework.

	\subsection{Outlook toward QED and gravity}

	A natural question is how DWPT should be generalized to relativistic quantum field theories such as QED and gravity. The main lesson is that the asymptotic Hamiltonian should not be taken to be the free Hamiltonian. For massive charged particles, the long-range interaction affects the asymptotic relative motion of each charged pair. At large separations, the center of mass moves freely, while the relative coordinate evolves in an effective Coulombic potential. Thus the appropriate asymptotic dynamics should be built from a multiparticle Hamiltonian containing Coulombic interactions within each pair of charged particles in the in-state, and similarly within each pair in the out-state.
	
	A relativistic version of the Coulomb Hamiltonian, its diagonalization, and the corresponding expansion of quantum fields in modes adapted to this Hamiltonian were studied in \cite{Lippstreu:2025jit} in the limit where one particle in the pair is infinitely massive. A key open problem is to extend this construction to a generic pair of masses. This requires identifying the appropriate relativistic relative coordinate and the corresponding effective Coulomb potential. Such a construction would provide the natural starting point for a relativistic version of DWPT. The effective-one-body framework of Todorov \cite{Todorov:1970gr}, and its recent formulation for gravity in \cite{Correia:2024}, may provide a useful starting point for this problem: in that framework, relativistic two-body scattering can be reorganized as a Born series for an effective one-body Schrödinger equation.
	
	There is also a second infrared issue in QED and gravity: the radiative infrared divergences associated with soft photon and graviton emission. The present paper addresses the Coulombic part of the long-range problem. However, the Coulombic and radiative sectors are not independent. They are related by the analytic properties of amplitudes, including crossing symmetry; see \cite{Lippstreu:2025jit} for a pedagogical discussion of this relationship. It is therefore plausible that a correct treatment of the Coulombic infrared problem will also give useful guidance for the radiative problem.

	\section*{Acknowledgments}
    I am grateful to Brando Bellazzini, Simon Caron-Huot, Miguel Correia, Andrea Guerrieri, Kelian H\"aring, Giulia Isabella, Julio Parra-Martinez, and Marcello Romano for helpful discussions and for valuable comments and feedback on earlier drafts of this manuscript.
	This work is supported by the enhanced research expenses
	\verb|RF\ERE\221030| associated with the Royal Society grant \verb|URF\R1\221233|.

    \textbf{AI usage:} The analytical calculations and writing of this paper are human-generated, with assistance from ChatGPT (OpenAI) and Claude (Anthropic) for brainstorming, literature searches, coding for numerical checks, cross-checking results, and grammar and typesetting. The author takes full responsibility for the manuscript.
	\appendix
	
	\section{Unitarity bounds in the purely short-range theory}\label{sect:bench}
	In Section~\ref{sect: unitarity bounds} we derived unitarity bounds on $\beta^2$ in the presence of a repulsive Coulomb interaction and found that they are most stringent for a repulsive Coulomb coupling $\alpha>0$ at the largest momentum in the EFT domain. When the Hamiltonian \eqref{hamil} is treated as the full UV theory, so that no cutoff restricts this momentum, the strongest bounds are obtained in the high-energy limit $\gamma\to0$, where they coincide with the bounds of the purely short-range theory in which only the inverse-square interaction is present. In this appendix we derive the unitarity bounds of this purely short-range theory, order by order in $\beta^2$; this is the computation that generates Table~\ref{tab:rapid}.

	We therefore set the Coulomb coupling to zero,
	\begin{gather}
		\alpha=0,
		\qquad
		V(r)=-\frac{\beta^2}{2m}\frac{1}{r^2}.
	\end{gather}
	The exact partial-wave expansion then takes the form
	\begin{gather}
		A(x)
		=
		\sum_{l=0}^{\infty}
		(2l+1)S_l P_l(x),
	\end{gather}
	with
	\begin{gather}
		S_l
		=
		\exp\left[
		i\pi\left(
		l+\frac12
		-
		\sqrt{\left(l+\frac12\right)^2-\beta^2}
		\right)
		\right].
		\label{short exact}
	\end{gather}
	For $\beta^2\leq1/4$, the square root is real for every partial wave, and hence $S_l$ is a pure phase. The exact amplitude is therefore unitary in this regime. However, for $\beta^2>\frac{1}{4}$ the $s$-wave ceases to be unitary, so the exact critical value is
	\begin{gather}
		\beta^2_{\rm crit}
		=
		\frac14 .
	\end{gather}
	
	Suppose we did not know the exact answer, but only had access to the low-order expansion of the amplitude in powers of $\beta^2$. What bound on $\beta^2$ would one infer from partial-wave unitarity?
	
	Elastic unitarity, $|S_l|=1$, with $S_l=1+2i a_l$, implies
	\begin{gather}
		(1+2i a_l)(1-2i a_l^{\star})=1
		\quad \Longrightarrow \quad
		|\operatorname{Re}a_l|\leq \frac12 .
		\label{eq: pure short elastic unitarity}
	\end{gather}
	We will apply this condition to the truncated expansion of $a_l$.
	
	Since the asymptotic dynamics is free when $\alpha=0$, standard perturbation theory reproduces the small-$\beta^2$ expansion of the exact partial waves. We can therefore bypass the explicit Feynman diagram calculation: the diagrams would simply reconstruct the same expansion obtained by expanding \eqref{short exact}. Let $a_l^{[n]}$ denote the expansion of $a_l$ through order $\beta^{2n}$. At order $\beta^{2n}$ we estimate the bound by imposing
	\begin{gather}
		|\operatorname{Re}a_l^{[n]}|
		\lesssim
		\frac12 .
	\end{gather}
	
	At leading order,
	\begin{gather}
		\operatorname{Re}a_l^{[1]}
		=
		\frac{\pi}{2}\frac{\beta^2}{2l+1}.\label{consistent betasquared}
	\end{gather}
	Thus
	\begin{gather}
		\frac{\pi}{2}\frac{\beta^2}{2l+1}
		\lesssim
		\frac12,
		\qquad\Longrightarrow\qquad
		\beta^2
		\lesssim
		\frac{2l+1}{\pi}.
	\end{gather}
	The strongest estimate comes from the $s$-wave,
	\begin{gather}
		\beta^2
		\lesssim
		\frac1{\pi}
		\simeq
		0.3183 .
	\end{gather}
	
	At second order,
	\begin{gather}
		\operatorname{Re}a_l^{[2]}
		=
		\frac{\pi}{2}
		\left(
		\frac{\beta^2}{2l+1}
		+
		\frac{\beta^4}{(2l+1)^3}
		\right).
	\end{gather}
	For $l=0$, this gives
	\begin{gather}
		\beta^2+\beta^4
		\lesssim
		\frac1{\pi},
		\qquad\Longrightarrow\qquad
		\beta^2
		\lesssim
		\frac{\sqrt{1+4/\pi}-1}{2}
		\simeq
		0.2539 .
	\end{gather}
	
	At third order,
	\begin{gather}
		\operatorname{Re}a_l^{[3]}
		=
		\frac{\pi}{2}
		\left(
		\frac{\beta^2}{2l+1}
		+
		\frac{\beta^4}{(2l+1)^3}
		+
		\frac{2\beta^6}{(2l+1)^5}
		\right)
		-
		\frac{\pi^3}{12}
		\frac{\beta^6}{(2l+1)^3}.
	\end{gather}
	For the $l=0$ mode, imposing $|\operatorname{Re}a_0^{[3]}|\leq \frac12$ gives
	\begin{gather}
		\beta^2
		\lesssim
		0.25017 .
	\end{gather}
	
	At fourth order,
	\begin{gather}
		\operatorname{Re}a_l^{[4]}
		=
		\frac{\pi}{2}
		\left(
		\frac{\beta^2}{2l+1}
		+
		\frac{\beta^4}{(2l+1)^3}
		+
		\frac{2\beta^6}{(2l+1)^5}
		+
		\frac{5\beta^8}{(2l+1)^7}
		\right)
		-
		\frac{\pi^3}{12}
		\left(
		\frac{\beta^6}{(2l+1)^3}
		+
		\frac{3\beta^8}{(2l+1)^5}
		\right).
	\end{gather}
	For the $l=0$ mode, the bound is determined by
	\begin{gather}
		\frac{\pi}{2}
		\left(
		\beta^2+\beta^4+2\beta^6+5\beta^8
		\right)
		-
		\frac{\pi^3}{12}
		\left(
		\beta^6+3\beta^8
		\right)
		\lesssim
		\frac12,
	\end{gather}
	which gives
	\begin{gather}
		\beta^2
		\lesssim
		0.25000465 .
	\end{gather}
	
	The resulting bounds are collected in Table~\ref{tab:rapid} and rapidly approach the exact critical value $\beta^2_{\rm crit}=\frac14$. By the high-energy coincidence established in Section~\ref{sect: unitarity bounds}, they are also the strongest unitarity bounds on $\beta^2$ when \eqref{hamil} is treated as the full UV theory.

	\section{One-loop amplitudes in the short-range expansion}
	\label{app:1loop amplitudes}
\begin{figure}[t]
		\centering
		\begin{subfigure}[t]{0.31\textwidth}
			\centering
			\begin{tikzpicture}[
				scale=0.88,
				transform shape,
				baseline={(current bounding box.center)}
				]
				\coordinate (left)  at (-1.15,0);
				\coordinate (right) at ( 1.15,0);
				
				\draw[one loop probe,one loop momentum] (-2.15,0.95) -- (left);
				\draw[one loop probe,one loop momentum] (left) -- (right);
				\draw[one loop probe,one loop outgoing momentum]
				(right) -- (2.15,0.95);
				
				\node[one loop momentum label,above left=1pt]
				at (-2.15,0.95) {$\vec p_1$};
				\node[one loop momentum label,above=3pt]
				at (0,0) {$\vec\ell$};
				\node[one loop momentum label,above right=1pt]
				at (2.15,0.95) {$\vec p_2$};
				
				\node[one loop source] (source-left)  at (-1.15,-1.35) {$\otimes$};
				\node[one loop source] (source-right) at ( 1.15,-1.35) {$\otimes$};
				\draw[one loop photon] (left)  -- (source-left.north);
				\draw[one loop photon] (right) -- (source-right.north);
			\end{tikzpicture}
			\caption{$I_{CC}$}
			\label{fig:one-loop-CC}
		\end{subfigure}
		\hfill
		\begin{subfigure}[t]{0.31\textwidth}
			\centering
			\begin{tikzpicture}[
				scale=0.88,
				transform shape,
				baseline={(current bounding box.center)}
				]
				\coordinate (left)  at (-1.15,0);
				\coordinate (right) at ( 1.15,0);
				
				\draw[one loop probe,one loop momentum] (-2.15,0.95) -- (left);
				\draw[one loop probe,one loop momentum] (left) -- (right);
				\draw[one loop probe,one loop outgoing momentum]
				(right) -- (2.15,0.95);
				
				\node[one loop momentum label,above left=1pt]
				at (-2.15,0.95) {$\vec p_1$};
				\node[one loop momentum label,above=3pt]
				at (0,0) {$\vec\ell$};
				\node[one loop momentum label,above right=1pt]
				at (2.15,0.95) {$\vec p_2$};
				
				\fill (left) circle (1.5pt);
				\node[one loop coupling] at (-0.88,0.28) {$\beta^2$};
				
				\node[one loop source] (source-left-a) at (-1.42,-1.35) {$\otimes$};
				\node[one loop source] (source-left-b) at (-0.88,-1.35) {$\otimes$};
				\node[one loop source] (source-right)  at ( 1.15,-1.35) {$\otimes$};
				
				\draw[one loop photon] (left)  -- (source-left-a.north);
				\draw[one loop photon] (left)  -- (source-left-b.north);
				\draw[one loop photon] (right) -- (source-right.north);
			\end{tikzpicture}
			\caption{$I_{CS}$}
			\label{fig:one-loop-CS}
		\end{subfigure}
		\hfill
		\begin{subfigure}[t]{0.31\textwidth}
			\centering
			\begin{tikzpicture}[
				scale=0.88,
				transform shape,
				baseline={(current bounding box.center)}
				]
				\coordinate (left)  at (-1.15,0);
				\coordinate (right) at ( 1.15,0);
				
				\draw[one loop probe,one loop momentum] (-2.15,0.95) -- (left);
				\draw[one loop probe,one loop momentum] (left) -- (right);
				\draw[one loop probe,one loop outgoing momentum]
				(right) -- (2.15,0.95);
				
				\node[one loop momentum label,above left=1pt]
				at (-2.15,0.95) {$\vec p_1$};
				\node[one loop momentum label,above=3pt]
				at (0,0) {$\vec\ell$};
				\node[one loop momentum label,above right=1pt]
				at (2.15,0.95) {$\vec p_2$};
				
				\fill (left)  circle (1.5pt);
				\fill (right) circle (1.5pt);
				\node[one loop coupling] at (-0.88,0.28) {$\beta^2$};
				\node[one loop coupling] at ( 0.88,0.28) {$\beta^2$};
				
				\node[one loop source] (source-left-a)  at (-1.42,-1.35) {$\otimes$};
				\node[one loop source] (source-left-b)  at (-0.88,-1.35) {$\otimes$};
				\node[one loop source] (source-right-a) at ( 0.88,-1.35) {$\otimes$};
				\node[one loop source] (source-right-b) at ( 1.42,-1.35) {$\otimes$};
				
				\draw[one loop photon] (left)  -- (source-left-a.north);
				\draw[one loop photon] (left)  -- (source-left-b.north);
				\draw[one loop photon] (right) -- (source-right-a.north);
				\draw[one loop photon] (right) -- (source-right-b.north);
			\end{tikzpicture}
			\caption{$I_{SS}$}
			\label{fig:one-loop-SS}
		\end{subfigure}
		\caption{The three contributions to the one-loop amplitude in
			short-range perturbation theory. A single wavy line denotes a Coulomb
			insertion, while the black vertex with two wavy lines denotes an
			inverse-square insertion proportional to $\beta^2$. The reflected
			ordering of the mixed diagram is included in $I_{CS}$.}
		\label{fig:one-loop-short-range-diagrams}
	\end{figure}
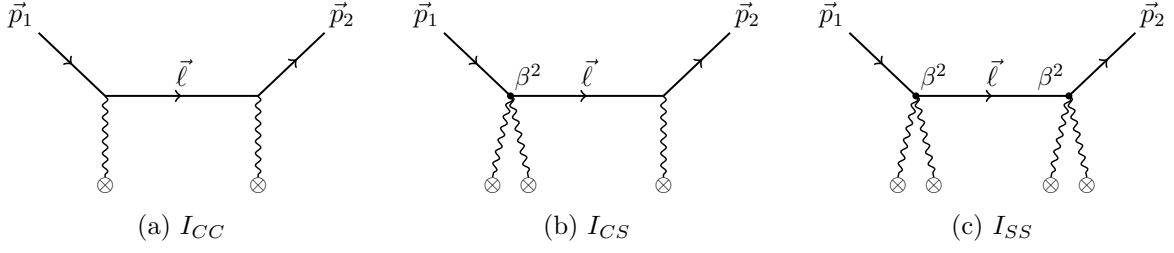

	In this appendix we compute the second Born terms obtained by applying short-range perturbation theory \eqref{short pert} to the Hamiltonian \eqref{hamil}. The one-loop amplitude is
	\begin{align}
		\mathcal{A}^{(1)}_{\rm short}(\vec p_1\rightarrow \vec p_2)
		&=
		-2\pi i\delta(E_1-E_2)
		\int \frac{\DD^3\vec{\ell}}{(2\pi)^3}\,
		\frac{
			\left(
			\frac{4\pi\alpha}{|\vec p_2-\vec \ell|^2}
			-
			\frac{\pi^2\beta^2}{m|\vec p_2-\vec \ell|}
			\right)
			\left(
			\frac{4\pi\alpha}{|\vec \ell-\vec p_1|^2}
			-
			\frac{\pi^2\beta^2}{m|\vec \ell-\vec p_1|}
			\right)
		}{
			E_1-\frac{|\vec \ell|^2}{2m}+i\epsilon
		} .
	\end{align}
	We decompose the integral according to the number of Coulomb and short-range
	insertions,
	\begin{gather}
		\mathcal{A}^{(1)}_{\rm short}(\vec p_1\rightarrow \vec p_2)
		=
		-2\pi i\delta(E_1-E_2)
		\left(
		I_{CC}
		+
		I_{CS}
		+
		I_{SS}
		\right),
	\end{gather}
	where $I_{CC}$ contains two Coulomb insertions, $I_{SS}$ contains two
	inverse-square insertions, and $I_{CS}$ contains the two mixed Coulomb--short
	terms. These three contributions are shown diagrammatically in
	Figure~\ref{fig:one-loop-short-range-diagrams}.
	
	The Coulomb--Coulomb integral requires an infrared regulator. We regulate it by
	giving the exchanged massless particle a mass $\lambda_{\text{I.R.}}$, and retain only the
	terms that survive as $\lambda_{\text{I.R.}}\to0$:
	\begin{align}
		I_{CC}
		&=
		(4\pi\alpha)^2
		\int \frac{\DD^3\vec{\ell}}{(2\pi)^3}\,
		\frac{1}{
			E_p-\frac{|\vec{\ell}|^2}{2m}+i\epsilon
		}
		\frac{1}{
			|\vec p_2-\vec{\ell}|^2
		}
		\frac{1}{
			|\vec{\ell}-\vec p_1|^2
		}
		\\
		&\overset{\mathrm{reg.}}{\longrightarrow}
		(4\pi\alpha)^2
		\int \frac{\DD^3\vec{\ell}}{(2\pi)^3}\,
		\frac{1}{
			E_p-\frac{|\vec{\ell}|^2}{2m}+i\epsilon
		}
		\frac{1}{
			|\vec p_2-\vec{\ell}|^2+\lambda_{\text{I.R.}}^2
		}
		\frac{1}{
			|\vec{\ell}-\vec p_1|^2+\lambda_{\text{I.R.}}^2
		}
		\\
		&=
		(4\pi\alpha)^2
		\left[
		-\frac{i m}{4\pi p |\vec q|^2}
		\log\!\left(\frac{|\vec q|^2}{\lambda_{\text{I.R.}}^2}\right)
		\right]
		+
		\mathcal{O}(\lambda_{\text{I.R.}})
		\\
		&=
		-\frac{4\pi i m\alpha^2}{p|\vec{q}|^2}
		\log\!\left(\frac{|\vec{q}|^2}{\lambda_{\text{I.R.}}^2}\right)
		+
		\mathcal{O}(\lambda_{\text{I.R.}}).
	\end{align}
	The Coulomb--short-range terms give
	\begin{align}
		I_{CS}
		&=
		-(4\pi\alpha)
		\left(
		\frac{\pi^2\beta^2}{m}
		\right)
		\int \frac{\DD^3\vec{\ell}}{(2\pi)^3}\,
		\nonumber\\
		&\quad\times
		\frac{1}{
			E_p-\frac{|\vec{\ell}|^2}{2m}+i\epsilon
		}
		\left[
		\frac{1}{
			|\vec p_2-\vec{\ell}|^2
		}
		\frac{1}{
			|\vec{\ell}-\vec p_1|
		}
		+
		\frac{1}{
			|\vec p_2-\vec{\ell}|
		}
		\frac{1}{
			|\vec{\ell}-\vec p_1|^2
		}
		\right]
		\\
		&\overset{\mathrm{reg.}}{\longrightarrow}
		-(4\pi\alpha)
		\left(
		\frac{\pi^2\beta^2}{m}
		\right)
		\int \frac{\DD^3\vec{\ell}}{(2\pi)^3}\,
		\nonumber\\
		&\quad\times
		\frac{1}{
			E_p-\frac{|\vec{\ell}|^2}{2m}+i\epsilon
		}
		\left[
		\frac{1}{
			|\vec p_2-\vec{\ell}|^2+\lambda_{\text{I.R.}}^2
		}
		\frac{1}{
			|\vec{\ell}-\vec p_1|
		}
		+
		\frac{1}{
			|\vec p_2-\vec{\ell}|
		}
		\frac{1}{
			|\vec{\ell}-\vec p_1|^2+\lambda_{\text{I.R.}}^2
		}
		\right]
		\\
		&=
		\frac{2\pi\alpha\beta^2}{p|\vec{q}|}
		\Bigg[
		i\pi\log\!\left(\frac{|\vec{q}|}{\lambda_{\text{I.R.}}}\right)
		+
		i\pi\log\!\left(\frac{4}{1+s}\right)
		\nonumber\\
		&\hspace{2cm}
		+
		\log(s)\log\!\left(\frac{1-s}{1+s}\right)
		+
		\operatorname{Li}_2(s)
		-
		\operatorname{Li}_2(-s)
		\Bigg]
		+
		\mathcal{O}(\lambda_{\text{I.R.}}),
	\end{align}
	where
	\begin{gather}
		|\vec{q}|=|\vec p_2-\vec p_1|,
		\qquad
		p=|\vec p_1|=|\vec p_2|,
		\qquad
		s=\sin\frac{\theta}{2}=\frac{|\vec{q}|}{2p}.
	\end{gather}
	
	These terms exhibit the same obstruction already seen at tree level: once the
	Coulomb interaction is treated in short-range perturbation theory, the amplitude
	depends on the infrared regulator $\lambda_{\text{I.R.}}$, and any unitarity bound extracted
	from it inherits this dependence. In addition, the loop terms contain logarithms
	that become large in the forward region $|\vec q|\to0$. Thus the forward
	region is not uniformly perturbative, which indicates that these logarithms
	must be resummed before using the amplitude to derive unitarity relations.

	Lastly, the short-short piece is more complicated:
	\begin{align}
		I_{SS}
		&=
		\left(
		\frac{\pi^2\beta^2}{m}
		\right)^2
		\int \frac{\DD^3\vec{\ell}}{(2\pi)^3}\,
		\frac{1}{
			E_p-\frac{|\vec{\ell}|^2}{2m}+i\epsilon
		}
		\frac{1}{
			|\vec p_2-\vec{\ell}|
		}
		\frac{1}{
			|\vec{\ell}-\vec p_1|
		}
		\\
		&=
		\frac{\pi i\beta^4}{mp}
		\Bigg\{
		i\pi
		\left[
		\frac{\pi}{4}
		K\!\left(\frac{1-x}{2}\right)
		-
		\frac{1}{2}
		\sqrt{\frac{1-x}{2}}\,
		{}_3F_2\!\left(
		1,1,1;
		\frac{3}{2},\frac{3}{2};
		\frac{1-x}{2}
		\right)
		\right]
		-
		\frac{\pi^2}{4}
		K\!\left(\frac{1+x}{2}\right)
		\Bigg\}.
	\end{align}
	Here $x=\cos\theta$ and $K$ is the complete elliptic integral of the first kind. We obtained this expression by performing the relevant partial-wave sum and then numerically confirmed that the result equals this integral.
	This purely short-range piece poses no obstacle to imposing unitarity bounds:
	it has no infrared-regulator dependence and no forward-region enhancement that
	invalidates a perturbative treatment.

	\input{Sections/first_order_dwpt_appendix}

\providecommand{\href}[2]{#2}\begingroup\raggedright\endgroup
\end{document}

%% file: Figures/dwpt_resummed_diagram.tex
\tikzset{
  dwpt probe/.style={line width=0.7pt},
  dwpt momentum/.style={
    postaction={decorate},
    decoration={
      markings,
      mark=at position 0.58 with {\arrow{>}}
    }
  },
  dwpt photon/.style={
    decorate,
    decoration={
      snake,
      amplitude=1pt,
      segment length=4.5pt
    },
    line width=0.55pt
  }
}

\begin{figure}[t]
\centering
\resizebox{\textwidth}{!}{%
$
\scalebox{1.5}{$\mathcal A^{(1)}$}
=
\mathop{\scalebox{1.7}{$\displaystyle\sum$}}\limits_{N_L,N_R\geq 0}\hspace{-0.3cm}
\begin{tikzpicture}[baseline={(current bounding box.center)}]
  \draw[dwpt probe,dwpt momentum] (-3.9,1.25) -- (-3.2,0.55);
  \draw[dwpt probe]               (-3.2,0.55) -- ( 3.2,0.55);
  \draw[dwpt probe,dwpt momentum] ( 3.2,0.55) -- ( 3.9,1.25);

  \node[above left=1pt]  at (-3.9,1.25) {$\vec p_1$};
  \node[above right=1pt] at ( 3.9,1.25) {$\vec p_2$};

  \foreach \x/\source in {
    -3.2/source-left-outer,
    -1.2/source-left-inner,
     1.2/source-right-inner,
     3.2/source-right-outer
  }{
    \node[font=\small,inner sep=0pt] (\source) at (\x,-1.02) {$\otimes$};
    \draw[dwpt photon] (\x,0.55) -- (\source.north);
  }

  \node[font=\Large] at (-2.2,-0.20) {$\cdots$};
  \node[font=\Large] at ( 2.2,-0.20) {$\cdots$};

  \fill (0,0.55) circle (1.5pt);
  \node[above=2pt] at (0,0.55) {$\beta^2$};

  \node[font=\small,inner sep=0pt]
    (source-quartic-left) at (-0.28,-1.02) {$\otimes$};
  \node[font=\small,inner sep=0pt]
    (source-quartic-right) at (0.28,-1.02) {$\otimes$};
  \draw[dwpt photon] (0,0.55) -- (source-quartic-left.north);
  \draw[dwpt photon] (0,0.55) -- (source-quartic-right.north);

  \draw[decorate,decoration={brace,mirror,amplitude=4pt}]
    (-3.4,-1.18) -- (-0.95,-1.18)
    node[midway,below=5pt] {$N_L$};

  \draw[decorate,decoration={brace,mirror,amplitude=4pt}]
    (0.95,-1.18) -- (3.4,-1.18)
    node[midway,below=5pt] {$N_R$};
\end{tikzpicture}
\;=
\;
\scalebox{1.3}{$
2\pi i\frac{\beta^2}{2m}\delta(E_1-E_2)
\left\langle
\phi_{\mathrm{out}}(\vec p_2)
\left|
\frac{1}{\hat r^2}
\right|
\phi_{\mathrm{in}}(\vec p_1)
\right\rangle
$}
\,.
$
}
\caption{The first-order DWPT correction \eqref{first-order-dwpt}, whose
partial-wave expansion is given in \eqref{first order distorted}, captures
the sum of Feynman diagrams containing one quartic vertex and an arbitrary
number of Coulomb exchanges.}
\label{fig:DWPT resummed diagram}
\end{figure}

%% file: Sections/first_order_dwpt_appendix.tex
\section{First-order DWPT integrals and summations}
\label{app:first order DWPT}
In this appendix we evaluate the first-order DWPT matrix element
\begin{align}
        \mathcal{A}^{(1)}
    &=
    2\pi i\frac{\beta^2}{2m}
    \delta(E_1-E_2)
    \left\langle
    \phi_{\t{out}}(\vec p_2)
    \left|
    \frac{1}{\hat r^2}
    \right|
    \phi_{\t{in}}(\vec p_1)
    \right\rangle
    \\
    &=
    2\pi i\frac{\beta^2}{2m}
    \delta(E_1-E_2)
    \int\DD^3\vec{x}\,
    \phi_{\t{out}}^{*}(\vec{x},\vec{p}_2)
    \frac{1}{|\vec{x}|^2}
    \phi_{\t{in}}(\vec{x},\vec{p}_1).
\end{align}
Let us first comment on the convergence properties of this integral. It is
ultraviolet finite: the regular Coulomb wavefunctions are
$\mathcal O(1)$ at the origin, so the factor $r^2$ in the measure cancels
the $1/r^2$ interaction, leaving a locally integrable radial measure
$\mathcal O(\DD r)$.

It is also infrared finite, as we now explain. At large distances, both
wavefunctions behave as distorted plane waves. The matrix element therefore has the same large-distance power
counting and oscillatory character as
\begin{gather}
    \int\DD^3r\,
    \frac{e^{i\vec q\cdot\vec r}}{r^2}
    =
    \frac{2\pi^2}{|\vec q|},
    \qquad
    \vec q\neq0 .
    \label{app Fourier transform inverse square}
\end{gather}
In particular, the position-space integral is not absolutely convergent:
after taking the absolute value, the radial integral diverges at infinity.
Rather, it is an oscillatory integral whose finite value arises from
cancellations among the large-distance oscillations.

We define this oscillatory integral in the Abel sense. For example, for
analytic or numerical evaluation we may introduce the convergence factor
\begin{gather}
    \mathcal I_\lambda(\vec p_2,\vec p_1)
    \defined
    \int\DD^3r\,
    e^{-\lambda r}
    \phi_{\t{out}}^*(\vec r,\vec p_2)
    \frac{1}{r^2}
    \phi_{\t{in}}(\vec r,\vec p_1),
    \qquad
    \lambda>0,
    \label{app Abel regulated first order matrix element}
\end{gather}
and define the original matrix element by
\begin{gather}
    \mathcal I(\vec p_2,\vec p_1)
    \defined
    \lim_{\lambda\to0^+}
    \mathcal I_\lambda(\vec p_2,\vec p_1).
    \label{app Abel limit first order matrix element}
\end{gather}
The limit \eqref{app Abel limit first order matrix element} is finite and
contains no residual dependence on $\lambda$.
The matrix element is therefore infrared finite, even though its
three-dimensional position-space representation is not absolutely
convergent. The Abel parameter $\lambda$ is removed completely after defining
the oscillatory integral: the limit $\lambda\to0^+$ exists and leaves no
residual $\lambda$-dependence. This differs from the infrared regulator
$\lambda_{\mathrm{I.R.}}$ in short-range perturbation theory, where
finite-order amplitudes and the resulting bounds retain terms such as
$\log(2p/\lambda_{\mathrm{I.R.}})$.

Throughout this appendix, all angular results are restricted to the open
interval $-1<\hat p_1\cdot\hat p_2<1$.

The position-space Coulomb wavefunctions \eqref{coulomb in} and
\eqref{coulomb out} may equivalently be written in partial-wave form as
follows. With $\rho\defined pr$,
\begin{align}
    \phi_{\mathrm{in/out}}(\vec x,\vec p)
    ={}&
    \sum_{l=0}^{\infty}
    i^l(2l+1)e^{\pm i\sigma_l(\gamma)}
    \frac{F_l(\gamma,\rho)}{\rho}
    P_l(\hat x\cdot\hat p),
    \label{app Coulomb partial wave functions}\\
    e^{2i\sigma_l(\gamma)}
    ={}&
    \frac{\Gamma(l+1+i\gamma)}
         {\Gamma(l+1-i\gamma)}.
    \label{app Coulomb phase}
\end{align}
The upper sign gives the in-state and the lower sign gives the out-state.
Here $F_l$ is the real regular Coulomb radial function,
\begin{align}
    F_l(\gamma,\rho)
    ={}&
    C_l(\gamma)\rho^{l+1}e^{i\rho}
    {}_1F_1\!\left(
        l+1+i\gamma;
        2l+2;
        -2i\rho
    \right),
    \label{app regular Coulomb function}\\
    C_l(\gamma)
    ={}&
    \frac{2^l e^{-\pi\gamma/2}
    \left|\Gamma(l+1+i\gamma)\right|}
    {\Gamma(2l+2)}.
    \label{app Coulomb normalization}
\end{align}
For every $R<\infty$, the series
\eqref{app Coulomb partial wave functions} converges absolutely and uniformly
in the angular variables and for $0\leq\rho\leq R$. Indeed,
$|P_l(x)|\leq1$, and the explicit representation
\eqref{app regular Coulomb function}--\eqref{app Coulomb normalization} bounds the magnitude of the $l$-th
term by $C_\gamma(2R)^l l!/(2l)!$. Since the ratio of successive bounds is
$R/(2l+1)\to0$, the result follows from the Weierstrass $M$-test. No Abel
factor is therefore needed for the partial-wave expansion itself on a finite
radial interval; the Abel factor introduced above concerns instead the
large-radius limit of the full three-dimensional matrix element.
\subsection{Evaluating in the partial-wave basis}
On the support of $\delta(E_1-E_2)$ we have
$|\vec p_1|=|\vec p_2|\defined p$.  Substituting
\eqref{app Coulomb partial wave functions} into the matrix element, and using
the reality of $F_l$, we obtain
\begin{align}
    \left\langle
    \phi_{\t{out}}(\vec p_2)
    \left|\frac{1}{\hat r^2}\right|
    \phi_{\t{in}}(\vec p_1)
    \right\rangle
    ={}&
    \frac{1}{p^2}
    \sum_{l,l'=0}^{\infty}
    (-i)^{l'}i^l
    (2l'+1)(2l+1)
    e^{i(\sigma_{l'}+\sigma_l)}
    \nonumber\\
    &\times
    \int_0^\infty\frac{\DD r}{r^2}\,
    F_{l'}(\gamma,pr)F_l(\gamma,pr)
    \int\DD^2\Omega_x\,
    P_{l'}(\hat p_2\cdot\hat x)
    P_l(\hat x\cdot\hat p_1).
    \label{app inserted partial waves}
\end{align}
The angular integral is
\begin{gather}
    \int\DD^2\Omega_x\,
    P_{l'}(\hat p_2\cdot\hat x)
    P_l(\hat x\cdot\hat p_1)
    =
    \delta_{ll'}\frac{4\pi}{2l+1}
    P_l(\hat p_1\cdot\hat p_2).
    \label{app Legendre angular integral}
\end{gather}
Consequently $(-i)^li^l=1$, while the two Coulomb phases combine into
$e^{2i\sigma_l}=S_{c,l}$.  The matrix element therefore reduces to
\begin{align}
    \left\langle
    \phi_{\t{out}}(\vec p_2)
    \left|\frac{1}{\hat r^2}\right|
    \phi_{\t{in}}(\vec p_1)
    \right\rangle
    &={}
    \frac{4\pi}{p}
    \sum_{l=0}^{\infty}(2l+1)S_{c,l}
    P_l(\hat p_1\cdot\hat p_2)
    \mathcal I_l(\gamma),
    \label{app reduced first order matrix element}
\end{align}
where the radial integral that remains to be evaluated is
\begin{gather}
    \boxed{
    \mathcal I_l(\gamma)
    \defined
    \int_0^\infty\DD\rho\,
    \frac{F_l(\gamma,\rho)^2}{\rho^2}.}
    \label{app first order radial integral}
\end{gather}
Thus the first-order correction is
\begin{gather}
    \mathcal A^{(1)}
    =
    i\frac{4\pi^2\beta^2}{mp}\,
    \delta(E_1-E_2)
    \sum_{l=0}^{\infty}(2l+1)S_{c,l}P_l(x)
    \mathcal I_l(\gamma)
    \label{app first order partial wave reduction}\\ 
    x\defined\hat p_1\cdot\hat p_2.
    \label{app scattering angle variable}
\end{gather}

\subsection{Evaluating the radial integral}
Unlike the full position-space matrix element, the fixed-$l$ radial integral
$\mathcal I_l(\gamma)$ defined in \eqref{app first order radial integral} is
absolutely convergent: its integrand behaves as $\rho^{2l}$ at the origin
and as a bounded oscillation times $\rho^{-2}$ at infinity. No Abel
convergence factor is therefore required to evaluate
$\mathcal I_l(\gamma)$.
The integral \eqref{app first order radial integral} can be
evaluated by the direct method described in Appendix~B of
\cite{Lippstreu:2023vvg}.  The following derivation makes a different and
useful point: the integral is determined entirely by radial data at
infinity. The mechanism is that the integrand can be written as the total
derivative of a Wronskian, so the radial integral reduces to boundary data.

The inverse-square interaction shifts the centrifugal coefficient
$l(l+1)\to l(l+1)-\beta^2$, which motivates continuing $l$ to a continuous
parameter $L$. Correspondingly, differentiating the radial equation with
respect to $L$ produces precisely the $1/\rho^2$ insertion appearing in the
first-order correction.

We continue the angular momentum from the non-negative integer $l$ to a real
parameter $L$. The regular Coulomb function then obeys
\begin{gather}
    \mathcal L_LF_L=0,
    \qquad
    \mathcal L_L
    \defined
    \frac{\DD^2}{\DD\rho^2}
    +1-\frac{2\gamma}{\rho}
    -\frac{L(L+1)}{\rho^2}.
    \label{app continuous degree Coulomb equation}
\end{gather}
Differentiating this equation with respect to $L$ gives
\begin{gather}
    \mathcal L_L\big(\partial_LF_L\big)
    =
    \frac{2L+1}{\rho^2}F_L.
    \label{app differentiated Coulomb equation}
\end{gather}
Consequently, the Wronskian
\begin{gather}
    W_L(\rho)
    \defined
    F_L\big(\partial_LF_L\big)'
    -F_L'\partial_LF_L,
    \label{app angular momentum Wronskian}
\end{gather}
where primes denote derivatives with respect to $\rho$, satisfies
\begin{gather}
    \frac{\DD W_L}{\DD\rho}
    =
    (2L+1)\frac{F_L(\gamma,\rho)^2}{\rho^2}.
    \label{app Wronskian derivative}
\end{gather}
For finite $0<\epsilon<R$, integration gives the exact identity
\begin{gather}
    (2L+1)
    \int_\epsilon^R\DD\rho\,
    \frac{F_L(\gamma,\rho)^2}{\rho^2}
    =
    W_L(R)-W_L(\epsilon).
    \label{app finite radial integral boundary terms}
\end{gather}
We therefore see that the integral is determined by boundary data alone. We
will now show that the Wronskian vanishes at the origin, so that the integral
is given entirely by the Wronskian at radial infinity, which we will show has
a well-defined limit.
At the origin, the regular solution and its $L$ derivative satisfy
\begin{align}
    F_L(\gamma,\rho)
    &=C_L(\gamma)\rho^{L+1}
    \left[1+\mathcal O(\rho)\right],
    \\
    \partial_LF_L(\gamma,\rho)
    &=F_L(\gamma,\rho)
    \left[
        \log\rho+\partial_L\log C_L(\gamma)
        +\mathcal O(\rho)
    \right].
    \label{app continuous degree Coulomb origin}
\end{align}
Consequently,
\begin{gather}
    W_L(\rho)=\mathcal O(\rho^{2L+1}),
    \qquad
    \lim_{\rho\to0^+}W_L(\rho)=0.
    \label{app Wronskian origin}
\end{gather}

Having shown that the contribution at $\rho=0$ vanishes, it remains to show
that the Wronskian at radial infinity has a finite limit. The large-$\rho$
asymptotics of the Coulomb wavefunction show that the four
quantities entering the Wronskian are
\begin{align}
    F_L
    &=
    \sin\vartheta_L+\mathcal O(\rho^{-1}),
    \\
    F_L'
    &=
    \left(1-\frac{\gamma}{\rho}\right)
    \cos\vartheta_L+\mathcal O(\rho^{-1}),
    \\
    \partial_LF_L
    &=
    a_L\cos\vartheta_L+\mathcal O(\rho^{-1}),
    \\
    (\partial_LF_L)'
    &=
    -a_L
    \left(1-\frac{\gamma}{\rho}\right)
    \sin\vartheta_L+\mathcal O(\rho^{-1}).
    \label{app differentiated Coulomb asymptotics}
\end{align}
Here
\begin{align}
    \vartheta_L(\rho)
    &\defined
    \rho-\gamma\log(2\rho)-\frac{\pi L}{2}
    +\sigma_L(\gamma),
    \\
    \sigma_L(\gamma)
    &\defined\arg\Gamma(L+1+i\gamma),
    \\
    \frac{\DD\vartheta_L}{\DD\rho}
    &=1-\frac{\gamma}{\rho},
    \\
    a_L
    &\defined
    \partial_L\vartheta_L
    =\partial_L\sigma_L-\frac{\pi}{2}.
    \label{app continuous degree Coulomb phase}
\end{align}
Substitution into \eqref{app angular momentum Wronskian} displays the
cancellation explicitly:
\begin{align}
    W_L(\rho)
    &={}
    -a_L\vartheta_L'
    \left(
        \sin^2\vartheta_L+\cos^2\vartheta_L
    \right)
    +\mathcal O(\rho^{-1})
    \nonumber\\
    &={}
    \left(
        \frac{\pi}{2}-\partial_L\sigma_L
    \right)
    \left(1-\frac{\gamma}{\rho}\right)
    +\mathcal O(\rho^{-1}).
    \label{app Wronskian cancellation}
\end{align}
Although the two terms in the Wronskian generally have no separate limits,
their leading oscillatory pieces combine into
$\sin^2\vartheta_L+\cos^2\vartheta_L=1$.  The asymptotic estimate then proves
that the Wronskian has the well-defined limit
\begin{gather}
    \lim_{\rho\to\infty}W_L(\rho)
    =
    \frac{\pi}{2}-\partial_L\sigma_L.
    \label{app Wronskian infinity}
\end{gather}
Taking $\epsilon\to0^+$ and then $R\to\infty$ in
\eqref{app finite radial integral boundary terms} therefore gives
\begin{gather}
    (2L+1)
    \int_0^\infty\DD\rho\,
    \frac{F_L(\gamma,\rho)^2}{\rho^2}
    =
    \lim_{\rho\to\infty}W_L(\rho).
    \label{app radial integral boundary terms}
\end{gather}
Finally,
\begin{gather}
    \partial_L\sigma_L(\gamma)
    =
    \operatorname{Im}\psi(L+1+i\gamma),
\end{gather}
where $\psi$ is the digamma function.  Setting $L=l$ in
\eqref{app radial integral boundary terms} therefore yields
\begin{align}
    \boxed{
    \mathcal I_l(\gamma)
    =
    \frac{
        \frac{\pi}{2}
        -\operatorname{Im}\psi(l+1+i\gamma)
    }{2l+1}
    =
    \frac{
        \pi-2\operatorname{Im}H_{l+i\gamma}
    }{2(2l+1)}.}
    \label{app evaluated first order radial integral}
\end{align}
Here $H_z=\psi(z+1)+\gamma_{\mathrm E}$ is the analytic continuation of
the harmonic numbers.  

Substituting \eqref{app evaluated first order radial integral} into
\eqref{app first order partial wave reduction} gives
\begin{align}
    \mathcal A^{(1)}
    &={}
    i\frac{2\pi^2\beta^2}{mp}\,
    \delta(E_1-E_2)
    \sum_{l=0}^{\infty}
    S_{c,l}
    \left(
        \pi-2\operatorname{Im}H_{l+i\gamma}
    \right)P_l(x)
    \nonumber\\
    &={}
    \frac{2\pi^2\beta^2}{mp}\,
    \delta(E_1-E_2)
    \sum_{l=0}^{\infty}
    S_{c,l}
    \left(
        i\pi-H_{l+i\gamma}+H_{l-i\gamma}
    \right)P_l(x).
    \label{app evaluated first order amplitude}
\end{align}
This precisely agrees with the order-$\beta^2$ term of the exact
non-perturbative amplitude \eqref{exact amp}--\eqref{Sl exact def}.
\subsection{Summing the partial-wave series}
The partial-wave representation \eqref{app evaluated first order amplitude}
is sufficient for deriving the unitarity bounds. A closed momentum-space form
is nevertheless useful because it exposes the function class and analytic
structure of the amplitude and facilitates direct comparison with short-range
perturbation theory. We therefore sum the partial-wave series. Define the
reduced sum
\begin{gather}
    \mathcal S_\gamma(x)
    \defined
    \sum_{l=0}^{\infty}
    \frac{\Gamma(l+1+i\gamma)}
         {\Gamma(l+1-i\gamma)}
    \left(
        i\pi+H_{l-i\gamma}-H_{l+i\gamma}
    \right)
    P_l(x),
    \qquad -1<x<1,
    \label{app reduced first order partial wave sum}
\end{gather}
so that
\begin{gather}
    \mathcal A^{(1)}
    =
    \frac{2\pi^2\beta^2}{mp}\,
    \delta(E_1-E_2)\,
    \mathcal S_\gamma(x).
    \label{app first order amplitude reduced sum}
\end{gather}

For fixed $-1<x<1$, the series
\eqref{app reduced first order partial wave sum} is conditionally convergent
but not absolutely convergent.  Indeed, writing $x=\cos\theta$ with
$0<\theta<\pi$, the large-$l$ asymptotics are
\begin{align}
    \frac{\Gamma(l+1+i\gamma)}
         {\Gamma(l+1-i\gamma)}
    &=
    l^{2i\gamma}
    \left[1+\mathcal O(l^{-1})\right],
    \\
    i\pi+H_{l-i\gamma}-H_{l+i\gamma}
    &=
    i\pi+\mathcal O(l^{-1}),
    \\
    P_l(\cos\theta)
    &=
    \sqrt{\frac{2}{\pi(l+\frac12)\sin\theta}}\,
    \cos\left[
        \left(l+\frac12\right)\theta-\frac{\pi}{4}
    \right]
    +\mathcal O(l^{-3/2}).
    \label{app first order summand large l}
\end{align}
The series therefore converges through cancellations among the Legendre
oscillations, whereas the sum of the absolute values diverges like a
weighted sum of $l^{-1/2}$.  An Abel factor $r^l$, followed by
$r\to1^-$, may be used to accelerate numerical convergence.  It is not
needed to define the series on the open interval $-1<x<1$, however, and we
will not introduce one below. This should be contrasted with the Coulomb
partial-wave amplitude \eqref{coulomb pw series}, which contains an additional
factor of $(2l+1)$ and hence is not conditionally convergent.

We instead require a temporary prescription in order to use Euler's beta
integral.  Introduce
\begin{gather}
    a_l\defined l+1+i\gamma,
    \qquad
    b_{+}\defined\epsilon_{+}-2i\gamma,
    \qquad
    \epsilon_{+}>0.
    \label{app beta continuation parameters}
\end{gather}
Euler's integral then gives the ordinary convergent representation
\begin{align}
    c_{l,\epsilon_{+}}(\gamma)
    &\defined
    \frac{\Gamma(l+1+i\gamma)}
         {\Gamma(l+1-i\gamma+\epsilon_{+})}
    \nonumber\\
    &=
    \frac{1}{\Gamma(b_{+})}
    \int_0^1\DD t\,
    t^{l+i\gamma}
    (1-t)^{-1-2i\gamma+\epsilon_{+}}.
    \label{app regulated beta coefficient}
\end{align}
The original Coulomb coefficient is recovered as
$\epsilon_{+}\to0^+$.  This continuation is necessary because the
unshifted beta parameter has real part
$\operatorname{Re}(-2i\gamma)=0$, exactly on the boundary of the domain
$\operatorname{Re}b>0$ in which Euler's integral is ordinarily
convergent.

Differentiating \eqref{app regulated beta coefficient} with respect to
$a_l$, while holding $b_{+}$ fixed, yields
\begin{align}
    &c_{l,\epsilon_{+}}(\gamma)
    \left(
        H_{l-i\gamma+\epsilon_{+}}-H_{l+i\gamma}
    \right)
    \nonumber\\
    &\hspace{1cm}={}
    -\frac{1}{\Gamma(b_{+})}
    \int_0^1\DD t\,
    t^{l+i\gamma}
    (1-t)^{-1-2i\gamma+\epsilon_{+}}
    \log t.
    \label{app regulated beta harmonic difference}
\end{align}
It is important to keep this term together with the $i\pi$ term.  Define
\begin{align}
    \mathcal S_{\gamma,\epsilon_{+}}(x)
    \defined
    \sum_{l=0}^{\infty}
    c_{l,\epsilon_{+}}(\gamma)
    \left(
        i\pi
        +H_{l-i\gamma+\epsilon_{+}}
        -H_{l+i\gamma}
    \right)
    P_l(x).
    \label{app regulated first order partial wave sum}
\end{align}
Using the Legendre generating function
\begin{gather}
    \sum_{l=0}^{\infty}t^lP_l(x)
    =
    \frac{1}{\sqrt{1-2xt+t^2}},
    \qquad 0\leq t<1,
    \label{app Legendre generating function}
\end{gather}
we obtain
\begin{align}
    \mathcal S_{\gamma,\epsilon_{+}}(x)
    =
    \frac{1}{\Gamma(b_{+})}
    \int_0^1\DD t\,
    \frac{
        t^{i\gamma}
        (1-t)^{-1-2i\gamma+\epsilon_{+}}
        \left(i\pi-\log t\right)
    }{
        \sqrt{1-2xt+t^2}
    }.
    \label{app regulated summed t integral}
\end{align}
No Abel factor is needed in this step.  For fixed $-1<x<1$, the partial
sums of $\sum_lP_l(x)$ are bounded, and summation by parts therefore bounds
$\sum_{l=0}^Nt^lP_l(x)$ uniformly for $0\leq t\leq1$.  Since the remaining
endpoint weight is integrable for every $\epsilon_{+}>0$, dominated
convergence justifies interchanging the sum with the $t$ integral.
The desired sum is the boundary value
\begin{gather}
    \mathcal S_\gamma(x)
    =
    \lim_{\epsilon_{+}\to0^+}
    \mathcal S_{\gamma,\epsilon_{+}}(x).
    \label{app epsilon boundary value partial wave sum}
\end{gather}

To evaluate this boundary value, set
\begin{gather}
    s\defined\sqrt{\frac{1-x}{2}},
    \qquad
    t=e^{-2\chi},
    \qquad
    0<s<1.
    \label{app first order summation variables}
\end{gather}
Since
\begin{gather}
    1-2xt+t^2
    =
    4e^{-2\chi}
    \left(\sinh^2\chi+s^2\right),
\end{gather}
equation \eqref{app regulated summed t integral} becomes
\begin{align}
    \mathcal S_{\gamma,\epsilon_{+}}(s)
    =
    \frac{2^{b_{+}}}{\Gamma(b_{+})}
    \int_0^\infty\DD\chi\,
    \frac{
        \left(\chi+\frac{i\pi}{2}\right)
        e^{-\epsilon_{+}\chi}
        \sinh(\chi)^{b_{+}-1}
    }{
        \sqrt{\sinh^2\chi+s^2}
    }.
    \label{app regulated summed chi integral}
\end{align}
The usefulness of retaining the two terms in parentheses together becomes
apparent after taking a Mellin transform in $s$.  Regarding the right-hand
side of \eqref{app regulated summed chi integral} as its continuation to
$s>0$, define
\begin{gather}
    \widehat{\mathcal S}_{\gamma,\epsilon_{+}}(u)
    \defined
    \int_0^\infty\DD s\,
    s^{u-1}\mathcal S_{\gamma,\epsilon_{+}}(s).
    \label{app regulated first order Mellin transform}
\end{gather}
For $0<\epsilon_{+}<1$, a fundamental strip is
\begin{gather}
    1-\epsilon_{+}<\operatorname{Re}u<1.
    \label{app regulated Mellin strip}
\end{gather}
Using
\begin{gather}
    \int_0^\infty\DD s\,
    \frac{s^{u-1}}{\sqrt{A^2+s^2}}
    =
    \frac{A^{u-1}}{2\sqrt{\pi}}\,
    \Gamma\left(\frac{u}{2}\right)
    \Gamma\left(\frac{1-u}{2}\right),
    \label{app elementary Mellin kernel}
\end{gather}
we are left with
\begin{gather}
    K_{\gamma,\epsilon_{+}}(u)
    \defined
    \int_0^\infty\DD\chi\,
    \left(\chi+\frac{i\pi}{2}\right)
    e^{-\epsilon_{+}\chi}
    \sinh(\chi)^{\kappa_{+}},
    \qquad
    \kappa_{+}\defined u+b_{+}-2.
    \label{app regulated auxiliary Mellin integral}
\end{gather}
Writing
\begin{gather}
    z\defined\frac{u}{2}-i\gamma
\end{gather}
and setting $q=e^{-2\chi}$ gives
\begin{align}
    \int_0^\infty\DD\chi\,
    e^{-\epsilon_{+}\chi}\sinh(\chi)^{\kappa_{+}}
    &=
    2^{-\kappa_{+}-1}
    B\left(1-z,2z+\epsilon_{+}-1\right),
    \\
    \int_0^\infty\DD\chi\,
    \chi e^{-\epsilon_{+}\chi}\sinh(\chi)^{\kappa_{+}}
    &=
    2^{-\kappa_{+}-2}
    B\left(1-z,2z+\epsilon_{+}-1\right)
    \left[
        \psi(z+\epsilon_{+})-\psi(1-z)
    \right].
    \label{app regulated auxiliary beta integrals}
\end{align}
Consequently,
\begin{align}
    K_{\gamma,\epsilon_{+}}(u)
    ={}&
    2^{-\kappa_{+}-2}
    B\left(1-z,2z+\epsilon_{+}-1\right)
    \nonumber\\
    &\times
    \left[
        \psi(z+\epsilon_{+})-\psi(1-z)+i\pi
    \right].
    \label{app regulated auxiliary Mellin result}
\end{align}
Combining this with \eqref{app elementary Mellin kernel} yields
\begin{align}
    \widehat{\mathcal S}_{\gamma,\epsilon_{+}}(u)
    ={}&
    \frac{2^{-u-1}}{
        \sqrt{\pi}\Gamma(b_{+})
    }
    \Gamma\left(\frac{u}{2}\right)
    \Gamma\left(\frac{1-u}{2}\right)
    B\left(1-z,2z+\epsilon_{+}-1\right)
    \nonumber\\
    &\times
    \left[
        \psi(z+\epsilon_{+})-\psi(1-z)+i\pi
    \right].
    \label{app regulated first order Mellin result}
\end{align}

We may now take the boundary value $\epsilon_{+}\to0^+$.  The digamma
reflection identity gives
\begin{align}
    \psi(z)-\psi(1-z)+i\pi
    &=
    \pi\left(i-\cot\pi z\right)
    \nonumber\\
    &=
    -e^{-i\pi z}\Gamma(z)\Gamma(1-z).
    \label{app first order reflection cancellation}
\end{align}
The factor $\Gamma(z)$ cancels the denominator of
$B(1-z,2z-1)$.  Thus
\begin{align}
    \widehat{\mathcal S}_\gamma(u)
    ={}&
    -\frac{
        2^{-u-1}e^{-i\pi(u/2-i\gamma)}
    }{
        \sqrt{\pi}\Gamma(-2i\gamma)
    }
    \Gamma\left(\frac{u}{2}\right)
    \Gamma\left(\frac{1-u}{2}\right)
    \nonumber\\
    &\times
    \Gamma(u-1-2i\gamma)
    \Gamma\left(1+i\gamma-\frac{u}{2}\right)^2.
    \label{app first order Mellin result}
\end{align}
This cancellation is the central simplification: the $i\pi$ term removes
the digamma function generated by the harmonic-number difference.

Inverting the Mellin transform, setting $u=2v$, and applying the duplication
formula to $\Gamma(2v-1-2i\gamma)$ gives
\begin{align}
    \mathcal S_\gamma(x)
    ={}&
    -\frac{
        e^{-\pi\gamma}2^{-2-2i\gamma}
    }{
        \pi\Gamma(-2i\gamma)
    }
    \frac{1}{2\pi i}
    \int_{\mathcal C}\DD v\,
    (-s^2+i0)^{-v}
    \nonumber\\
    &\times
    \Gamma(v)
    \Gamma\left(v-\frac12-i\gamma\right)
    \Gamma(v-i\gamma)
    \Gamma\left(\frac12-v\right)
    \Gamma(1+i\gamma-v)^2.
    \label{app first order Mellin Barnes integral}
\end{align}
The contour and the indicated boundary value are inherited from
$\epsilon_{+}>0$.  For $0<s<1$, closing the contour to the left encloses
the three pole families
\begin{gather}
    v=-n,
    \qquad
    v=\frac12+i\gamma-n,
    \qquad
    v=i\gamma-n,
    \qquad
    n=0,1,2,\ldots.
    \label{app first order Mellin Barnes poles}
\end{gather}
Writing $x=\cos\theta$, summing their residues gives
\begin{equation}
    \boxed{
    \begin{aligned}
    \mathcal S_\gamma(\cos\theta)
    ={}&
    \frac{i\pi}{2}
    \frac{\Gamma\left(\frac12+i\gamma\right)}
         {\Gamma\left(\frac12-i\gamma\right)}
    \left(\sin\frac{\theta}{2}\right)^{-1-2i\gamma}
    {}_2F_1\left(
        \frac12,-i\gamma;
        \frac12-i\gamma;
        \sin^2\frac{\theta}{2}
    \right)
    \\
    &+
    \frac{
        e^{-\pi\gamma}\Gamma(1+i\gamma)^2
    }{
        1+2i\gamma
    }
    {}_2F_1\left(
        \frac12,1+i\gamma;
        \frac32+i\gamma;
        \sin^2\frac{\theta}{2}
    \right)
    \\
    &+
    \frac{\Gamma(i\gamma)}{\Gamma(-i\gamma)}
    \left(\sin\frac{\theta}{2}\right)^{-2i\gamma}
    {}_3F_2\left(
        \begin{matrix}
            \frac12-i\gamma,1,1\\
            1-i\gamma,\frac32
        \end{matrix};
        \sin^2\frac{\theta}{2}
    \right).
    \end{aligned}}
    \label{app summed first order partial wave series}
\end{equation}
No partial-wave sum or auxiliary integral remains.  As a check, in the
zero-Coulomb limit the second and third lines cancel and
\begin{gather}
    \mathcal S_0(x)
    =
    \frac{i\pi}{2s},
\end{gather}
which is the expected Fourier-transform result \eqref{app Fourier transform inverse square} for the inverse-square
interaction.  In the forward limit $s\to0$, the first line of
\eqref{app summed first order partial wave series} dominates and has
magnitude proportional to $1/s$, dressed by the Coulomb phase
$s^{-2i\gamma}$.

\paragraph{Numerical verification.}
We have also verified \eqref{app summed first order partial wave series}
numerically by inserting an Abel regulator $r^l$ into the original
partial-wave series and extrapolating the regulated sum to $r\to1^-$. At
$\gamma=0.3$ and $x=0.2$, the extrapolated result agrees with the closed
hypergeometric expression to about $2.5\times10^{-11}$, providing an
independent check of the resummation.

\subsection{Comparison with short-range perturbation theory}
The first-order amplitude computed in this appendix,
\eqref{app first order amplitude reduced sum}, contains exactly one
inverse-square interaction but treats the Coulomb interaction
non-perturbatively.  When re-expanded in $\alpha$, it therefore represents
the sum of the short-range perturbative diagrams containing one
inverse-square vertex and an arbitrary number of Coulomb insertions,
\begin{gather}
    \beta^2,\qquad
    \alpha\beta^2,\qquad
    \alpha^2\beta^2,\qquad\ldots .
    \label{app one inverse square perturbative tower}
\end{gather}
In this subsection we make this statement explicit at the first
nontrivial order in $\alpha$.

We test this interpretation by expanding the DWPT result through first
order in the Coulomb coupling $\alpha$ and comparing it directly with
short-range perturbation theory.

In Appendix~\ref{app:1loop amplitudes}, the terms of order $\beta^2$ in
short-range perturbation theory were computed through one loop.  The
tree-level term in \eqref{tree short} is
\begin{gather}
    \mathcal A_{\beta^2}^{(0)}
    =
    2\pi i\delta(E_1-E_2)
    \frac{\pi^2\beta^2}{m|\vec q|},
    \label{app short range tree inverse square amplitude}
\end{gather}
while the two mixed Coulomb--inverse-square diagrams give
\begin{gather}
    \mathcal A_{\rm short}^{(\alpha\beta^2)}
    =
    -2\pi i\delta(E_1-E_2)I_{CS}^{\rm short},
    \label{app short range mixed amplitude definition}
\end{gather}
where
\begin{equation}
    I_{CS}^{\rm short}
    =
    \frac{2\pi\alpha\beta^2}{p|\vec q|}
    \left[
        i\pi\left(
            \log\frac{|\vec q|}{\lambda_{\text{I.R.}}}
            +\log\frac{4}{1+s}
        \right)
        +R_2(s)
    \right]
    +\mathcal O(\lambda_{\text{I.R.}})
    \label{app short range mixed Coulomb short term}
\end{equation}
and
\begin{align}
    R_2(s)
    &\defined
    \log s\log\left(\frac{1-s}{1+s}\right)
    +\operatorname{Li}_2(s)
    -\operatorname{Li}_2(-s)
    \label{app weight two remainder}
\end{align}
Thus the first Coulomb correction contains weight-two logarithms and
dilogarithms, together with the infrared logarithm associated with the
screening mass $\lambda_{\text{I.R.}}$.

We expand the hypergeometric functions in
\eqref{app summed first order partial wave series} through first order in
$\gamma$ using the Mathematica package \texttt{HypExp}~2.0
\cite{Huber:2007dx}.  We find
\begin{equation}
    \begin{aligned}
    \mathcal S_\gamma(s)
    ={}&
    \frac{i\pi}{2s}
    +
    \frac{\gamma}{s}
    \left[
        \pi\left(
            \gamma_{\rm E}
            +\log s
            +\log\frac{4}{1+s}
        \right)
        -iR_2(s)
    \right]
    +
    \mathcal O(\gamma^2).
    \end{aligned}
    \label{app weak Coulomb expansion}
\end{equation}
When inserted into \eqref{app first order amplitude reduced sum}, the leading term reproduces
\eqref{app short range tree inverse square amplitude}.  Defining the
coefficient of $\alpha\beta^2$ by
\begin{gather}
    \mathcal A_{\rm DWPT}^{(\alpha\beta^2)}
    =
    -2\pi i\delta(E_1-E_2)I_{CS}^{\rm DWPT},
\end{gather}
and substituting \eqref{app weak Coulomb expansion} into
\eqref{app first order amplitude reduced sum}, we obtain
\begin{equation}
    I_{CS}^{\rm DWPT}
    =
    \frac{2\pi\alpha\beta^2}{p|\vec q|}
    \left[
        i\pi\left(
            \gamma_{\rm E}
            +\log s
            +\log\frac{4}{1+s}
        \right)
        +R_2(s)
    \right].
    \label{app DWPT mixed Coulomb short term}
\end{equation}
The difference between the two results is
\begin{align}
    I_{CS}^{\rm short}-I_{CS}^{\rm DWPT}
    =
    \frac{2\pi\alpha\beta^2}{p|\vec q|}
    i\pi
    \left(
        \log\frac{2p}{\lambda_{\text{I.R.}}}
        -\gamma_{\rm E}
    \right),
    \label{app mixed term screening difference}
\end{align}
where $|\vec q|=2ps$ has been used.  This difference is the tree-level
inverse-square amplitude multiplied by the order-$\alpha$ term in the
universal Coulomb screening phase.  Indeed, defining
\begin{gather}
    Z_{\lambda_{\text{I.R.}}}(\gamma)
    =
    \exp\left[
        -2i\gamma
        \left(
            \log\frac{2p}{\lambda_{\text{I.R.}}}
            -\gamma_{\rm E}
        \right)
    \right].
    \label{app Coulomb screening phase}
\end{gather}
the two amplitudes are related, in the sector containing one
inverse-square insertion, by
\begin{gather}
    \mathcal A_{{\rm short},\lambda_{\text{I.R.}}}
    =
    Z_{\lambda_{\text{I.R.}}}(\gamma)\,
    \mathcal A_{\rm DWPT}
    +
    \mathcal O(\lambda_{\text{I.R.}}).
    \label{app screened and DWPT amplitude relation}
\end{gather}

This provides a first nontrivial check that either the partial-wave sum
\eqref{app reduced first order partial wave sum} or, equivalently, its
hypergeometric form
\eqref{app summed first order partial wave series}, resums the short-range
perturbative diagrams containing one inverse-square vertex.  The short-range
and DWPT formulations differ only in that the short-range amplitude is related to
the DWPT amplitude by the universal soft factor $Z_{\lambda_{\text{I.R.}}}$, and in that
short-range perturbation theory contains a disconnected identity
component whereas DWPT does not.

%% file: main.bbl
\begin{thebibliography}{10}

\bibitem{Pham:1985cr}
T.~N. Pham and T.~N. Truong, {\it {Evaluation of the Derivative Quartic Terms
  of the Meson Chiral Lagrangian From Forward Dispersion Relation}},  {\em
  Phys. Rev. D} {\bf 31} (1985) 3027.

\bibitem{Ananthanarayan:1994hf}
B.~Ananthanarayan, D.~Toublan, and G.~Wanders, {\it {Consistency of the chiral
  pion pion scattering amplitudes with axiomatic constraints}},  {\em Phys.
  Rev. D} {\bf 51} (1995) 1093--1100,
  [\href{http://arxiv.org/abs/hep-ph/9410302}{{\tt hep-ph/9410302}}].

\bibitem{Adams:2006sv}
A.~Adams, N.~Arkani-Hamed, S.~Dubovsky, A.~Nicolis, and R.~Rattazzi, {\it
  {Causality, analyticity and an IR obstruction to UV completion}},  {\em JHEP}
  {\bf 10} (2006) 014, [\href{http://arxiv.org/abs/hep-th/0602178}{{\tt
  hep-th/0602178}}].

\bibitem{Camanho:2014apa}
X.~O. Camanho, J.~D. Edelstein, J.~Maldacena, and A.~Zhiboedov, {\it {Causality
  Constraints on Corrections to the Graviton Three-Point Coupling}},  {\em
  JHEP} {\bf 02} (2016) 020, [\href{http://arxiv.org/abs/1407.5597}{{\tt
  arXiv:1407.5597}}].

\bibitem{Bellazzini:2014waa}
B.~Bellazzini, L.~Martucci, and R.~Torre, {\it {Symmetries, Sum Rules and
  Constraints on Effective Field Theories}},  {\em JHEP} {\bf 09} (2014) 100,
  [\href{http://arxiv.org/abs/1405.2960}{{\tt arXiv:1405.2960}}].

\bibitem{Cheung:2016yqr}
C.~Cheung and G.~N. Remmen, {\it {Positive Signs in Massive Gravity}},  {\em
  JHEP} {\bf 04} (2016) 002, [\href{http://arxiv.org/abs/1601.04068}{{\tt
  arXiv:1601.04068}}].

\bibitem{deRham:2017avq}
C.~de~Rham, S.~Melville, A.~J. Tolley, and S.-Y. Zhou, {\it {Positivity bounds
  for scalar field theories}},  {\em Phys. Rev. D} {\bf 96} (2017), no.~8
  081702, [\href{http://arxiv.org/abs/1702.06134}{{\tt arXiv:1702.06134}}].

\bibitem{deRham:2017zjm}
C.~de~Rham, S.~Melville, A.~J. Tolley, and S.-Y. Zhou, {\it {UV complete me:
  Positivity Bounds for Particles with Spin}},  {\em JHEP} {\bf 03} (2018) 011,
  [\href{http://arxiv.org/abs/1706.02712}{{\tt arXiv:1706.02712}}].

\bibitem{Bellazzini:2020cot}
B.~Bellazzini, J.~Elias~Mir{\'o}, R.~Rattazzi, M.~Riembau, and F.~Riva, {\it
  {Positive moments for scattering amplitudes}},  {\em Phys. Rev. D} {\bf 104}
  (2021), no.~3 036006, [\href{http://arxiv.org/abs/2011.00037}{{\tt
  arXiv:2011.00037}}].

\bibitem{Tolley:2020gtv}
A.~J. Tolley, Z.-Y. Wang, and S.-Y. Zhou, {\it {New positivity bounds from full
  crossing symmetry}},  {\em JHEP} {\bf 05} (2021) 255,
  [\href{http://arxiv.org/abs/2011.02400}{{\tt arXiv:2011.02400}}].

\bibitem{Caron-Huot:2020cmc}
S.~Caron-Huot and V.~Van~Duong, {\it {Extremal Effective Field Theories}},
  {\em JHEP} {\bf 05} (2021) 280, [\href{http://arxiv.org/abs/2011.02957}{{\tt
  arXiv:2011.02957}}].

\bibitem{Arkani-Hamed:2020blm}
N.~Arkani-Hamed, T.-C. Huang, and Y.-t. Huang, {\it {The EFT-Hedron}},  {\em
  JHEP} {\bf 05} (2021) 259, [\href{http://arxiv.org/abs/2012.15849}{{\tt
  arXiv:2012.15849}}].

\bibitem{Caron-Huot:2021rmr}
S.~Caron-Huot, D.~Mazac, L.~Rastelli, and D.~Simmons-Duffin, {\it {Sharp
  boundaries for the swampland}},  {\em JHEP} {\bf 07} (2021) 110,
  [\href{http://arxiv.org/abs/2102.08951}{{\tt arXiv:2102.08951}}].

\bibitem{Caron-Huot:2022ugt}
S.~Caron-Huot, Y.-Z. Li, J.~Parra-Martinez, and D.~Simmons-Duffin, {\it
  {Causality constraints on corrections to Einstein gravity}},  {\em JHEP} {\bf
  05} (2023) 122, [\href{http://arxiv.org/abs/2201.06602}{{\tt
  arXiv:2201.06602}}].

\bibitem{deRham:2022hpx}
C.~de~Rham, S.~Kundu, M.~Reece, A.~J. Tolley, and S.-Y. Zhou, {\it {Snowmass
  White Paper: UV Constraints on IR Physics}},  in {\em {Snowmass 2021}}, 3,
  2022.
\newblock \href{http://arxiv.org/abs/2203.06805}{{\tt arXiv:2203.06805}}.

\bibitem{Yafaev:1998LongRangeAmplitude}
D.~Yafaev, {\it The scattering amplitude for the {Schr{\"o}dinger} equation
  with a long-range potential},  {\em Communications in Mathematical Physics}
  {\bf 191} (1998) 183--218.

\bibitem{Yafaev:2000ScatteringTheory}
D.~R. Yafaev, {\em Scattering Theory: Some Old and New Problems}, vol.~1735 of
  {\em Lecture Notes in Mathematics}.
\newblock Springer, Berlin, Heidelberg, 2000.

\bibitem{herbst1974connectedness}
I.~W. Herbst, {\it On the connectedness structure of the {Coulomb}
  {$S$}-matrix},  {\em Commun. Math. Phys.} {\bf 35} (1974) 181--191.
  \href{https://doi.org/10.1007/BF01646192}{doi:10.1007/BF01646192}.

\bibitem{Haring:2022cyf}
K.~H{\"a}ring and A.~Zhiboedov, {\it {Gravitational Regge bounds}},  {\em
  SciPost Phys.} {\bf 16} (2024), no.~1 034,
  [\href{http://arxiv.org/abs/2202.08280}{{\tt arXiv:2202.08280}}].

\bibitem{Bethe}
H.~A. Bethe and L.~C. Maximon, {\it Theory of bremsstrahlung and pair
  production. i. differential cross section},  {\em Phys. Rev.} {\bf 93} (Feb,
  1954) 768--784.

\bibitem{Glauber}
P.~C. Martin and R.~J. Glauber, {\it Relativistic theory of radiative orbital
  electron capture},  {\em Phys. Rev.} {\bf 109} (Feb, 1958) 1307--1325.

\bibitem{CROTHERS1992287}
D.~S. Crothers and L.~J. Dubé, {\it Continuum distorted wave methods in
  ion—atom collisions},  in {\em Advances In Atomic, Molecular, and Optical
  Physics} (D.~Bates and B.~Bederson, eds.), vol.~30, pp.~287--337.
\newblock Academic Press, 1992.

\bibitem{Lippstreu:2023vvg}
L.~Lippstreu, {\it {A perturbation theory for the Coulomb phase
  infrared-divergence}},  \href{http://arxiv.org/abs/2312.08455}{{\tt
  arXiv:2312.08455}}.

\bibitem{Lippstreu:2025jit}
L.~Lippstreu, {\it {Analytic Properties of Infrared-Finite Amplitudes in
  Theories with Long-Range Forces}},
  \href{http://arxiv.org/abs/2505.04702}{{\tt arXiv:2505.04702}}.

\bibitem{weinberg_2015}
S.~Weinberg, {\em {Lectures on Quantum Mechanics}}.
\newblock Cambridge University Press, Cambridge, 2~ed., 2015.

\bibitem{Weinberg:1965nx}
S.~Weinberg, {\it {Infrared photons and gravitons}},  {\em Phys. Rev.} {\bf
  140} (1965) B516--B524.

\bibitem{Chang:2025cxc}
C.-H. Chang and J.~Parra-Martinez, {\it {Graviton loops and negativity}},  {\em
  JHEP} {\bf 08} (2025) 175, [\href{http://arxiv.org/abs/2501.17949}{{\tt
  arXiv:2501.17949}}].

\bibitem{FuentesZamoro:2025exp}
M.~Fuentes~Zamoro, B.~Grinstein, and P.~Qu{\'\i}lez, {\it {Taming forward
  scattering singularities in partial waves}},
  \href{http://arxiv.org/abs/2510.08784}{{\tt arXiv:2510.08784}}.

\bibitem{bellazzini2026positivitylongrangeinteractions}
B.~Bellazzini, J.~Berman, G.~Isabella, F.~Riva, M.~Romano, and F.~Sciotti, {\it
  Positivity with long-range interactions},  2026.

\bibitem{Henriksson_2022}
J.~Henriksson, B.~McPeak, F.~Russo, and A.~Vichi, {\it Bounding violations of
  the weak gravity conjecture},  {\em Journal of High Energy Physics} {\bf
  2022} (Aug., 2022).

\bibitem{Alberte:2020bdz}
L.~Alberte, C.~de~Rham, S.~Jaitly, and A.~J. Tolley, {\it {QED positivity
  bounds}},  {\em Phys. Rev. D} {\bf 103} (2021), no.~12 125020,
  [\href{http://arxiv.org/abs/2012.05798}{{\tt arXiv:2012.05798}}].

\bibitem{Alberte_2020}
L.~Alberte, C.~de~Rham, S.~Jaitly, and A.~J. Tolley, {\it Positivity bounds and
  the massless spin-2 pole},  {\em Physical Review D} {\bf 102} (Dec., 2020).

\bibitem{Caron-Huot:2021enk}
S.~Caron-Huot, D.~Mazac, L.~Rastelli, and D.~Simmons-Duffin, {\it {AdS bulk
  locality from sharp CFT bounds}},  {\em JHEP} {\bf 11} (2021) 164,
  [\href{http://arxiv.org/abs/2106.10274}{{\tt arXiv:2106.10274}}].

\bibitem{Bellazzini:2019xts}
B.~Bellazzini, M.~Lewandowski, and J.~Serra, {\it {Positivity of Amplitudes,
  Weak Gravity Conjecture, and Modified Gravity}},  {\em Phys. Rev. Lett.} {\bf
  123} (2019), no.~25 251103, [\href{http://arxiv.org/abs/1902.03250}{{\tt
  arXiv:1902.03250}}].

\bibitem{blas2021unitarizationinfiniterangeforcesgravitongraviton}
D.~Blas, J.~Mart{\'i}n~Camalich, and J.~A. Oller, {\it Unitarization of
  infinite-range forces: graviton-graviton scattering},  {\em JHEP} {\bf 08}
  (2022) 266, [\href{http://arxiv.org/abs/2010.12459}{{\tt arXiv:2010.12459}}].

\bibitem{Oller:2022Coulomb}
J.~A. Oller, {\it Unitarizing non-relativistic {Coulomb} scattering},  {\em
  Phys. Lett. B} {\bf 835} (2022) 137568,
  [\href{http://arxiv.org/abs/2207.08784}{{\tt arXiv:2207.08784}}].

\bibitem{Oller:2022Review}
J.~A. Oller, {\it Unitarizing infinite-range forces: Graviton-graviton
  scattering, the graviball, and {Coulomb} scattering},  {\em EPJ Web Conf.}
  {\bf 274} (2022) 08011, [\href{http://arxiv.org/abs/2211.02084}{{\tt
  arXiv:2211.02084}}].

\bibitem{plestid2026partialwaveunitaritylongrangeinteractions}
R.~Plestid and P.~Q. Lasanta, {\it Partial-wave unitarity and long-range
  interactions},  2026.

\bibitem{Taylor}
J.~R. Taylor, {\it A new rigorous approach to coulomb scattering},  {\em Il
  Nuovo Cimento B (1971-1996)} {\bf 23} (1974), no.~2 313--334.

\bibitem{Landau:1991wop}
L.~D. Landau and E.~M. Lifshitz, {\em {Quantum Mechanics}: {Non-Relativistic
  Theory}}, vol.~v.3 of {\em Course of Theoretical Physics}.
\newblock Butterworth-Heinemann, Oxford, 1991.
\newblock See sections 36, 135, 136, 137.

\bibitem{Eden:1966dnq}
R.~J. Eden, P.~V. Landshoff, D.~I. Olive, and J.~C. Polkinghorne, {\em {The
  analytic S-matrix}}.
\newblock Cambridge Univ. Press, Cambridge, 1966.

\bibitem{Zwanziger}
D.~Zwanziger, {\it Reduction formulas for charged particles and coherent states
  in quantum electrodynamics},  {\em Phys. Rev. D} {\bf 7} (Feb, 1973)
  1082--1099.

\bibitem{dolan2012conformalpartialwavesmathematical}
F.~A. Dolan and H.~Osborn, {\it Conformal partial waves: Further mathematical
  results},  \href{http://arxiv.org/abs/1108.6194}{{\tt arXiv:1108.6194}}.

\bibitem{Symanzik:1972wj}
K.~Symanzik, {\it {On Calculations in conformal invariant field theories}},
  {\em Lett. Nuovo Cim.} {\bf 3} (1972) 734--738.

\bibitem{guevara2021celestialopeblocks}
A.~Guevara, {\it Celestial {OPE} blocks},
  \href{http://arxiv.org/abs/2108.12706}{{\tt arXiv:2108.12706}}.

\bibitem{gradshteyn2014table}
I.~S. Gradshteyn and I.~M. Ryzhik, {\em Table of integrals, series, and
  products}.
\newblock Academic press, 2014.

\bibitem{kang1962higher}
I.-J. Kang and L.~M. Brown, {\it Higher born approximations for the coulomb
  scattering of a spinless particle},  {\em Physical Review} {\bf 128} (1962),
  no.~6 2828.

\bibitem{pilkuhn2013relativistic}
H.~M. Pilkuhn, {\em Relativistic particle physics}.
\newblock Springer Science \& Business Media, 2013.

\bibitem{EssinGriffiths2006}
A.~M. Essin and D.~J. Griffiths, {\it Quantum mechanics of the {$1/x^2$}
  potential},  {\em American Journal of Physics} {\bf 74} (2006), no.~2
  109--117.

\bibitem{BawinCoon2003}
M.~Bawin and S.~A. Coon, {\it Singular inverse square potential, limit cycles,
  and self-adjoint extensions},  {\em Physical Review A} {\bf 67} (2003)
  042712, [\href{http://arxiv.org/abs/quant-ph/0302199}{{\tt
  quant-ph/0302199}}].

\bibitem{BarfordBirse2005}
T.~Barford and M.~C. Birse, {\it {Effective theories of scattering with an
  attractive inverse-square potential and the three-body problem}},  {\em J.
  Phys. A} {\bf 38} (2005) 697--720,
  [\href{http://arxiv.org/abs/nucl-th/0406008}{{\tt nucl-th/0406008}}].

\bibitem{Case1950}
K.~M. Case, {\it Singular potentials},  {\em Physical Review} {\bf 80} (1950),
  no.~5 797--806.

\bibitem{Zeldovich:1972SuperheavyAtoms}
Y.~B. Zel'dovich and V.~S. Popov, {\it Electronic structure of superheavy
  atoms},  {\em Soviet Physics Uspekhi} {\bf 14} (1972), no.~6 673--694.

\bibitem{Pieper:1969InteriorShells}
W.~Pieper and W.~Greiner, {\it Interior electron shells in superheavy nuclei},
  {\em Zeitschrift f{\"u}r Physik A} {\bf 218} (1969) 327--340.

\bibitem{GreinerMuellerRafelski1985}
W.~Greiner, B.~M{\"u}ller, and J.~Rafelski, {\em Quantum Electrodynamics of
  Strong Fields}.
\newblock Springer, Berlin, 1985.

\bibitem{bawin1981instability}
M.~Bawin and J.~Cugnon, {\it Instability of point nuclei},  {\em Physics
  Letters B} {\bf 107} (1981), no.~4 257--258.

\bibitem{HammerSwingle2006}
H.-W. Hammer and B.~G. Swingle, {\it On the limit cycle for the {$1/r^2$}
  potential in momentum space},  {\em Annals of Physics} {\bf 321} (2006),
  no.~2 306--317, [\href{http://arxiv.org/abs/quant-ph/0503074}{{\tt
  quant-ph/0503074}}].

\bibitem{BarfordBirse2003}
T.~Barford and M.~C. Birse, {\it {A renormalisation group approach to two-body
  scattering in the presence of long-range forces}},  {\em Phys. Rev. C} {\bf
  67} (2003) 064006, [\href{http://arxiv.org/abs/hep-ph/0206146}{{\tt
  hep-ph/0206146}}].

\bibitem{Long_2008}
B.~Long and U.~van Kolck, {\it Renormalization of singular potentials and power
  counting},  {\em Annals of Physics} {\bf 323} (June, 2008) 1304--1323.

\bibitem{Mukunda:1978CoulombCompleteness}
N.~Mukunda, {\it Completeness of the {Coulomb} wave functions in quantum
  mechanics},  {\em American Journal of Physics} {\bf 46} (1978), no.~9
  910--913.

\bibitem{Mukhamedzhanov:2008CoulombCompleteness}
A.~M. Mukhamedzhanov and M.~Akin, {\it Completeness of the {Coulomb} scattering
  wave functions},  {\em European Physical Journal A} {\bf 37} (2008) 185--192,
  [\href{http://arxiv.org/abs/nucl-th/0602006}{{\tt nucl-th/0602006}}].

\bibitem{michel2008direct}
N.~Michel, {\it Direct demonstration of the completeness of the eigenstates of
  the schr{\"o}dinger equation with local and nonlocal potentials bearing a
  coulomb tail},  {\em Journal of mathematical physics} {\bf 49} (2008), no.~2.

\bibitem{Kabat_1992}
D.~Kabat and M.~Ortiz, {\it Eikonal quantum gravity and planckian scattering},
  {\em Nuclear Physics B} {\bf 388} (dec, 1992) 570--592.

\bibitem{Lee:1977eg}
B.~W. Lee, C.~Quigg, and H.~B. Thacker, {\it {Weak Interactions at Very
  High-Energies: The Role of the Higgs Boson Mass}},  {\em Phys. Rev. D} {\bf
  16} (1977) 1519.

\bibitem{Bagger:1995mk}
J.~Bagger, V.~D. Barger, K.-m. Cheung, J.~F. Gunion, T.~Han, G.~A. Ladinsky,
  R.~Rosenfeld, and C.-P. Yuan, {\it {CERN LHC Analysis of the Strongly
  Interacting $WW$ System: Gold-Plated Modes}},  {\em Phys. Rev. D} {\bf 52}
  (1995) 3878--3889, [\href{http://arxiv.org/abs/hep-ph/9504426}{{\tt
  hep-ph/9504426}}].

\bibitem{Todorov:1970gr}
I.~T. Todorov, {\it {Quasipotential equation corresponding to the relativistic
  eikonal approximation}},  {\em Phys. Rev. D} {\bf 3} (1971) 2351--2356.

\bibitem{Correia:2024}
M.~Correia and G.~Isabella, {\it {The Born regime of gravitational
  amplitudes}},  {\em JHEP} {\bf 03} (2025) 144,
  [\href{http://arxiv.org/abs/2406.13737}{{\tt arXiv:2406.13737}}].

\bibitem{Huber:2007dx}
T.~Huber and D.~Ma{\^\i}tre, {\it {HypExp 2, Expanding Hypergeometric Functions
  about Half-Integer Parameters}},  {\em Comput. Phys. Commun.} {\bf 178}
  (2008) 755--776, [\href{http://arxiv.org/abs/0708.2443}{{\tt
  arXiv:0708.2443}}].

\end{thebibliography}
